\documentclass[]{aastex62}
\usepackage{chemformula}
\usepackage{gensymb}
\let\ce\ch
\usepackage[caption=false]{subfig}
\usepackage{hyperref}
\definecolor{medium-blue}{rgb}{0,0,1}
\hypersetup{colorlinks, urlcolor={medium-blue}}

\received{December 18 2020}
\revised{April 13 2021}
\accepted{April 13 2021}
\published{June 8 2021}
\submitjournal{Astronomical Journal}
\shorttitle{Ultraviolet Mapping of Cometary Activity from 67P/Churyumov-Gerasimenko}
\shortauthors{Noonan et al.}
\begin{document}
\title{Spatial Distribution of Ultraviolet Emission from Cometary Activity at 67P/Churyumov-Gerasimenko}
\author[0000-0003-2152-6987]{John W. Noonan}
\affiliation{Department of Space Studies,Southwest Research Institute, Suite 300, 1050 Walnut Street,Boulder, Colorado 80302,USA}
\affiliation{Lunar and Planetary Laboratory,University of Arizona, 1629 E University Blvd, Tucson, Arizona 85721-0092, USA }

\author{Dominique Bockel\'ee-Morvan}
\affiliation{LESIA, Observatoire de Paris, Universit\'e PSL, CNRS, Sorbonne Universit\'e, Universit\'e de Paris, 5 place Jules Janssen, 92195 Meudon, France}

\author[0000-0002-9318-259X]{Paul D. Feldman}
\affiliation{Department of Physics and Astronomy, The Johns Hopkins University, 3400 N. Charles Street, Baltimore, Maryland 21218, USA }

\author[0000-0001-5018-7537]{S. Alan Stern}
\affiliation{Department of Space Studies, Southwest Research Institute, Suite 300, 1050 Walnut Street, Boulder, Colorado 80302,USA}
 
 \author[0000-0003-0797-5313]{Brian A. Keeney}
\affiliation{Department of Space Studies, Southwest Research Institute, Suite 300, 1050 Walnut Street, Boulder, Colorado 80302,USA}
 
\author[0000-0002-3672-0603]{Joel Wm. Parker}
\affiliation{Department of Space Studies, Southwest Research Institute, Suite 300, 1050 Walnut Street, Boulder, Colorado 80302,USA}
 
\author{Nicolas Biver}
\affiliation{LESIA, Observatoire de Paris, Universit\'e PSL, CNRS, Sorbonne Universit\'e, Universit\'e de Paris, 5 place Jules Janssen, 92195 Meudon, France}
 
\author[0000-0003-2781-6897]{Matthew M. Knight}
\affiliation{Astronomy Department, University of Maryland, College Park, Maryland 20742, USA}
\affiliation{Department of Physics, United States Naval Academy, 572C Holloway Rd, Annapolis MD 21402, USA}

\author[0000-0002-4230-6759]{Lori M. Feaga}
\affiliation{Astronomy Department, University of Maryland, College Park, Maryland 20742, USA}

\author{Mark D. Hofstadter}
\affiliation{Jet Propulsion Laboratory, California Institute of Technology, Pasadena, CA 91109, USA}

\author{Seungwon Lee}
\affiliation{Jet Propulsion Laboratory, California Institute of Technology, Pasadena, CA 91109, USA}

\author[0000-0002-8227-9564]{Ronald J. Vervack Jr.}
\affiliation{Johns Hopkins University Applied Physics Laboratory, 11100 Johns Hopkins Road, Laurel, Maryland 20723-6099,USA}

\author[0000-0002-5358-392X]{Andrew J. Steffl}
\affiliation{Department of Space Studies, Southwest Research Institute, Suite 300, 1050 Walnut Street, Boulder, Colorado 80302,USA}

\author{Rebecca N. Schindhelm}
\affiliation{Department of Space Studies, Southwest Research Institute, Suite 300, 1050 Walnut Street, Boulder, Colorado 80302,USA}
\affiliation{Ball Aerospace and Technology Corp., 1600 Commerce Street, Boulder, CO 80301, USA}

\author{Jon Pineau}
\affiliation{Stellar Solutions, Inc., Palo Alto, California 94306}

\author{Richard Medina}
\affiliation{Department of Space Operations, Southwest Research Institute, Suite 300, 1050 Walnut Street, Boulder, Colorado 80302,USA}

\author[0000-0003-0951-7762]{Harold A. Weaver}
\affiliation{Johns Hopkins University Applied Physics Laboratory, 11100 Johns Hopkins Road, Laurel, Maryland 20723-6099,USA}

\author{Jean-Loup Bertaux}
\affiliation{LATMOS, CNRS/UVSQ/IPSL, 11 Boulevard d'Alembert, 78280 Guyancourt, France}

\author{Michael F. A'Hearn}
\altaffiliation{Deceased}
\affiliation{Astronomy Department, University of Maryland, College Park, Maryland 20742, USA}

\correspondingauthor{John Noonan}
\email{noonan@lpl.arizona.edu}

\begin{abstract}
 The Alice ultraviolet spectrograph on board the \textit{Rosetta} orbiter provided the first near-nucleus ultraviolet observations of a cometary coma from arrival at comet 67P/Churyumov-Gerasimenko in 2014 August through 2016 September.  The characterization of atomic and molecular emissions in the coma revealed the unexpected contribution of dissociative electron impact emission at large heliocentric distances and during some outbursts. This mechanism also proved useful for compositional analysis, and Alice observed many cases that suggested elevated levels of the supervolatile \ce{O2}, identifiable in part to their emissions resulting from dissociative electron impact. In this paper we present the first two-dimensional UV maps constructed from Alice observations of atomic emission from 67P during an increase in cometary activity on 2015 November 7-8. Comparisons to observations of background coma and of an earlier collimated jet are used to describe possible changes to the near-nucleus coma and plasma. To verify the mapping method and place the Alice observations in context, comparisons to images derived from the MIRO and VIRTIS-H instruments are made. The spectra and maps we present show an increase in dissociative electron impact emission and an \ce{O2}/\ce{H2O} ratio of $\sim$0.3 for the activity; these characteristics have been previously identified with cometary outbursts seen in Alice data. Further, UV maps following the increases in activity show the spatial extent and emission variation experienced by the near-nucleus coma, informing future UV observations of comets that lack the same spatial resolution. 

\end{abstract}

\section{Introduction}\label{Intro}
From August of 2014 through September of 2016 the European Space Agency's \textit{Rosetta} spacecraft performed escort operations around the comet 67P/Churyumov-Gerasimenko, observing changes to the comet's nucleus, coma, and plasma environment at a range of heliocentric and comet-centric distances. The comprehensive survey carried out by \textit{Rosetta} instruments has provided valuable insight into comet outgassing and outbursts. In particular, observations by the Alice ultraviolet (UV) spectrograph \citep{stern2007alice} revealed the prevalence of dissociative electron impact emission at heliocentric distances greater than 2 au \citep{feldman2018fuv,galand2020far,stephenson2021multi} that was also observed by the OSIRIS instrument \citep{bodewits2016changes}, and that observations of outbursts with different local origins, observing geometries, outgassing rates, and compositions displayed substantial increases in dissociative electron impact emission \citep{feldman2016nature,Noonan2021hybrid}.

These transient events highlight the variable nature of comets and provide insight into how the near-nucleus coma is affected by injections of gas within a short period of time. Given the consistent presence of dissociative electron impact emission near the nucleus at the larger heliocentric distances observed by \textit{Rosetta} and during outburst \citep{feldman2016nature,feldman2018fuv,Noonan2021hybrid}, it is critical to understand how perturbations to the near-nucleus coma affect the electron impact emission features and overall UV spectrum. Previous studies have shown a one dimensional spatial profile along the Alice slit, detailing the sunward and anti-sunward asymmetry of emission features and the radial extent for dissociative electron impact emission during outbursts \citep{feldman2016nature}. Due to the typical stare-type observations of Alice these one dimensional spatial profiles were the best representation of the spatial behavior of coma emission features. 

However, inner coma mapping has been implemented with both the Visible InfraRed Thermal Imaging Spectrometer (VIRTIS) \citep{coradini2007virtis} and Microwave Instrument on the Rosetta Orbiter (MIRO) \citep{gulkis2007miro}. \cite{migliorini2016water} used VIRTIS to simultaneously map the \ce{H2O} and \ce{CO2} column densities in April of 2015, while \cite{fink2016investigation} obtained \ce{H2O} and \ce{CO2} production rates for a period in February 28 as well as April 27 of 2015.  Outbursts that occurred around the comet's perihelion in August 2015 were mapped by \cite{rinaldi2018summer} using VIRTIS high resolution data, showing an increase in dust production and change of dust color. The MIRO instrument implemented a raster technique to measure the distribution of abundant cometary molecules, including H$_{2}^{18}$O at 3.4 au \citep{biver2015distribution}, when activity levels were low due to the relatively large heliocentric distance. The same technique was extended to other isotopologues of water, \ce{CH3OH}, \ce{NH3}, and \ce{CO} in over 100 maps to determine production rates and radial dependencies \citep{Biver2019}. 

Using coordinated pointings we have created atomic emission maps and compare Alice UV maps to calibrated maps from VIRTIS and MIRO. The additional spatial dimension from this technique provides detail on the influence of outbursts on the near-nucleus coma and plasma environment, and in particular how the spatial distribution of dissociative electron impact changes during the period. Given the expected gap in time between the \textit{Rosetta} mission and the next UV-enabled comet spacecraft, it is of critical importance to understand what processes could be observed from Earth-based observatories under optimal conditions, especially if a spectroscopic identifier for small outbursts can be determined in Alice data. The relationship between UV emissions and the cometary plasma environment is better characterized now thanks to \textit{Rosetta} \citep{galand2020far,stephenson2021multi}, but is still far from completely understood. The work of \cite{galand2020far} and \cite{stephenson2021multi} showed that for comet/spacecraft distances less than 100 km the observed Alice UV brightness near the nucleus could be well modeled with in-situ electron population measurements from the Rosetta Plasma Consortium Ion and Electron Spectrograph (RPC-IES) \citep{burch2007rpc}, water column density measurements from MIRO or VIRTIS, and abundances of \ce{CO2}, \ce{O2}, and \ce{CO} relative to water from the ROSINA mass spectrometer \citep[]{balsiger2007rosina}. Critically, \citet{galand2020far} found that solar wind electrons accelerated along the ambipolar electric field of the nucleus were responsible for generating these FUV emissions observed by Alice at large heliocentric distances, making them auroral in nature. If electron impact is detected in future UV comet observations then it may be possible to not only continue compositional characterization via atomic emission lines but expand into remote plasma characterization of excitation regions of the inner coma with improved modeling of dissociative electron impact excitation \citep{galand2020far,stephenson2021multi}. 

This paper discusses a unique period of activity that occurred while the Alice instrument made ``ride-along'' observations during VIRTIS-led pointings. The VIRTIS and MIRO instruments mapped the distribution of \ce{H2O} and \ce{CO2} in the coma using raster scans at this time, providing context maps to compare to the new Alice maps. Furthermore, analysis of these maps shows evidence of an expanding zone of influence where electron impact dominates over fluorescence for short time scales, even at a relatively low heliocentric distance. This paper reviews data taken between 21:16 UTC on 2015 November 7 and 02:55 UTC on November 8 and is intended to be a companion paper to \citet{Noonan2021hybrid} which reviews data taken on November 7 between 12:00 UTC and 19:18 UTC.  Section 2 will discuss the instrument, the relevant observations, and their geometry. Section 3 details the method used to map the Alice data. Section 4 describes the three data products that can be derived from Alice data for this time period: spectra, spatial profiles, and 2-D maps. Maps from the VIRTIS-H and MIRO instruments are presented for comparison. Section 5 then discusses the morphology, composition, and excitation processes of the activity and their implications. A summary is presented in Section 6. 

\section{Observations}\label{Obs}

\subsection{Instrument Description}\label{Instrument}

The Alice instrument was a light-weight and low-power imaging spectrograph on-board the \textit{Rosetta} spacecraft tasked with characterizing the surface properties, coma composition, and coupling of the coma and nucleus. To fulfill these goals, the instrument was designed to be sensitive from 700 - 2050 \AA\ with a filled-slit spectral resolution determined in flight to be 11 \AA\ in the center of the ``dog bone" shaped slit, which was narrower in the middle than on the bottom or top. The narrow center is $0.05^{\circ}$ (100 $\mu$m) wide in the middle $2.0^{\circ}$ of the slit, while the upper and lower portions are both $0.10^{\circ}$ (210 $\mu$m) wide. The lower and middle sections of the slit measure 2.0$\degree$ long, with the upper section measuring 1.53$\degree$. The Alice detector was a microchannel plate with 1024 columns in the spectral dimension and 32 rows in the spatial dimension. Of the 32 spatial rows, only rows 5 through 23 (zero indexed) were exposed to light from the slit. Each row subtended $0.30^{\circ}$ on the sky. One detector effect that can affect results is the odd/even effect, which was the tendency of the Alice detector to push counts to odd rows over even rows \citep{feldman2011rosetta,chaufray2017rosetta}. The instrument is described in detail in \citet{stern2007alice}.

\subsection{Geometry and Spacecraft Pointing}\label{geometry}
On 2015 November 7-8, \textit{Rosetta} and 67P were 1.61 au away from the Sun, outbound in their orbit after passing through perihelion in August 2015. The first observations used in this analysis were taken when \textit{Rosetta} was 240 km from the nucleus of 67P on 15:04 UTC November 7. The spacecraft then slowly approached 67P, reaching as close as 215 km by the end of the VIRTIS raster observations at 10:30 UTC on November 8. The phase angle decreased from 64$\degree$ to 61.5$\degree$ over the same time period. More information on the pointing geometry of each block is presented in Table \ref{table:observations}, and further information on Alice observation modes can be found in \cite{pineau2018flight}.

\subsubsection{Stable Pointing}\label{Stable}
\begin{figure}
\centering
\includegraphics[angle=-90,width=0.5\linewidth]{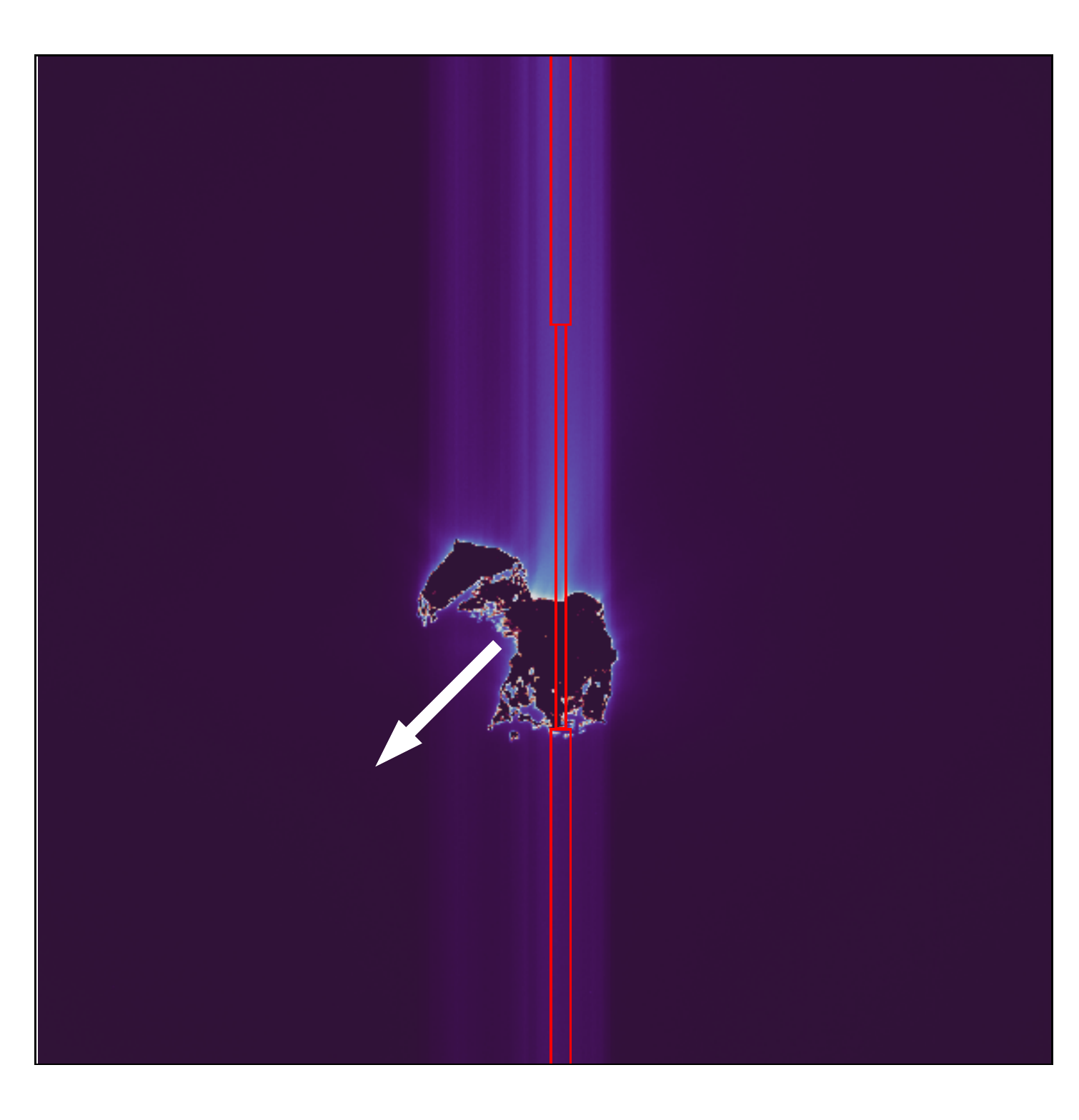}
\caption{NAVCAM image from November 7 at 19:18 UTC with nucleus pixels masked to highlight faint activity.The rotation axis of 67P is marked with a white arrow. The Alice slit is overlaid in red and the Sun is to the right. With the exception of faint jets emanating from the neck there is no evidence of background activity.}
\label{fig:NAVCAM_Jet}
\label{fig:11_07_3dT}
\end{figure}

From 15:04 UTC until 19:18 UTC on November 7 the Alice instrument was in a stable pointing mode, where there was minimal motion of the instrument's line of sight for a long period of time. At this time the Alice slit was centered on the nucleus of 67P, with the upper rows in the sunward direction and the lower rows in the anti-sunward direction (Figure \ref{fig:11_07_3dT}). From this pointing geometry Alice observed an increase in activity starting at 16:00 UTC that was recorded until the end of scheduled observations at 19:18 UTC. At this time Alice observations ceased until the ride-along observations. A more complete analysis of this period of interest is presented in \citet{Noonan2021hybrid} but several individual observations from this period are referenced here. 

\subsubsection{Raster Pointing}\label{Raster}
Following an approximately 1.5 hour gap in observations, the Alice instrument resumed exposures during a VIRTIS-driven pointing mode designed to map the inner coma of 67P. The raster scans began at approximately 20:45 UTC November 7 and continued until 10:30 UTC on November 8, with Alice observations starting at 21:16 UTC November 7. The scan rate of the Alice slit during the raster varies throughout the period, so for the 300 s exposures the area scanned by the slit varies from 0.1 to 1.4 degrees (0.38-5.32 km) with an average of 0.6 degrees (2.28 km). This limits the spatial resolution of the Alice data, which is affected by ``smearing", which can spread the signal from one area of the coma over 2-3 spatial pixels depending on the individual observations scan rate. These inner coma raster scans for this particular time period moved the Alice boresight, centered in row 15 of the detector, within 5 km of the nucleus center in both the X and Y spacecraft directions. 

\section{Mapping Methods}\label{Mapping}
\begin{figure}
  \centering
   \includegraphics[width=1.0\linewidth]{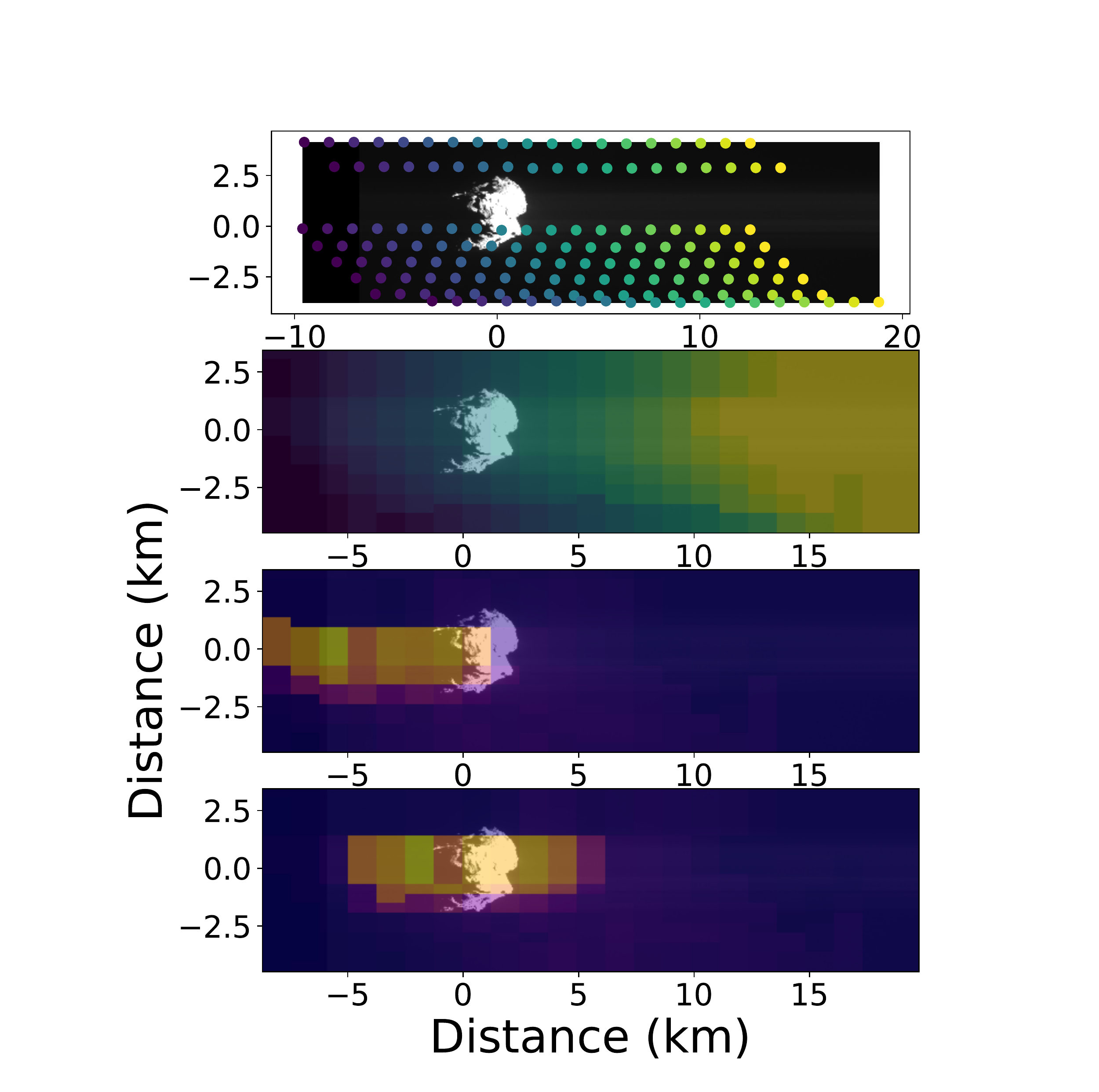}
  \caption{Series of plots depicting the mapping method described in Section \ref{Mapping}. The top figure depicts the average location of each row of the Alice detector for each observation in block 1 from Table \ref{table:observations}, with the color scheme detailing the row number (row 5 is dark purple, on the left, row 23 is yellow, on the right). The figure second from the top shows the interpolation of the rows to a grid with a pixel size defined by the pixel size of each row and the scan rate of the Alice detector. The figure third from the top details the initial map that is generated by the program before any correction is applied to shift the measured reflected solar signal to center on the nucleus. The bottom figure shows the solar reflectance map for block 1 after the weighted barycenter shift has been applied. The Sun is to the right ($+$X) in this and all maps presented.}
    \label{fig:pixel_map}
\end{figure}

We developed processing software to create emission maps from the Alice spectral images taken during raster ride-along observations driven by VIRTIS and MIRO. Observations from the period between 20:45 UTC on November 7 and 05:18 UTC November 8 were split into groups of ten where possible, further detailed in Table \ref{table:observations}. Groups of roughly ten are chosen where possible to balance temporal resolution ($\sim$50 minutes) with spatial coverage of the inner coma (Figure \ref{fig:pixel_map}). For each individual observation within a block, a Python routine is used to iterate through each row of each observation, to filter for the presence of any stellar continuum or presence of the anomalistic ``Chameleon" feature that may contaminate the measurements \citep{noonan2016investigation}. Following this check the brightnesses of four strong diagnostic atomic emission features at 67P are determined: Lyman-$\beta$, the \ion{O}{1} 1304 triplet, \ion{O}{1}] 1356 \AA\, and \ion{C}{1} 1657 \AA. This is done by integrating the flux in the 3-$\sigma$ range of each emission feature's center wavelength, assuming FWHM of 11 \AA\ for each feature. We note that for \ion{C}{1} 1657 \AA\ emission this would lead to a blending with the CO Fourth Positive Band emission at 1653 \AA\ (Figure \ref{fig:stable_spectra}). However the abundance of CO at this time is not significant enough to contaminate the atomic carbon substantially; based on the strength of the CO Fourth Positive (0-1) emission feature at 1600 \AA\, which is approximately equivalent to the strength of the CO Fourth Positive (0-2) emission at 1653 \AA\, we would expect less than 5\% of the integrated flux for \ion{C}{1} 1657 \AA\ to be attributable to CO Fourth Positive Group emissions. Each row of the Alice spectral image, which corresponds to a different spatial coordinate relative to the nucleus at the start and end of each observation, is tagged with the X/Y pointing of the row in spacecraft coordinates from mid-exposure taken from the FITS headers, and the brightness is calculated in Rayleighs. Based on the SPICE kernels of the reconstructed spacecraft trajectory, pointing, and scan rate the size that an Alice pixel subtends at the nucleus is calculated for the middle of each exposure. This allows a dataset of coordinates, brightnesses, and the observation's pixel height and width at the nucleus to be generated for each observing block. 

Due to the scan motion care must be taken to correctly account for the true location of the nucleus in the maps compared to the average position of the Alice rows. To accomplish this the weighted intensity barycenter of reflected solar continuum emissions between 1850 and 1950 \AA\ is computed to find the center of the nucleus in the Alice data in X,Y space. The barycenter is then shifted until it aligns with the center of the illuminated nucleus in the nearest NAVCAM image, which may or may not have been taken within the set of Alice observations that the map is derived from and thus have a slightly different geometry. This process is shown in Figure \ref{fig:pixel_map}. The FUV continuum also includes reflected sunlight off of dust near the nucleus, but manual inspection of each produced map reduces any effects of systematic nucleus offset between observing block maps. The large pixel size of the resulting images allows general activity to be distinguished but prevents fine structures from being resolved.

After the brightnesses for each row of every observation in an observing block are obtained, the individual row brightnesses are mapped to their X/Y locations relative to the comet nucleus. This array of points is then interpolated to generate a map of each emission feature, with pixel sizes no smaller than the Alice detector pixel subtended at the largest cometocentric distance within the observing block plus the distance covered by the slit while scanning during the observation, assuming a constant scan rate for the five minute exposure.

\subsection{NAVCAM Correlation}
Of the three available NAVCAM images taken during the period in question only one has meaningful activity context information for the Alice data and is shown in Figure \ref{fig:NAVCAM_Jet}. In the NAVCAM image at 19:18 UTC, a series of jets can be clearly seen extending across the Alice slit. Comparing the activity, captured in projection, to the geomorphological areas presented in \cite{el2016regional}, it seems likely that the activity can be linked to solar illumination of the Geb, Neith, or Sobek regions, which are experiencing the highest intensity sunlight around this time. The NAVCAM image allows us to place the nearest spectrum taken by Alice, the 19:18 UTC spectrum plotted in Figure \ref{fig:stable_spectra}, into the context of southern hemisphere jet activity. The increase in overall emission that occurs between the last stable pointing spectrum at 19:18 UTC and the first raster observations a 21:16 UTC would suggest that a new jet or jets became active or an outburst occurred in the 87 minutes between the observations, altering both the observed direction and intensity of emission. Due to the uncertainty of the source of the newly introduced emissions, we will refer to this as an activity increase. 

\begin{table*}
\begin{center}
\begin{tabular}{c c c c c c c c}
\hline
Observing & Number & Time Range & Comet & Sub-Spacecraft & Sub-Spacecraft& Pointing \\
Block & of Observations & (UTC) &  Distance (km) & Latitude ($\degree$) & Longitude ($\degree$) & Scheme  \\
\hline
1 & 25 & 15:04-19:18 & $+$233.5 -- $+$228.8 & -2.29 -- $-$0.99 & 15.1 -- $-$109.0 & Stare \\
2 & 10 & 21:16-22:07 & $+$226.7 -- $+$225.8 & -0.36 -- $-$0.08  & -166.8 -- $+$168.2 & Raster \\
3 & 9 & 22:13-22:59 & $+$225.7 -- $+$225.0 & -0.05 -- $+$0.20  & $+$165.4 -- $+$142.9 & Raster\\
4 & 10 & 23:04-23:56 & $+$224.9 -- $+$224.0 & $+$0.23 -- $+$0.51 & $+$140.1 -- $+$114.9 & Raster\\
5 & 5 & 00:02-00:32 & $+$223.9 -- $+$223.6 & $+$0.54 --  $+$0.71 & $+$112.1 -- $+$97.0 & Raster\\
6 & 10 & 01:13-01:58 & $+$222.9 -- $+$222.1 & $+$0.94 -- $+$1.19 & $+$77.4 -- $+$55.2 & Raster\\
7 & 10 & 02:04-02:55 & $+$222.0 -- $+$221.1 & $+$1.22 -- $+$1.51 & $+$52.4 -- $+$27.1 & Raster\\
8 & 25 & 03:01-05:18 & $+$221.0 -- $+$218.9 & $+$1.55 -- $+$2.33 & $+$24.5 -- $-$42.8 & Raster \\ 
	\end{tabular}
\caption{ Alice observations taken 2015 November 7-8. A more detailed analysis of observations in observing block 1 can be found in \citet{Noonan2021hybrid}. Observing block 8 is only used for the light curve in Figure \ref{fig:time_vs_brightness}.}	
\label{table:observations}
\end{center}    
\end{table*}

\section{Results}
Due to the nature of the observing schemes, several different data products can be made to better understand the near-nucleus coma and the effects of the strengthened activity. The two observing schemes of November 7-8 provide spectra, spatial profiles, and maps to draw comparisons between activity periods and instrument maps. 

\subsection{Spectra}\label{Spectra}
Alice observations taken immediately prior to the mapping scheme, between 15:04 and 19:18 UTC (Table \ref{table:observations}), are stationary relative to the nucleus and provide a higher signal-to-noise ratio and little uncertainty in the pointing of the Alice boresight. These initial spectra yield no unique features compared to other previously published outbursts or activity \citep{feldman2016nature,feldman2018fuv} but spectra taken during this period display contribution from dissociative electron impact of the common volatiles \ce{H2O} and \ce{CO2}, as well as two outbursts containing \ce{O2}, evident from a \ion{O}{1}] 1356/\ion{O}{1} 1304 \AA\ ratio that is above 1. Analysis of these spectra is presented in \citet{Noonan2021hybrid}, and it is recommended that the reader begin with that article to familiarize themselves with the events prior to the raster observations. 

\begin{figure*}
\centering
\includegraphics[width=1.0\linewidth]{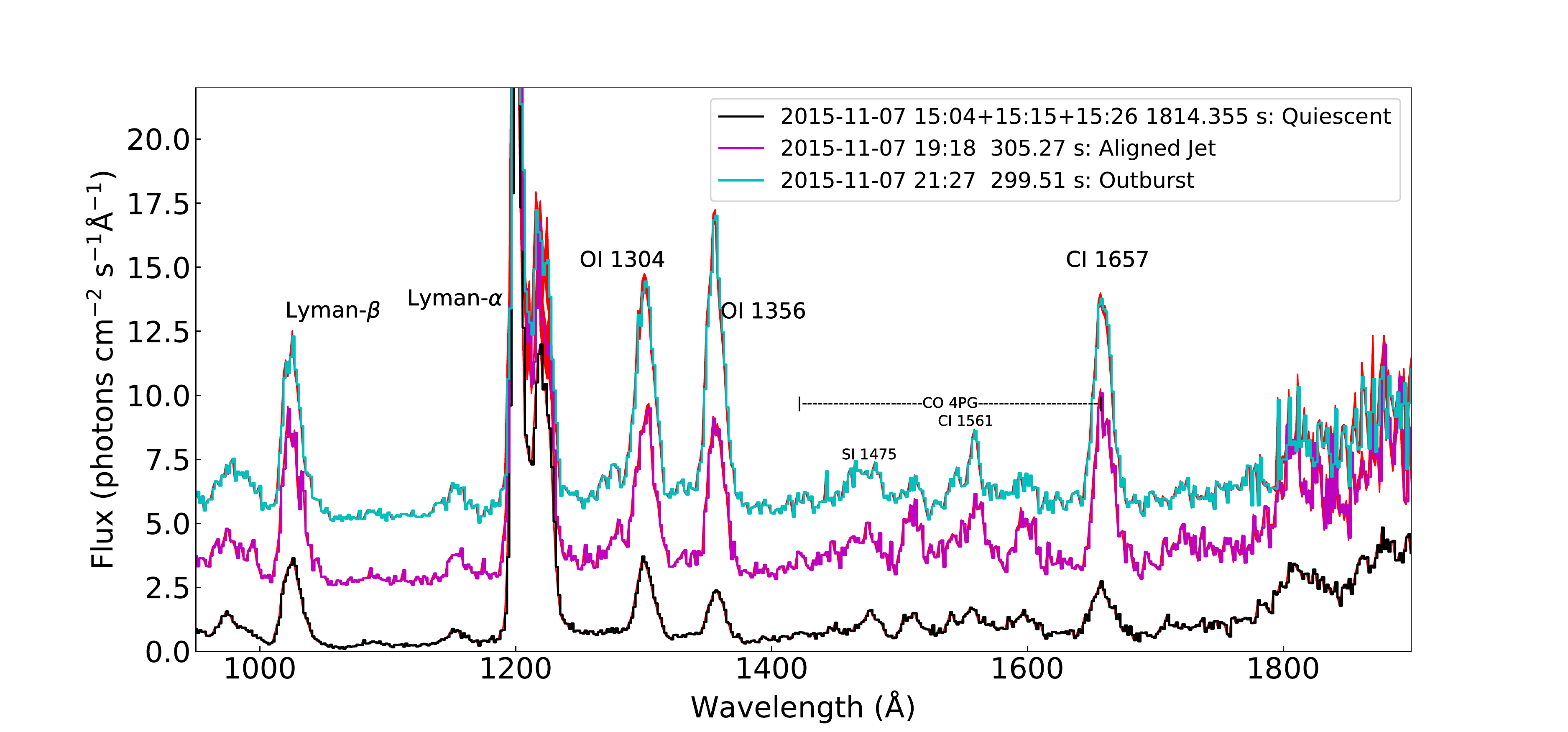} 
\caption{Stable pointing spectra from rows 18-23 taken during a quiescent period (black), a jet observation (magenta), and just following a large activity increase on November 7 (cyan). Flux error is plotted but is smaller than the plotted line width. Of particular interest is the \ion{O}{1}] 1356 and \ion{O}{1} 1304 ratio, indicative of dissociative electron impact. Spectra are offset by 3 photons cm$^{-2}$ s$^{-1}$ \AA$^{-1}$. Rows 18-23 are between 5 and 11 km in the sunward direction from the nucleus center for these times. There is also faint CO Fourth Positive emission in the 1400 \AA\ to 1600 \AA\ region mixed with emission from dissociative electron impact excitation of \ce{CO2}, discussed further in Section \ref{Discussion}. Note the non-Gaussian shape of Lyman-$\alpha$ due to gain sag of the Alice detector, rendering it inadequate for diagnostic purposes. Other notable features not analyzed in this paper are labeled with a smaller font. }
\label{fig:stable_spectra}
\end{figure*}

A quiescent spectrum is shown from an Alice observation taken at 15:04 UTC, the last time on November 7 prior to the activity detailed in \citet{Noonan2021hybrid} (see Fig. \ref{fig:stable_spectra}, 15:04 UTC spectrum). Atomic emissions from H, O, C and S are all clearly visible, evidence for photodissociation and dissociative electron impact on \ce{H2O} and \ce{CO2}. The \ion{O}{1}] 1356/\ion{O}{1} 1304 \AA\ ratio is less than 1, indicating a presence of e+\ce{H2O}/\ce{CO2}, but below the ratio $\geq$ 1 that would indicate substantial dissociative electron impact of \ce{O2}.  There also appears to be weak emission of the \ce{CO} Fourth Positive group between 1400 and 1600 \AA\, which is likely the result of resonance fluorescence and e+\ce{CO2} \citep{ajello1971dissociative,ajello1971emission,ajello2019uv}.

\begin{figure*}
\centering
\includegraphics[width=1.0\linewidth]{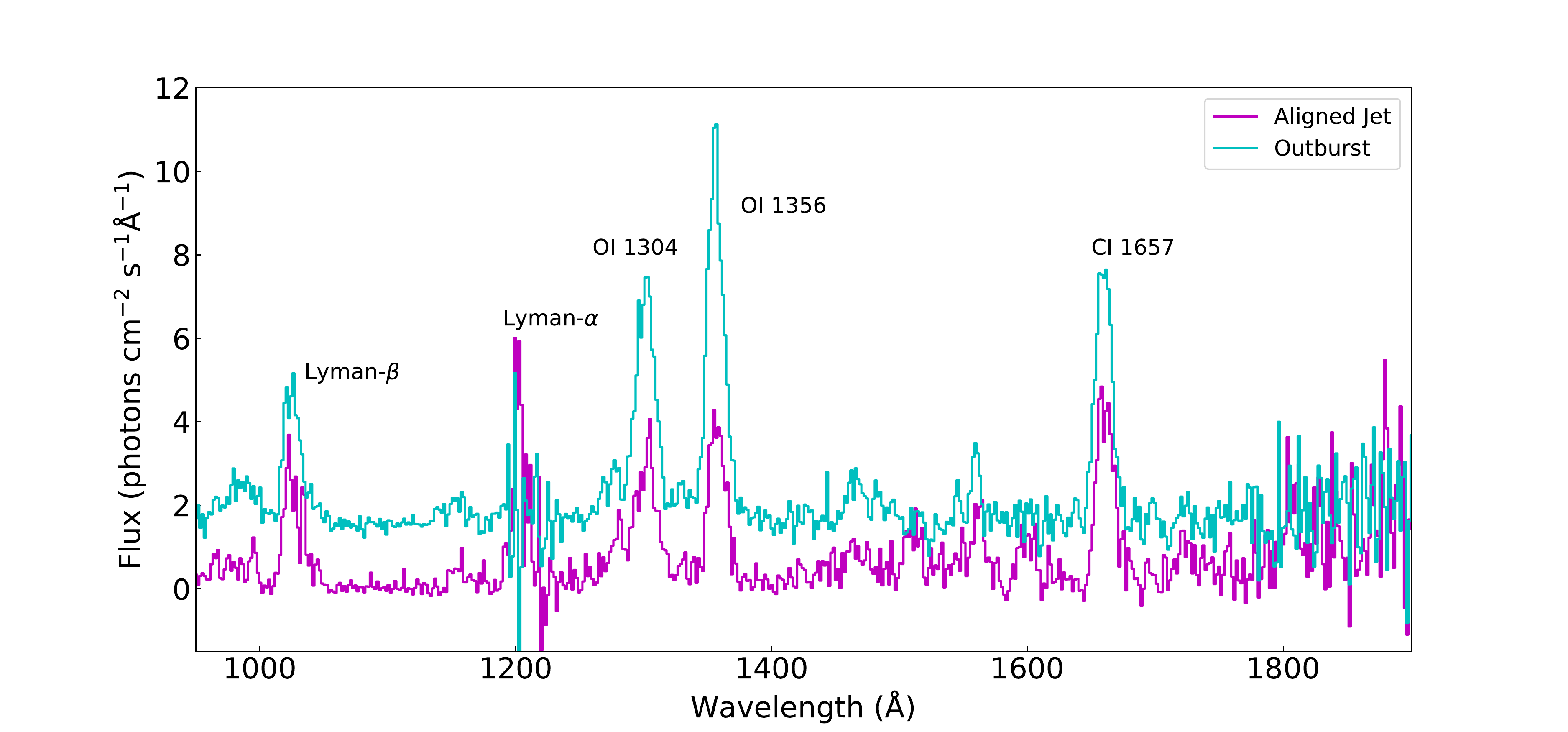} 
\caption{Difference spectra resulting from the subtraction of the 15:04 UTC spectrum from the 19:18 and 21:27 UTC spectra on November 7. Notice the substantial difference in \ion{O}{1}] 1356/\ion{O}{1} 1304 \AA\ emission, as well as the minimal contribution from the CO Fourth Positive group despite an increase in \ion{C}{1} emission. }
\label{fig:diff_spec}
\end{figure*}

For comparison to this early quiescent period, we have chosen two times that correspond to two unique points in the activity; the first corresponds to  emissions seen at 19:18 UTC on November 7, the second taken at 21:27 UTC, right after the start of Alice raster observations (See Figure \ref{fig:stable_spectra}). These spectra detail two unique activity types, allowing a spectral comparison of both cometary jets following outbursts and a later activity increase. To better compare the changes between the activity and quiescent observations a difference spectrum is produced for each exposure, seen in Figure \ref{fig:diff_spec}. Lyman-$\beta$ and the \ion{C}{1} 1561 and 1657 \AA\ features appear to show little change between the jets and activity increase observations, possibly indicating that the \ce{H2O} and carbon-bearing species maintained an elevated production rate for the hour and a half duration between the observations. The \ion{O}{1} 1304 and 1356 \AA\ emissions do not exhibit the same characteristics; the jet shows an \ion{O}{1}] 1356/\ion{O}{1} 1304 ratio of $\sim$1 while at the start of raster observations this ratio is greater than 1. This latter ratio is expected from electron impact dissociation of \ce{O2} \citep{kanik2003electron} which we will discuss further in Section \ref{Discussion}, and suggests that the jet active at 19:18 UTC is depleted of the super-volatile \ce{O2}.

There are several additional points to be made about these UV spectra that can complicate their analysis, which we outline here.  Evidence of \ion{S}{1} emission in Figure \ref{fig:stable_spectra} is present in each spectrum as a weak triplet at 1807, 1820, and 1826 \AA\ but not evident in the quiescent-subtracted spectra in Figure \ref{fig:diff_spec}. The presence of \ion{S}{1} emission in the 1800 \AA\ region suggests that there are weaker \ion{S}{1} multiplets in the 1473 and 1425 \AA\ regions as well (approximately in 1:2 and 2:5 ratios relative to the \ion{S}{1} 1807 \AA\ feature; see \citet{kaufman1982spectrum,roettger1989iue,meier1997atomic}), though likely blended with weak \ce{CO} Fourth Positive emission that appears to be present at a low level (\textless2 Rayleighs) in Figure \ref{fig:stable_spectra} and does not appear strongly in a difference spectrum (Figure \ref{fig:diff_spec}). The line ratios between the strong \ion{C}{1} multiplets at 1657 and 1561 \AA, and to an extent the weaker 1597 \AA\ CO Fourth Positive Group feature, are in disagreement with that expected
from pure dissociative electron impact on \ce{CO} or \ce{CO2} \citep{ajello1971dissociative,ajello1971emission, ajello2019uv}. We also note that Ly-$\beta$ is contaminated with \ion{O}{1} 1025.72 \AA\ emission, which is not resolved. The \ion{O}{1} contribution to the Ly-$\beta$ + \ion{O}{1} 1025.72 \AA\ blend can be estimated from e+\ce{O2} cross sections at 200 eV of \cite{ajello1985study} provided the \ce{O2} column density is known, though this electron energy is significantly above the expected mean in the near-nucleus environment \citep{clark2015suprathermal}. As shown in \citet{ajello1985study} the line ratio for the \ion{O}{1} 1356/1025.72 \AA\ features resulting from dissociative electron impact of \ce{O2} is approximately 35, so the e+\ce{O2} contribution to Ly-$\beta$ is assumed to be negligible and ignored.

\begin{figure}
\centering
\includegraphics[width=1.0\linewidth]{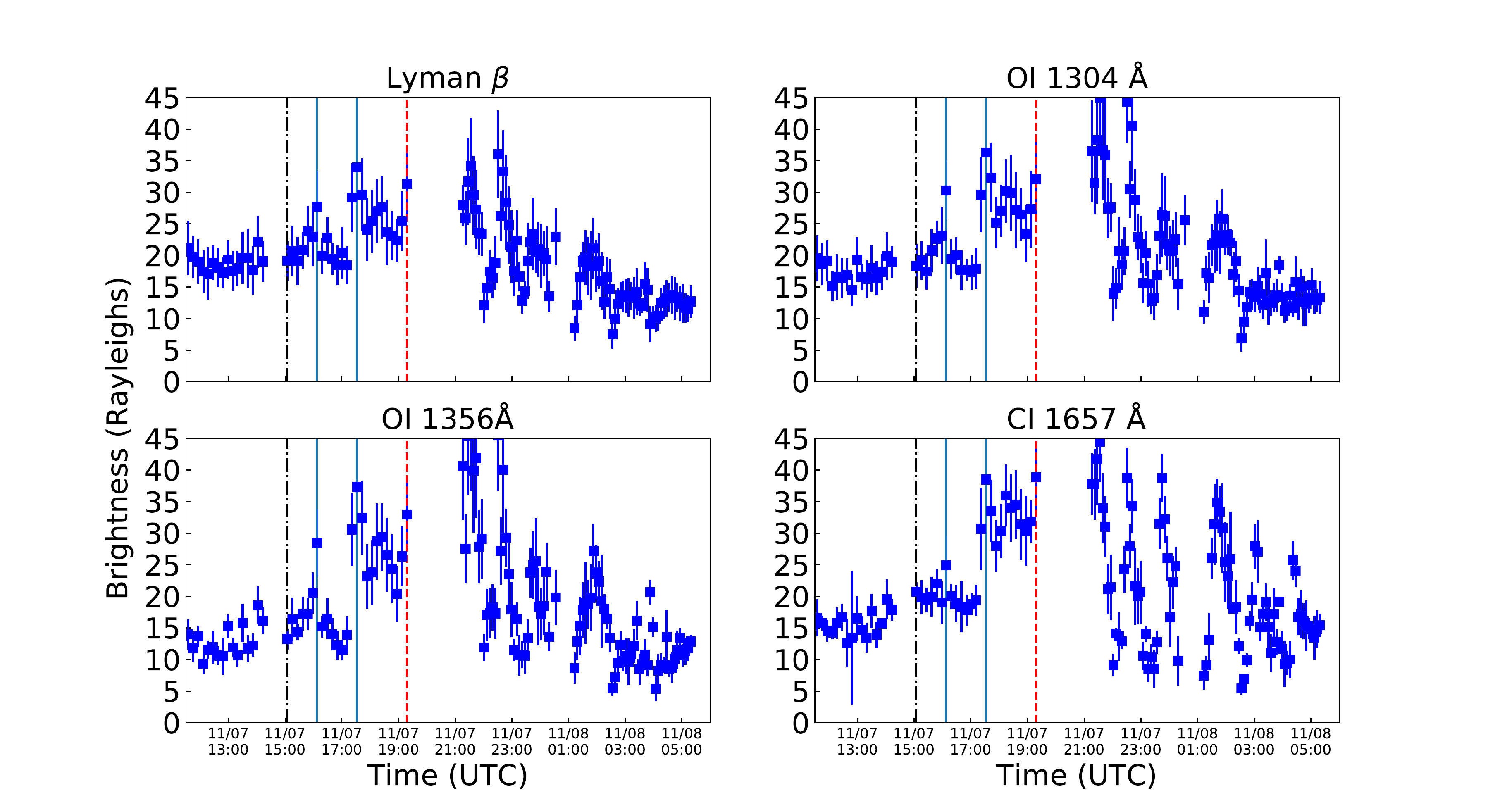} 
\caption{Light curves for dominant atomic emission features during all observing blocks listed in Table \ref{table:observations}, plotted by rows over which the brightness is calculated. Only emissions in rows 18 through 23 are co-added to maximize coma signal. The displayed errorbars are largely the result of brightness variations between rows, in addition to the odd-even detector effect.} The dotted and dashed vertical line marks the quiescent spectrum, the blue vertical lines mark two outbursts discussed in \citet{Noonan2021hybrid}, and the red dashed vertical line marks the nearly-aligned jet observation that is referenced here and discussed further in \citet{Noonan2021hybrid}.
\label{fig:time_vs_brightness}
\end{figure}

\subsection{Light Curves}\label{lightcurves}
By integrating each emission feature in a spectrum over a set number of rows and plotting these integrated emission strengths as a function of time a light curve can be made, detailing the rise and fall of atomic emissions from photodissociation or dissociative electron impact excitation of \ce{H2O}, \ce{CO2}, \ce{CO}, and \ce{O2}. Light curves for Lyman-$\beta$, \ion{O}{1} 1304, \ion{O}{1}] 1356, and \ion{C}{1} 1657 \AA\ , the four strongest emission features except for Lyman-$\alpha$, are shown for the relevant time period in Figure \ref{fig:time_vs_brightness}. The general trend is the same for all emission features. An initial increase in activity commences around 16:00 UTC, with two outbursts spiking the emissions up to 4$\times$ that of the quiescent emission. These outbursts finished by 17:32 UTC, and emissions are elevated relative to the quiescent activity until the end of Alice observations. When Alice observations resume at 21:18 UTC, emissions are almost a factor of two stronger than at the peak of the previous activity for \ion{O}{1} emissions, while Lyman-$\beta$ and \ion{C}{1} 1657 \AA\ emissions remain in a similar elevated state as they were at 19:18 UTC.  Over the next two hours a decrease in emissions is seen. This indicates that at some point during the observing gap there was a substantial increase in activity. In one possible case, the activity began to occur immediately before the 21:18 UTC observation and produced a maximum emission immediately after. On the other extreme, the increase in activity could have occurred immediately after observations ceased at 19:18 UTC and decreased to the values measured at 21:18 UTC. In this case the maximum emissions can be estimated by approximating the slope of the light curves and working backwards. For slopes of $-$10 Rayleighs hr$^{-1}$ this would imply that the upper limit activity case would have peaked at between 60 and 80 Rayleighs for the atomic emissions. However, without additional information to support either option we will avoid further speculation.  

These light curves are complicated by the change in observing modes. Prior to 19:18 UTC the pointing is stable and centered on the nucleus with no change. After 21:18 UTC the raster scans had begun, changing the previously consistent relative location of each row relative to the nucleus. This effect manifests in the light curves taken after 21:18 UTC as scatter in the data points, as the rows capture different areas of the near-nucleus coma with each raster observation, keeping the center of the Alice slit within 3 km off of the nucleus in any direction. We note that the atomic emission is best characterized from rows 18-23 on the detector, which are not contaminated by reflected sunlight from the nucleus of the solar emission features in this stable pointing. 

\subsection{Emission Mapping}\label{Emission}
For the period described in Section \ref{Raster}, four sets of four maps were made to illustrate near-nucleus emission following the initial activity that occurred in observing block 1 (Table \ref{table:observations}). These emission maps are generated from the observations contained in blocks 2, 3, 4, and 7 described in Table \ref{table:observations} in an attempt to discern the morphology of the initial activity observed in block 1. The varied spatial coverage due to the change in location of the Alice slit helps to mitigate the contribution from the instrumental odd/even effect, though the effect is still visible at the edges of some maps. The maps produced from Blocks 5 and 6 are omitted owing to limited spatial coverage. The intensity barycenter of each emission map is calculated by weighting each pixel by intensity, multiplying each grid of X and Y locations by the intensity of pixels over 30 Rayleighs, summing the X and Y weighted averages and dividing by the sum of the intensity weights. For \ion{C}{1} 1657 \AA\ the additional masking of pixels brighter than 90 Rayleighs prevents the solar reflectance off the nucleus from dominating the barycenter.  This produces an intensity barycenter in the cometocentric coordinates that can be used to calculate an angle relative to the nucleus.

\begin{figure*}
\centering
\includegraphics[clip,width=1.0\linewidth]{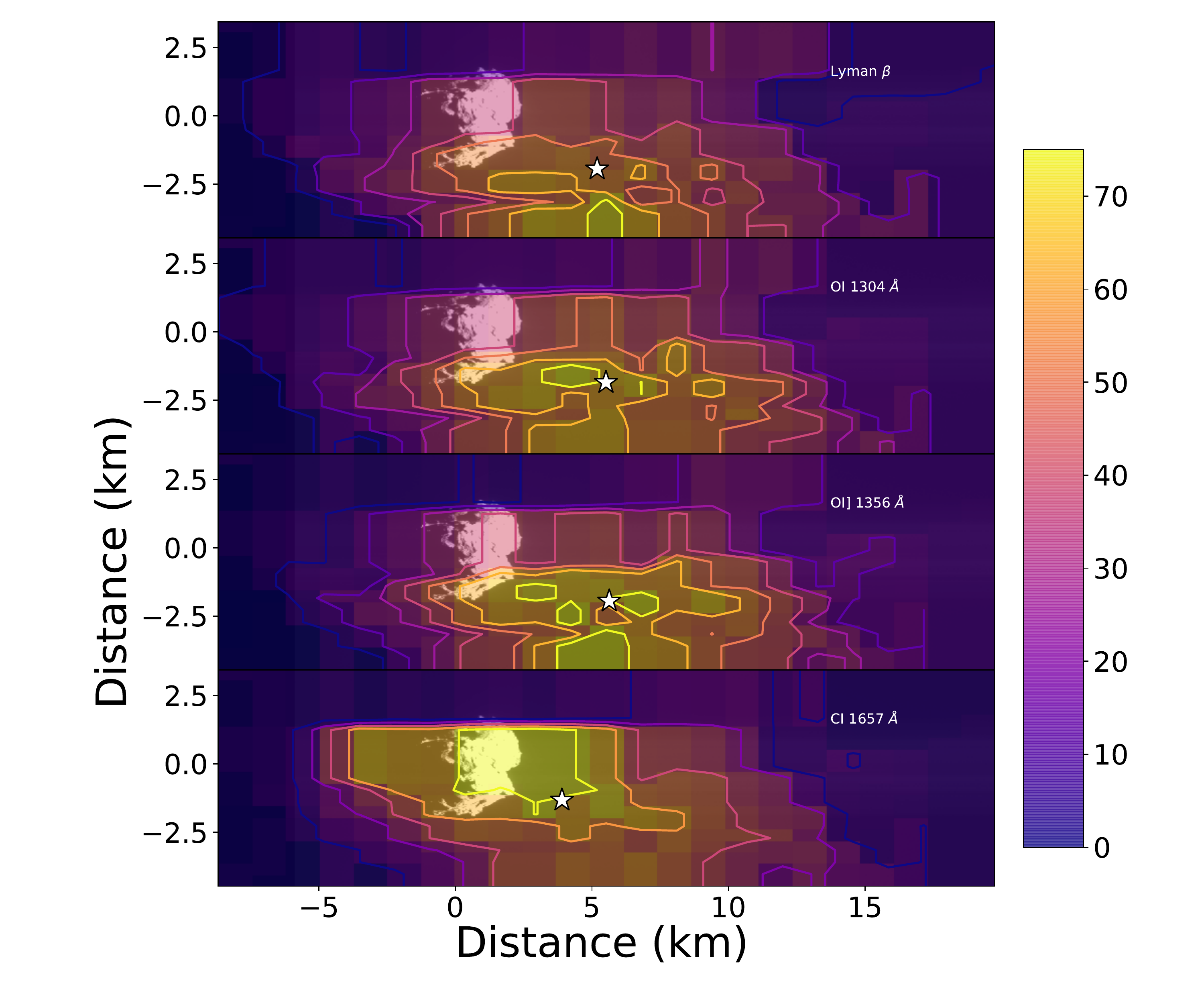} 
\caption{Emission maps created from Alice observations taken between 21:16 and 22:07 UTC on November 7 (observing block 2). The sunward direction is to the right. The weighted intensity barycenter of off-nucleus emission is marked with a white star in each map. The color bar marks 0 to 75 Rayleighs. Contours on the plot mark 10 Rayleigh isophotes. The overlaid NAVCAM image was taken at 21:18 UTC November 7. Error on brightnesses is at most $\sim$5 Rayleighs for displayed brightnesses, error on pixel position is $\pm$ 1.2 km, approximately the size of an Alice pixel subtended at the nucleus.}
\label{fig:map_1}
\end{figure*}

The first map (Figure \ref{fig:map_1}) is derived from the first 10 observations taken after observations resumed at 21:18 UTC on November 7 and shows the strongest presence of the four emission features, with the highest intensities located between $-45^{\circ}$ and $-60^{\circ}$, or clockwise, relative to the Sun-comet line exhibiting strengths up to 70-80 Rayleighs, notably higher than the brightnesses reported in the spectra at 19:18 UTC (Figure \ref{fig:stable_spectra}).  Each emission feature shows substantial strength on the sunward side of the nucleus (on the right of the nucleus in the overlaid NAVCAM images). The resolution of the Alice instrument for this period is unable to discern a jet from more general increased activity, but the difference between the opposite sides of the near-nucleus coma is greater than 50 Rayleighs on average. The distance between the intensity barycenter and the nucleus is largest in Figure \ref{fig:map_1} for all emission features and is between 35$\degree$ and 39$\degree$ clockwise of the Sun-comet line that extends along the X-axis. 

\begin{figure*}
\centering
\includegraphics[clip,width=1.0\linewidth]{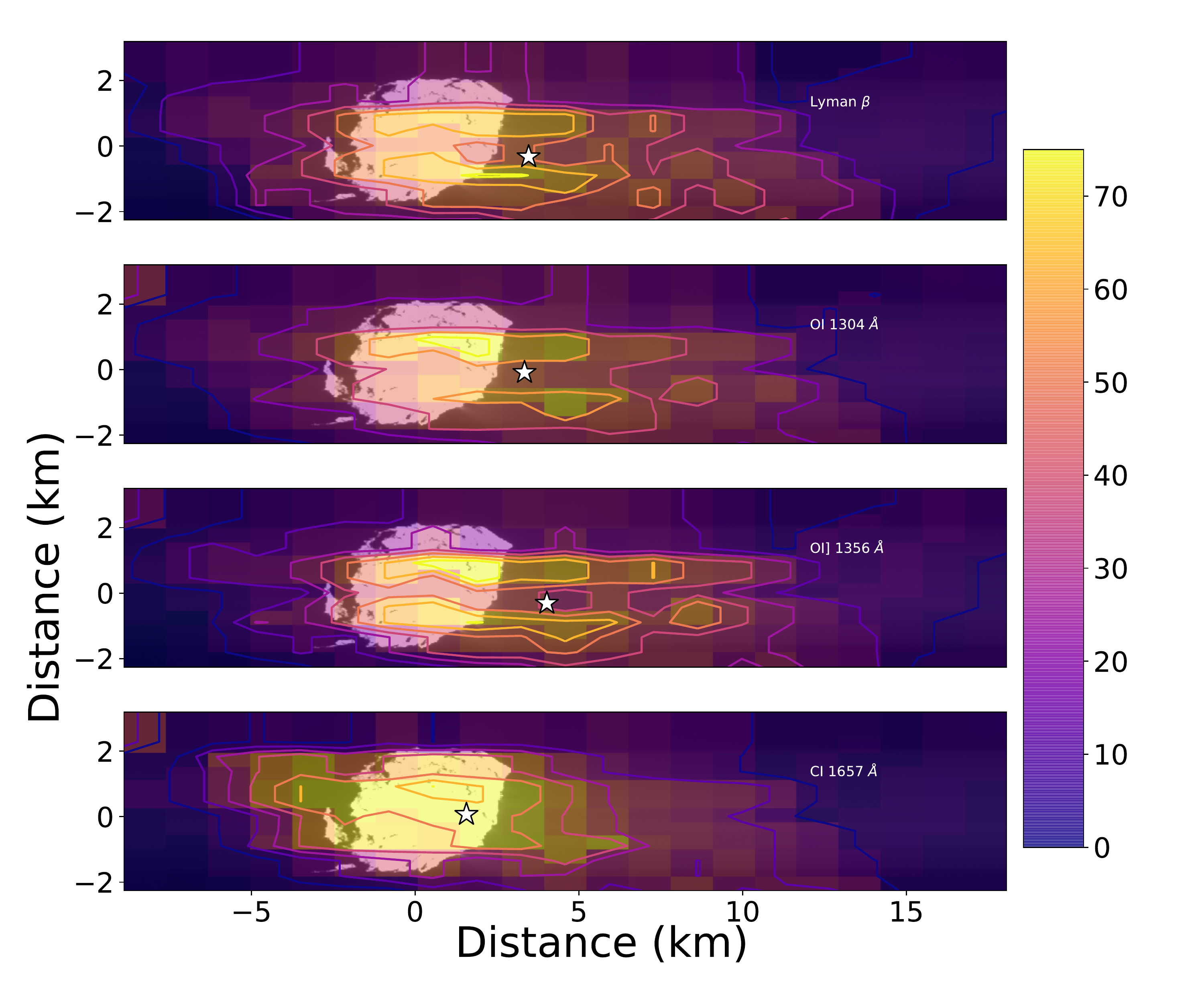} 
\caption{Emission maps created from Alice observations taken between 22:13 and 22:59 UTC on November 7 (observing block 3). The sunward direction is to the right. The weighted intensity barycenter of off-nucleus emission is marked with a white star in each map. The color bar marks 0 to 75 Rayleighs. Contours on the plot mark 10 Rayleigh isophotes. The overlaid NAVCAM image was taken at 23:40 UTC November 7. Error on brightnesses is at most $\sim$5 Rayleighs for displayed brightnesses, error on pixel position is $\pm$ 1.2 km, approximately the size of an Alice pixel subtended at the nucleus.}
\label{fig:map_2}
\end{figure*}

\begin{figure*}
\centering
\includegraphics[clip,width=1.0\linewidth]{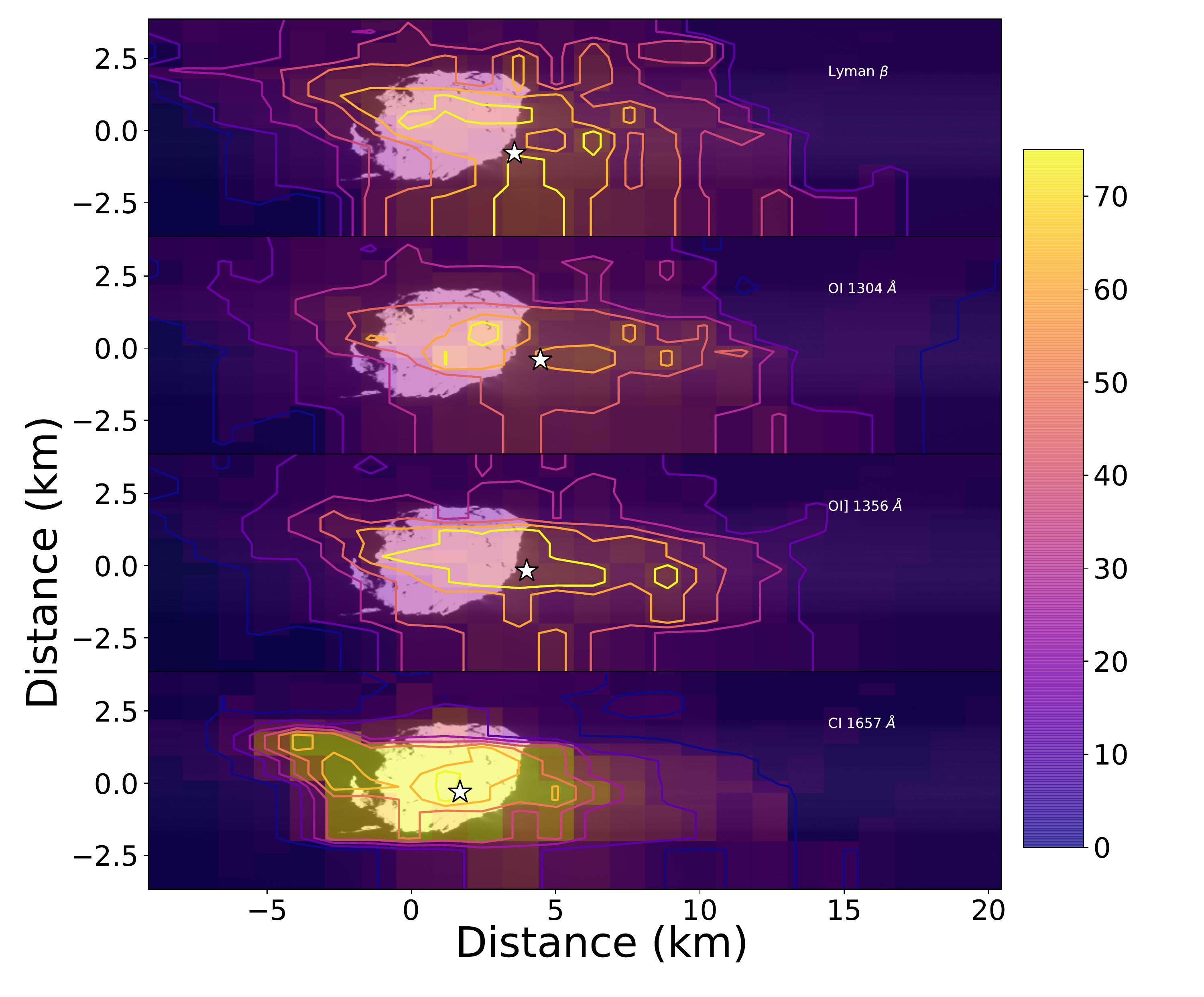} 
\caption{Emission maps created from Alice observations taken between 23:04 and 23:56 UTC on November 7 (observing block 4). The sunward direction is to the right. The weighted intensity barycenter of off-nucleus emission is marked with a white star in each map. The color bar marks 0 to 75 Rayleighs. Contours on the plot mark 10 Rayleigh isophotes. The overlaid NAVCAM image was taken at 23:40 UTC November 7. Error on brightnesses is at most $\sim$5 Rayleighs for displayed brightnesses, error on pixel position is $\pm$ 1.2 km, approximately the size of an Alice pixel subtended at the nucleus.}
\label{fig:map_3}
\end{figure*}

Maps made from observing block 3 showcase the behavior of the emissions as the nucleus rotates beneath \textit{Rosetta}. (Compare Figure \ref{fig:map_1} to Figures \ref{fig:map_2} and \ref{fig:map_3}). In this period, Alice spectra have better coverage in the +90$\degree$ sector (X$\geq$ 0, Y$\geq$0), leading to a sharp cutoff at the 0.0 km mark in Y. One hour after the observations used in Figure \ref{fig:map_1}, there is already a substantial decrease in the brightness of the emission features, specifically hydrogen and oxygen features that are less susceptible to reflected sunlight. The rotation of the nucleus, the scanning motion of the Alice spectrograph, and the near-nucleus dust reflecting solar emission during these observations may produce the noticeable ``smear" in the maps like Figure \ref{fig:map_3}. This extension appears in both the \ion{C}{1} 1657 \AA\ and FUV continuum integrated channel used to correct the position of data, and is treated as a maximum positional uncertainty in the data when comparing to the VIRTIS-H and MIRO maps. For Ly-$\beta$, \ion{O}{1} 1304 and 1356 \AA\ the intensity barycenter has moved much closer to the nucleus but remains at a similar angle relative to the Sun-comet line, varying between 38-39$\degree$ for the emission features. However, for \ion{C}{1} 1657 \AA\ the intensity barycenter falls upon the nucleus. This trend continues in Figure \ref{fig:map_3}, where each intensity barycenter is calculated to be within one map pixel of the nucleus location. 

\begin{figure*}
\centering
\includegraphics[clip,width=1.0\linewidth]{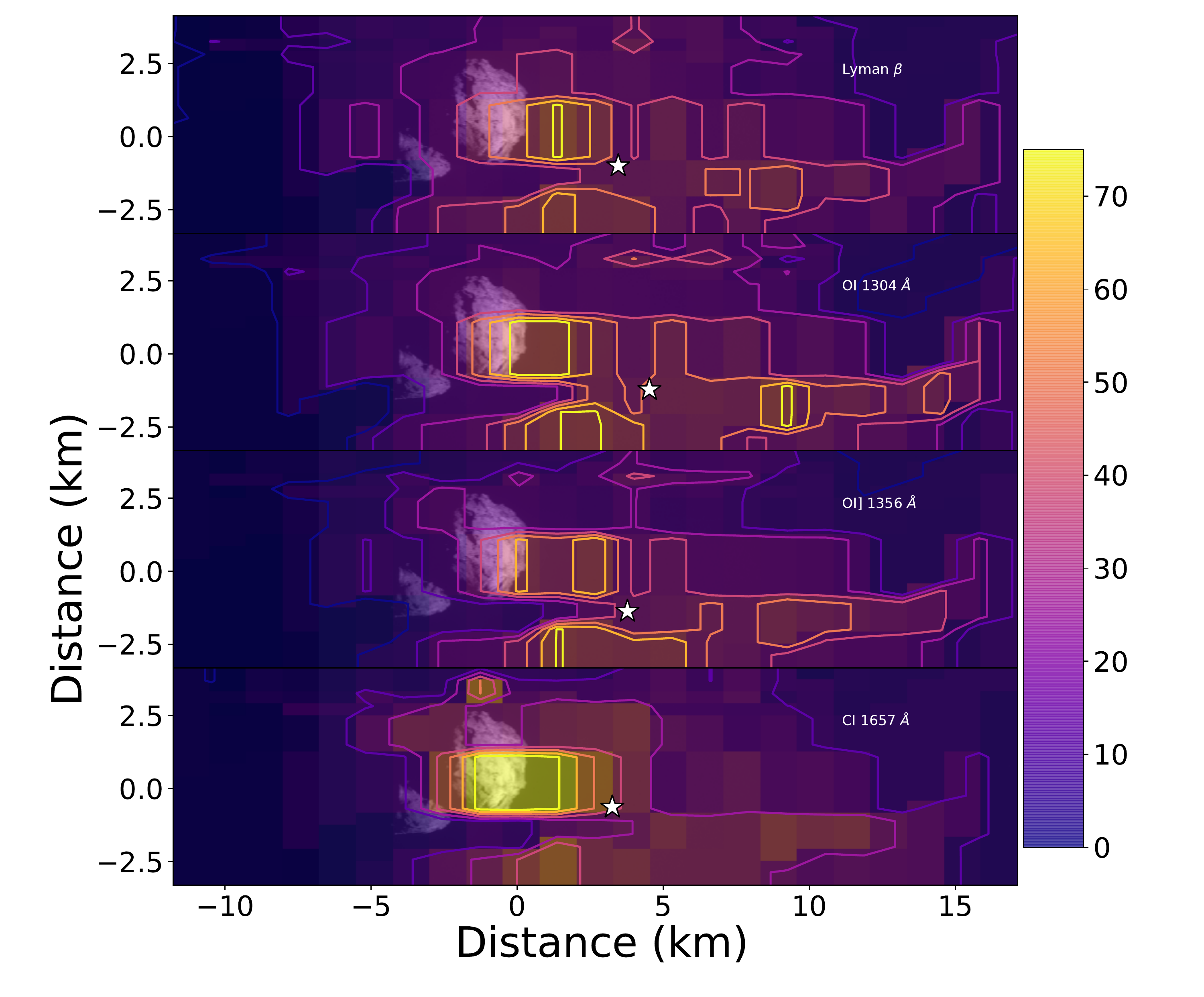} 
\caption{Emission maps created from Alice observations taken between 02:04 and 02:55 UTC on November 8 (observing block 7). The sunward direction is to the right. The weighted intensity barycenter of off-nucleus emission is marked with a white star in each map. The color bar marks 0 to 75 Rayleighs. Contours on the plot mark 10 Rayleigh isophotes. The overlaid NAVCAM image was taken at 01:20 UTC November 7. Error on brightnesses is at most $\sim$5 Rayleighs for displayed brightnesses, error on pixel position is $\pm$ 1.2 km, approximately the size of an Alice pixel subtended at the nucleus.}
\label{fig:map_6}
\end{figure*}

By observing block 7 the emission strength had continued to drop, approaching a nearly uniform, though still elevated, level in  Figure \ref{fig:map_6}, the last full coverage map available from Alice data. Almost ten hours after the first increase in activity at approximately 16:07 UTC on November 7, the UV maps no longer exhibit spatial heterogeneity in the coma, suggesting a decrease or cease to the activity/process that elevated the emissions initially, and a gradual return to a steady state near-nucleus coma. The intensity barycenters remain clustered close between 35-38$\degree$ clockwise of the Sun-comet line, but very near the limb of the nucleus. 


\subsection{Correlation to VIRTIS and MIRO}
Comparing the morphology of the Alice maps to those of VIRTIS-H and MIRO, instruments that were designed to map molecular emission and absorption at 67P, we can determine the validity of the UV mapping method. Each instrument is sensitive to different wavelengths of light, so each map will have slightly different relevant scales and morphologies but in general for a period experiencing a significant increase of activity we would expect all remote instruments that were observing to observe emission increases \citep{grun20162016}. Using this simple assumption we can qualitatively and quantitatively compare activity maps between Alice, VIRTIS-H, and MIRO, which were all observing during the same raster observations. 
\subsubsection{VIRTIS-H Maps}
Data gathered from the VIRTIS-H instrument enables mapping of both the \ce{H2O} and \ce{CO2} near-nucleus environment via the 2.7 $\mu m$ and 4.27 $\mu m$ emission bands, respectively. Owing to an excess of scattered light during these observations, the processing of VIRTIS-H data to convert from intensity to column density is difficult. Additionally, deriving total column densities from Alice for direct comparison becomes difficult owing to the number of emission processes that become relevant along the Alice line of sight when not centered on the nucleus \citep{feldman2018fuv,chaufray2017rosetta}. However, for the purposes of determining the viability of the Alice maps a relative comparison from VIRTIS-H is quite useful.

\begin{figure*}[htb]
\centering
\subfloat{%
\begin{tabular}{c c c}
\includegraphics[width=0.3\textwidth,trim=4.5cm 9cm 4.5cm 9cm,clip]{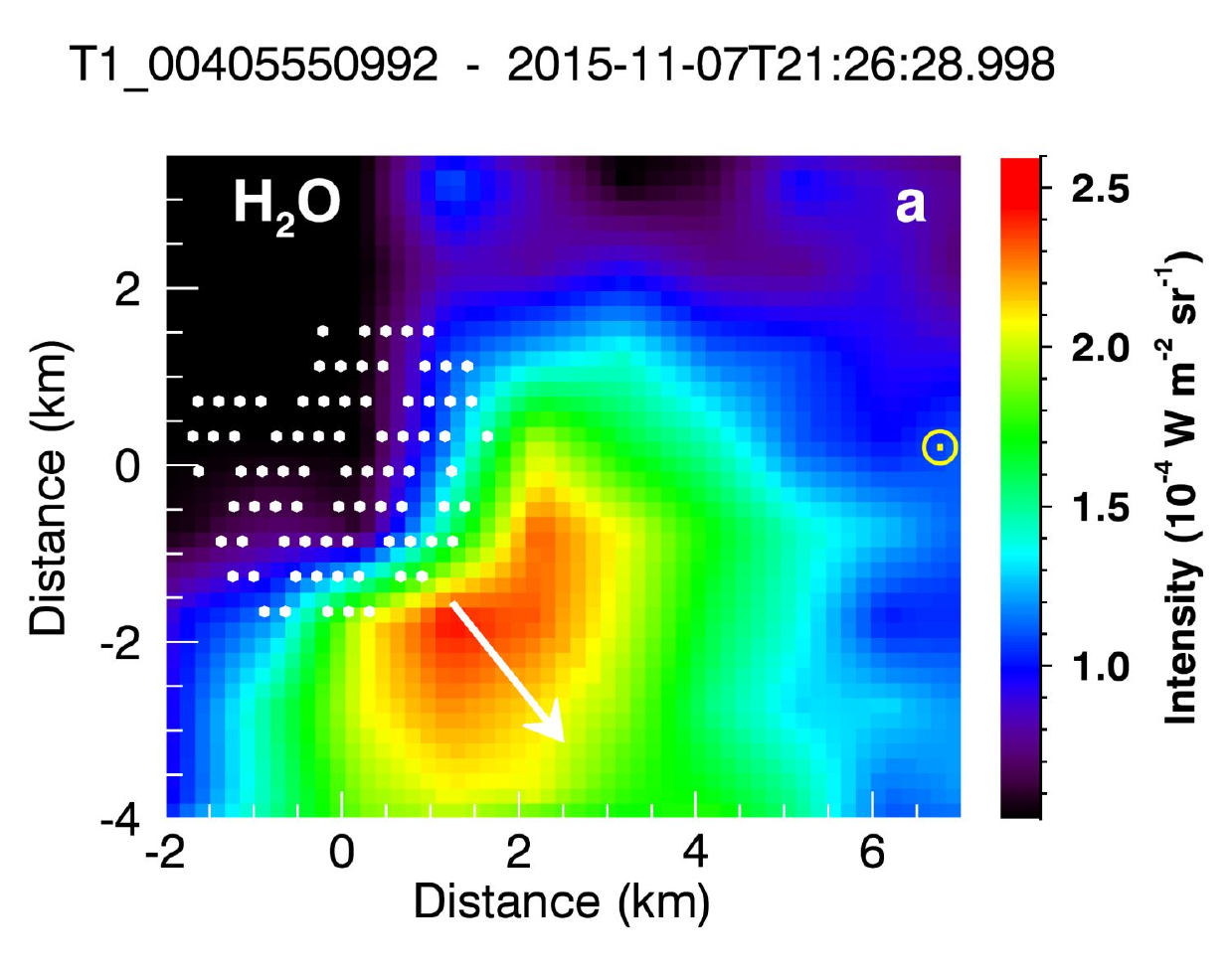}    & \includegraphics[width=0.3\textwidth,trim=4.5cm 9cm 4.5cm 9cm,clip]{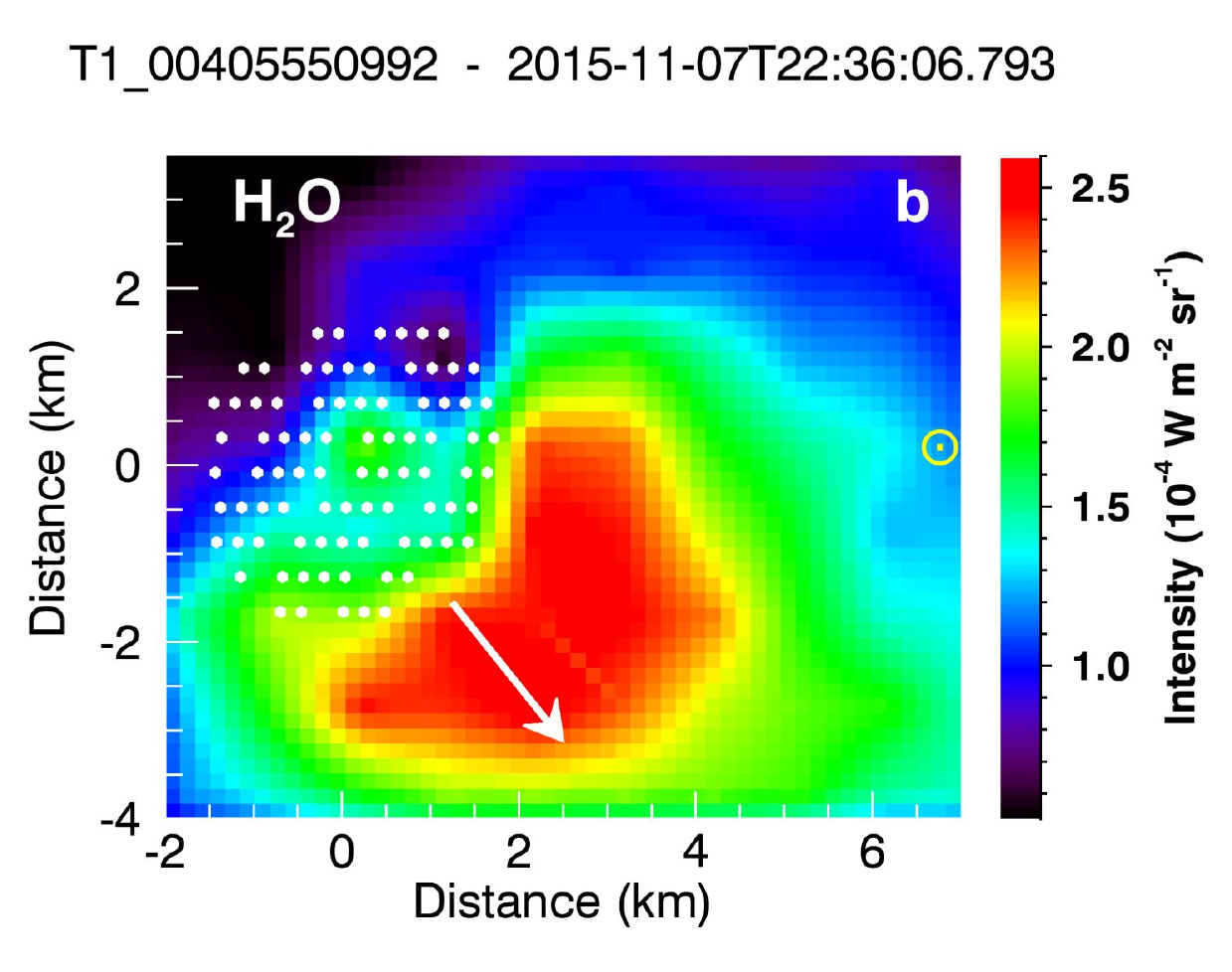}  &  \includegraphics[width=0.3\textwidth,trim=4.5cm 9cm 4.5cm 9cm,clip]{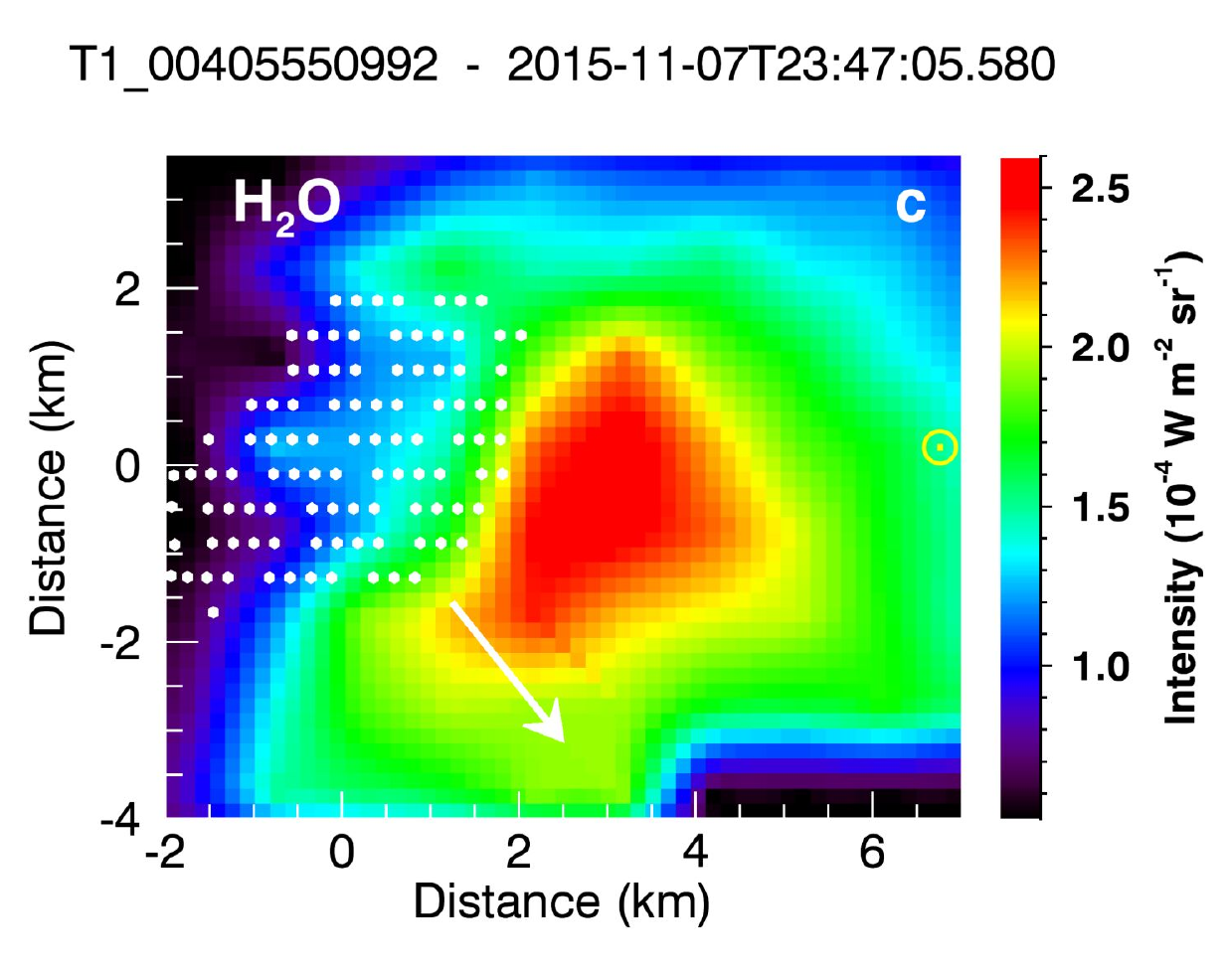} \\
\includegraphics[width=0.3\textwidth,trim=4.5cm 9cm 4.5cm 9cm,clip]{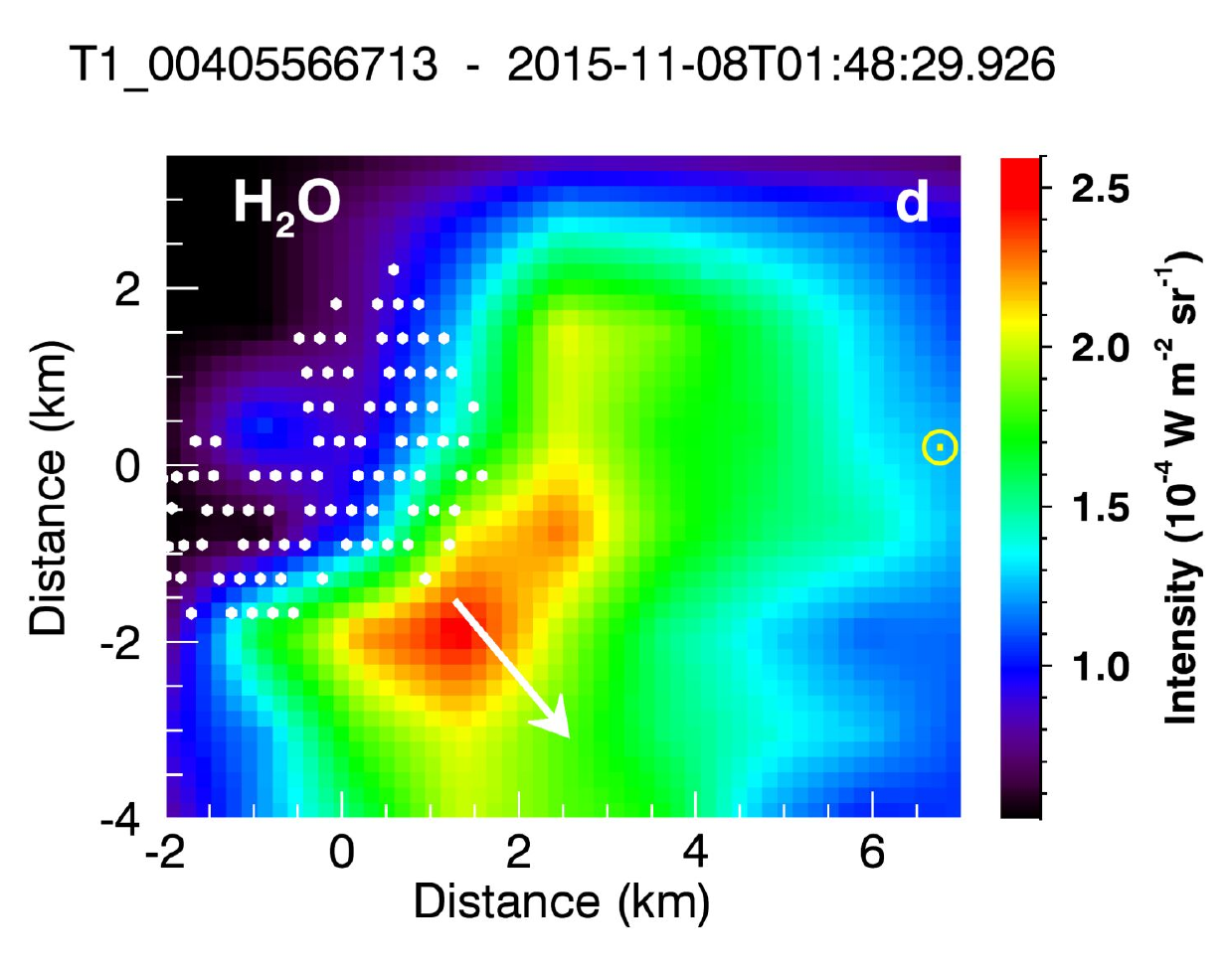} & \includegraphics[width=0.3\textwidth,trim=4.5cm 9cm 4.5cm 9cm,clip]{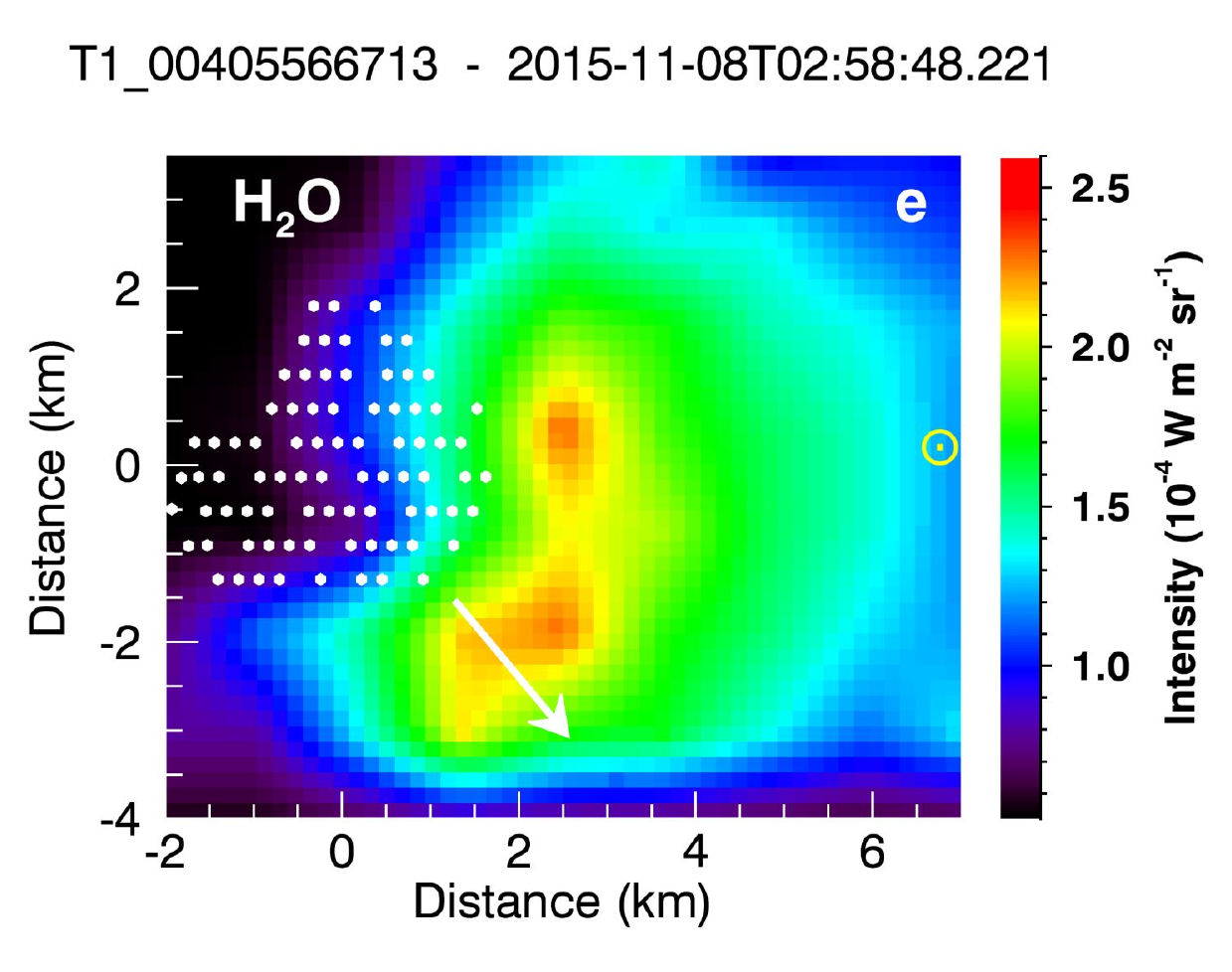}    & 
\includegraphics[width=0.3\textwidth,trim=4.5cm 9cm 4.5cm 9cm,clip]{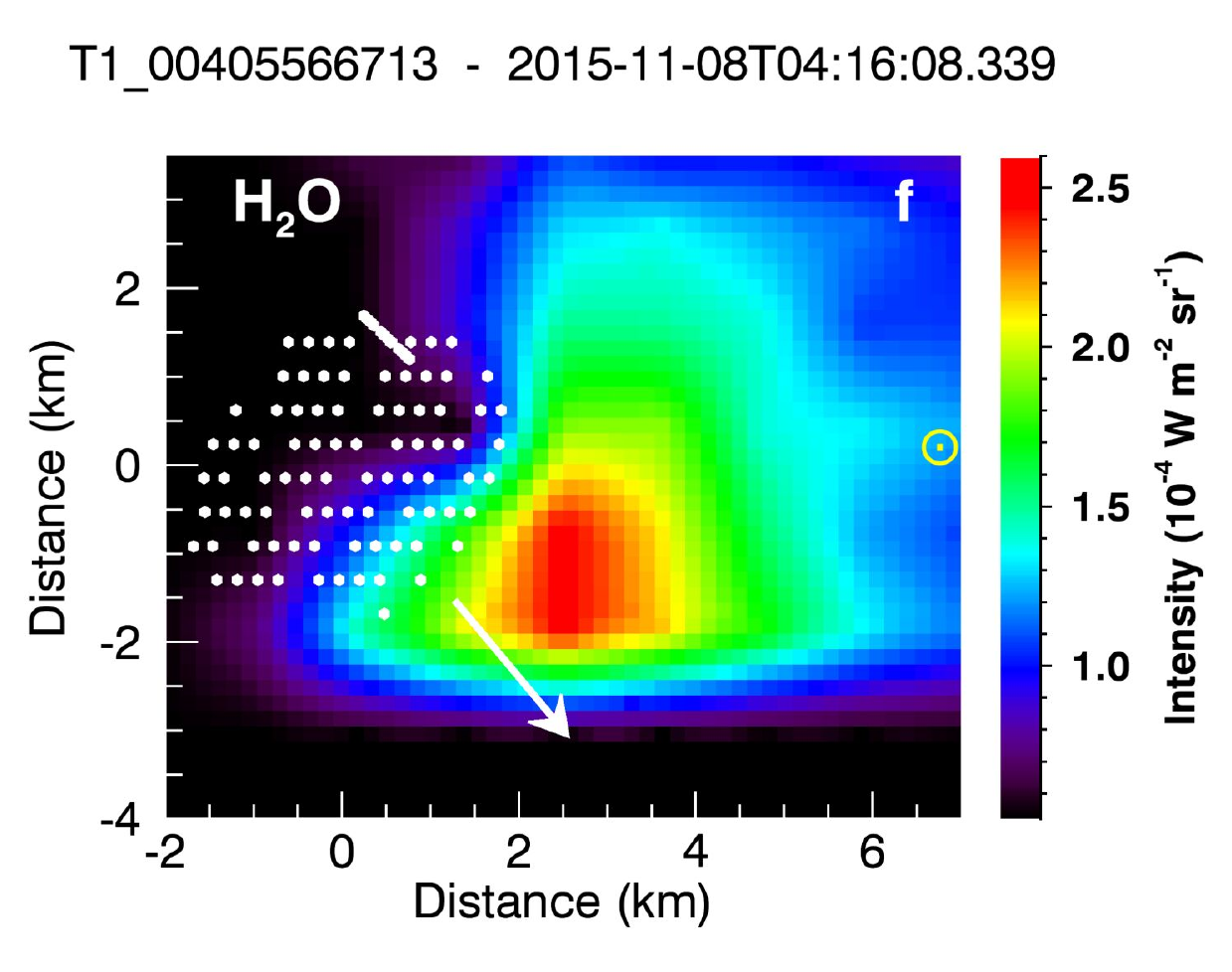} \\
\includegraphics[width=0.3\textwidth,trim=4.5cm 9cm 4.5cm 9cm,clip]{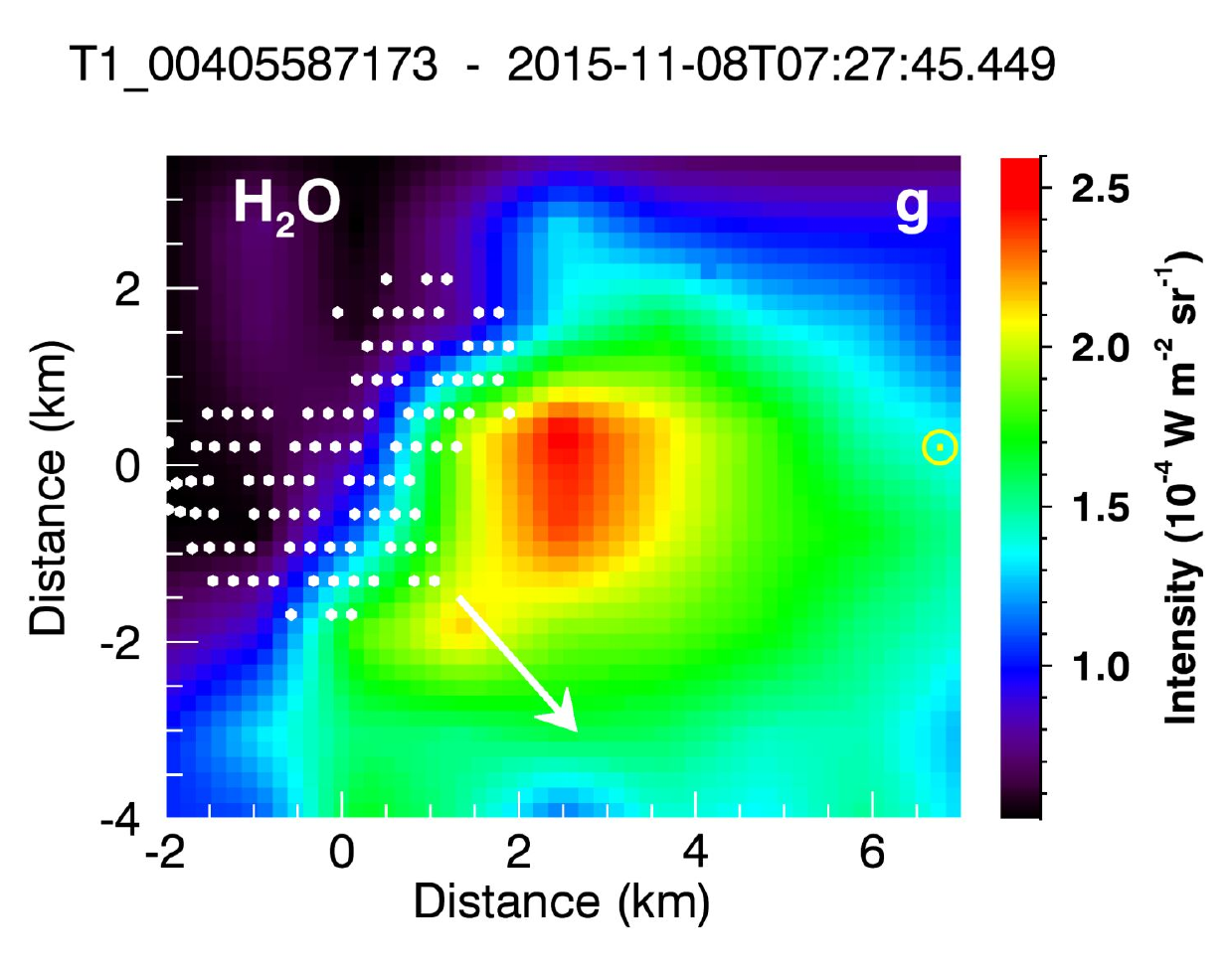}  &
\includegraphics[width=0.3\textwidth,trim=4.5cm 9cm 4.5cm 9cm,clip]{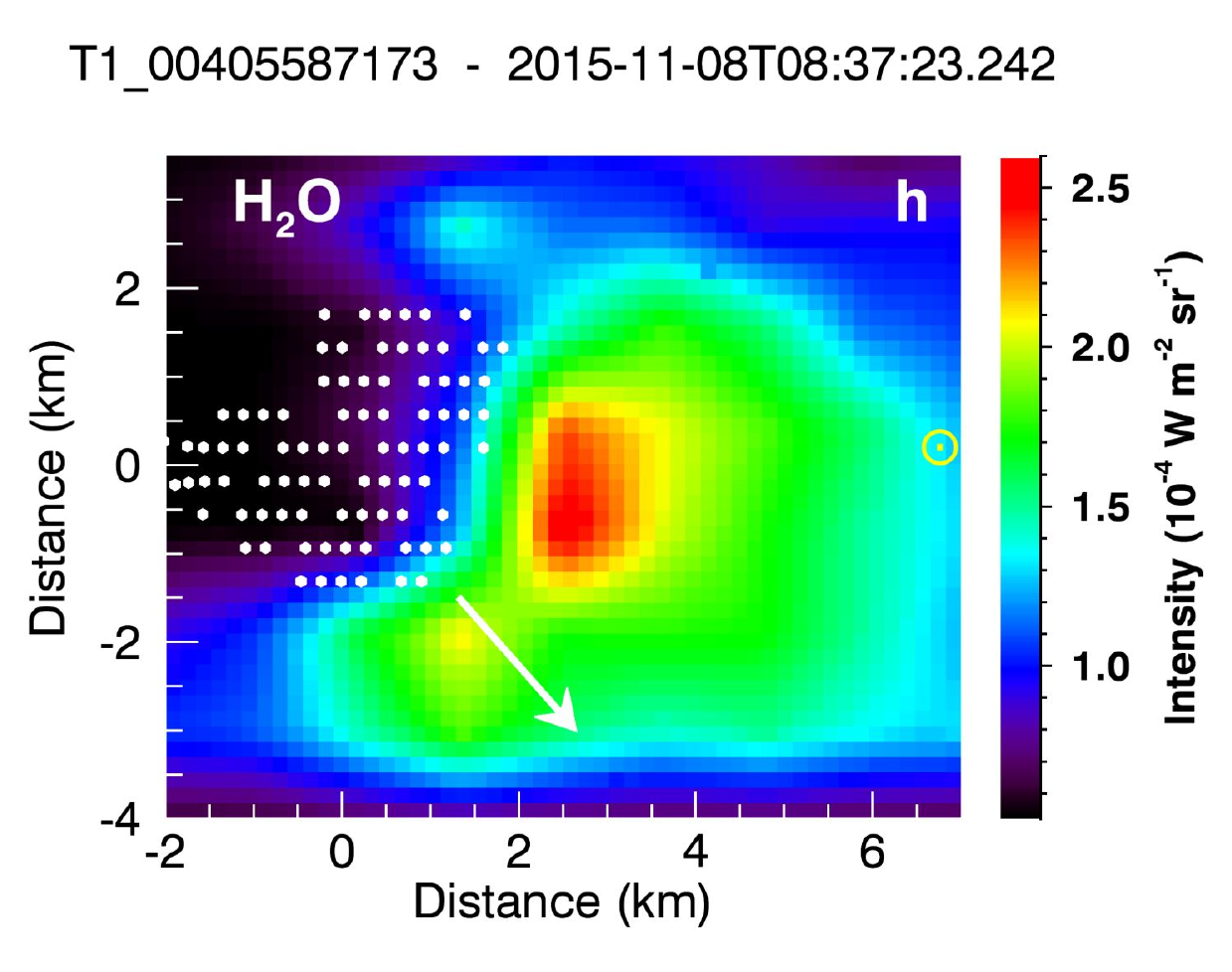} &
\includegraphics[width=0.3\textwidth,trim=4.5cm 9cm 4.5cm 9cm,clip]{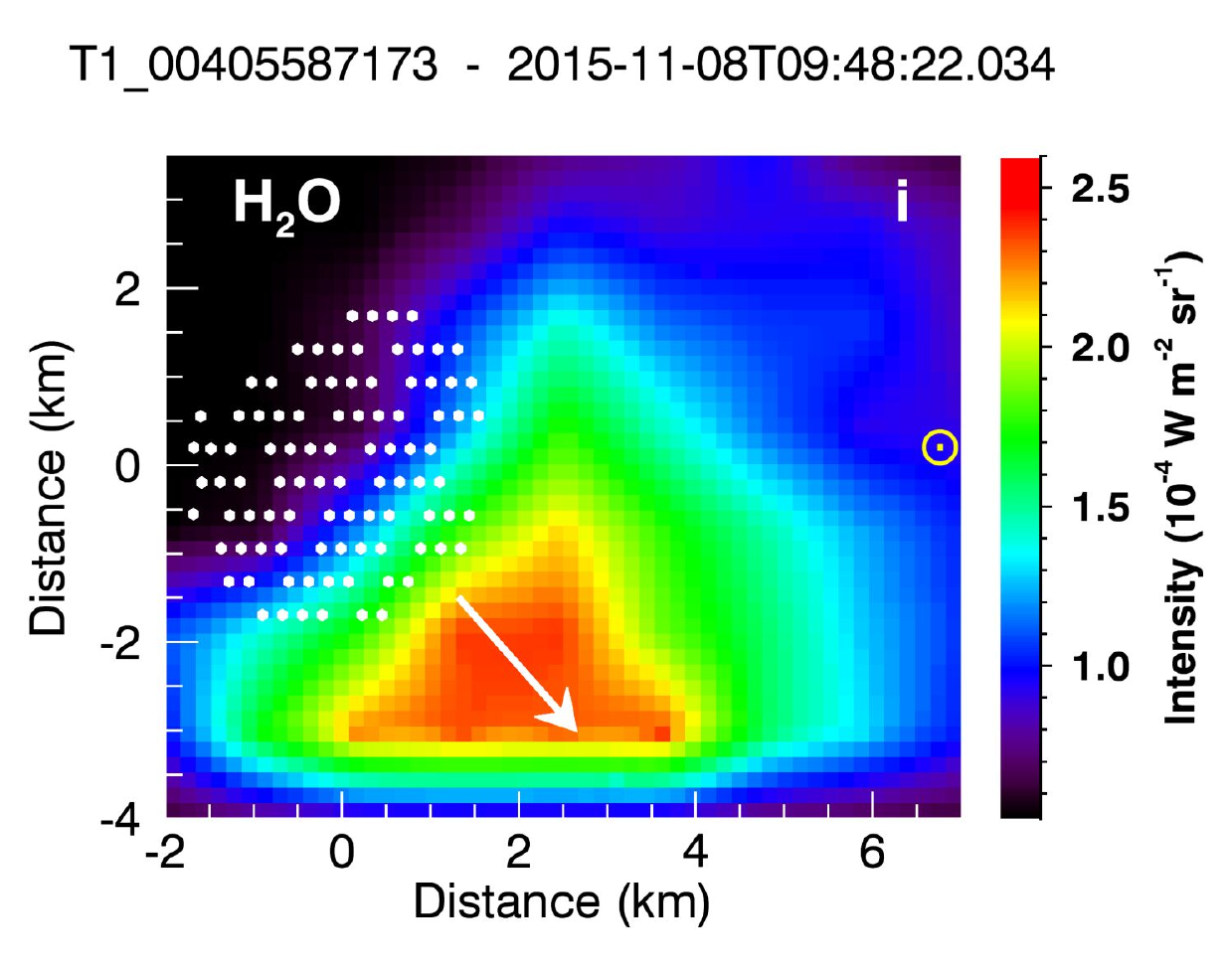} \\
\end{tabular}
}\caption{VIRTIS-H - \ce{H2O} 2.7 $\mu m$ band maps. Data on nucleus acquisitions are shown as white dots to outline the nucleus shape and are used in map generation. The colorbar shows the intensity of the water emission feature. The white line 
extends the rotation axis from the south pole. The sunward direction is to the right (see yellow sun). Note the difference in field of view (FOV) compared to Alice maps in Figures \ref{fig:map_1}-\ref{fig:map_6}.}
\label{fig:VIRTIS H2O Maps}
\end{figure*}

\begin{figure*}
\centering
\subfloat{%
\begin{tabular}{c c c}
\includegraphics[width=0.3\textwidth,trim=4.5cm 9cm 4.5cm 9cm,clip]{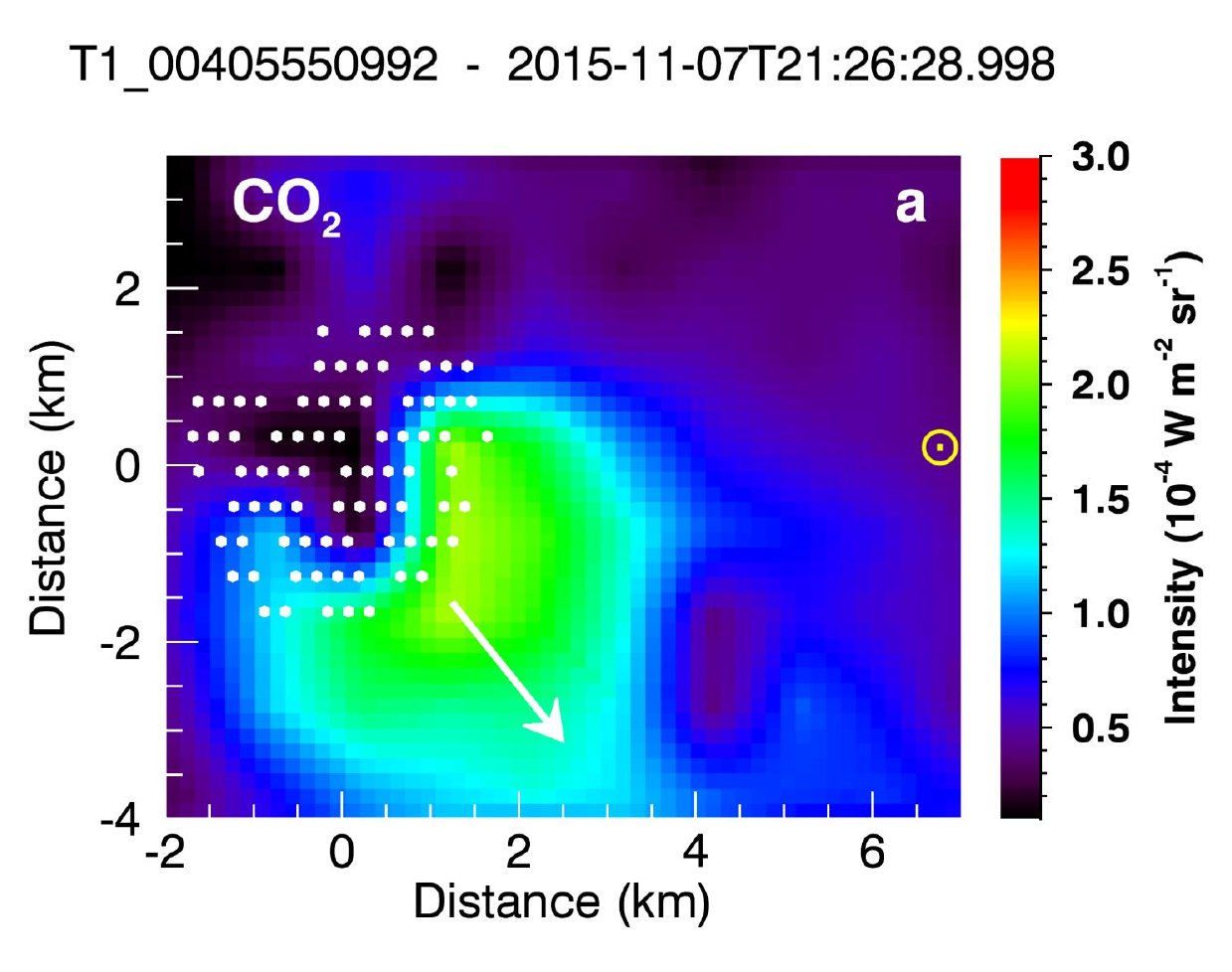} &
\includegraphics[width=0.3\textwidth,trim=4.5cm 9cm 4.5cm 9cm,clip]{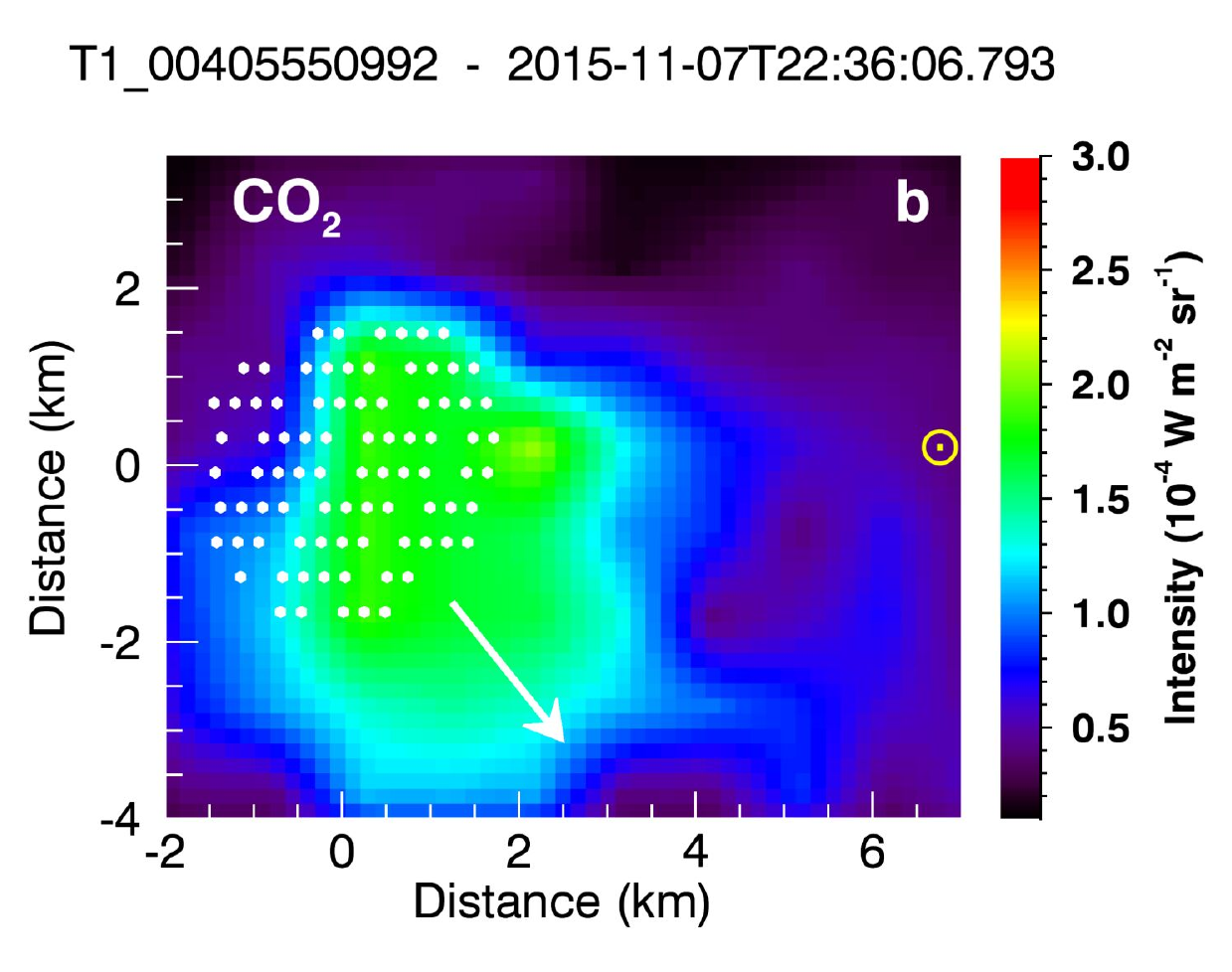} &
\includegraphics[width=0.3\textwidth,trim=4.5cm 9cm 4.5cm 9cm,clip]{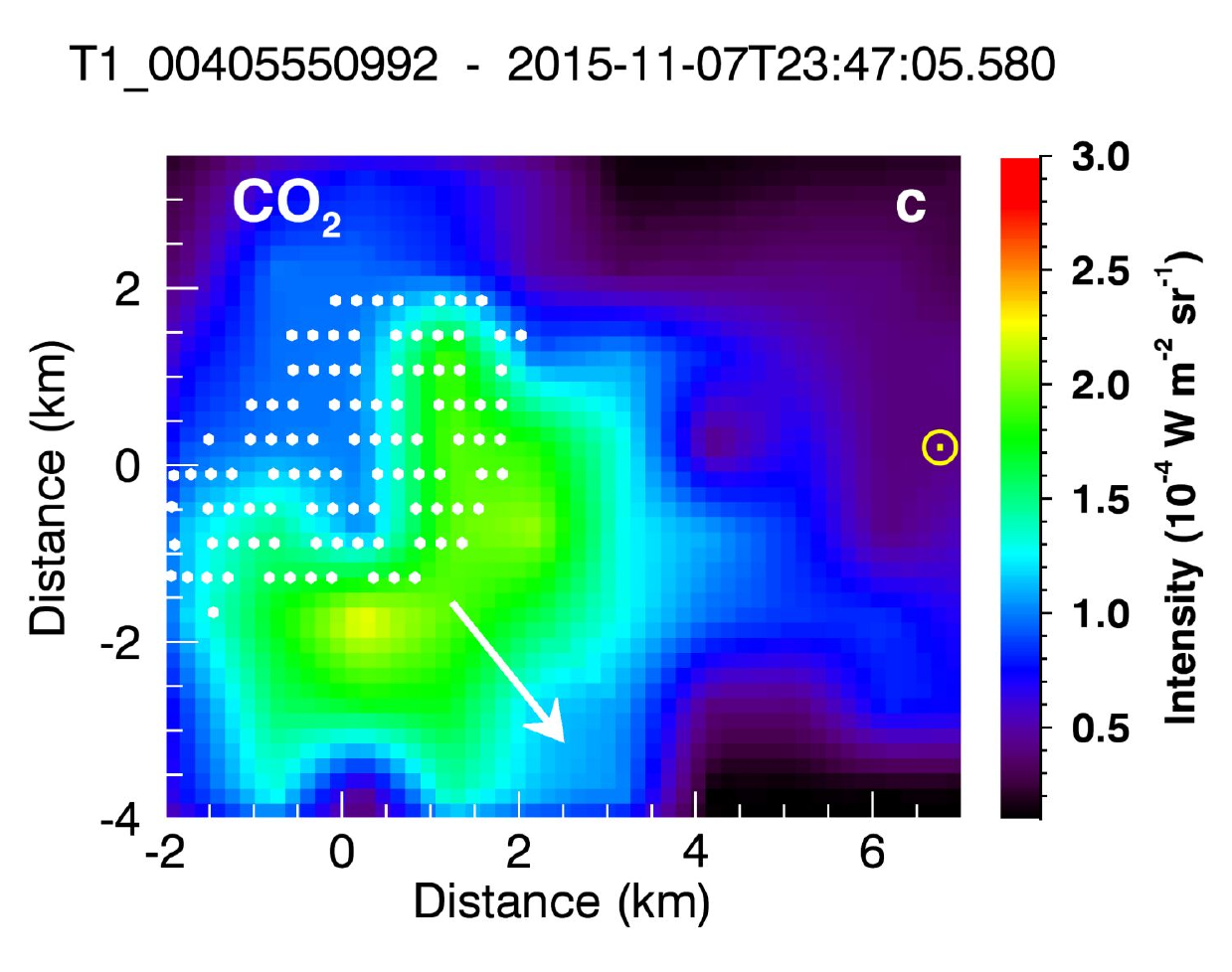} \\
\includegraphics[width=0.3\textwidth,trim=4.5cm 9cm 4.5cm 9cm,clip]{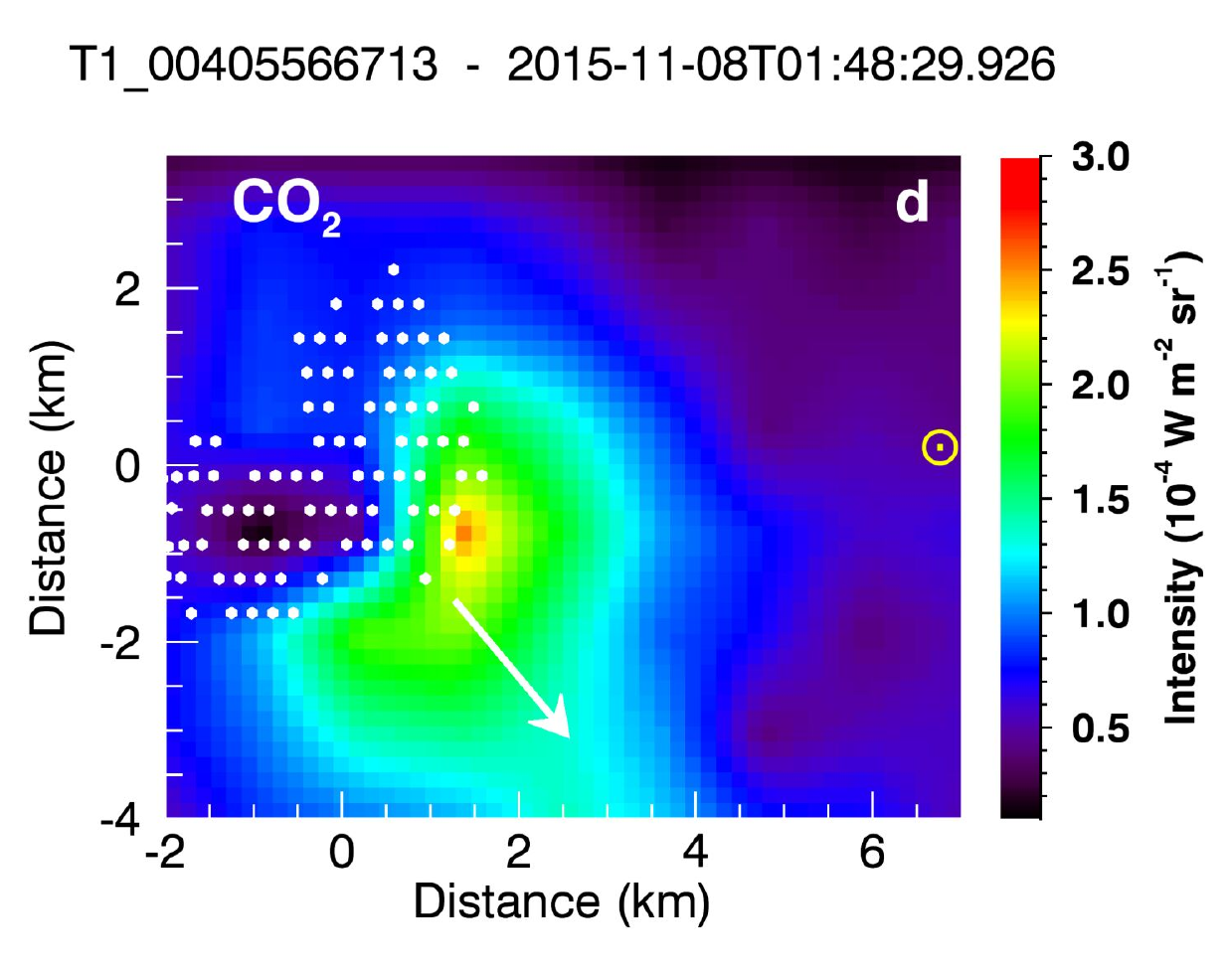}  &
\includegraphics[width=0.3\textwidth,trim=4.5cm 9cm 4.5cm 9cm,clip]{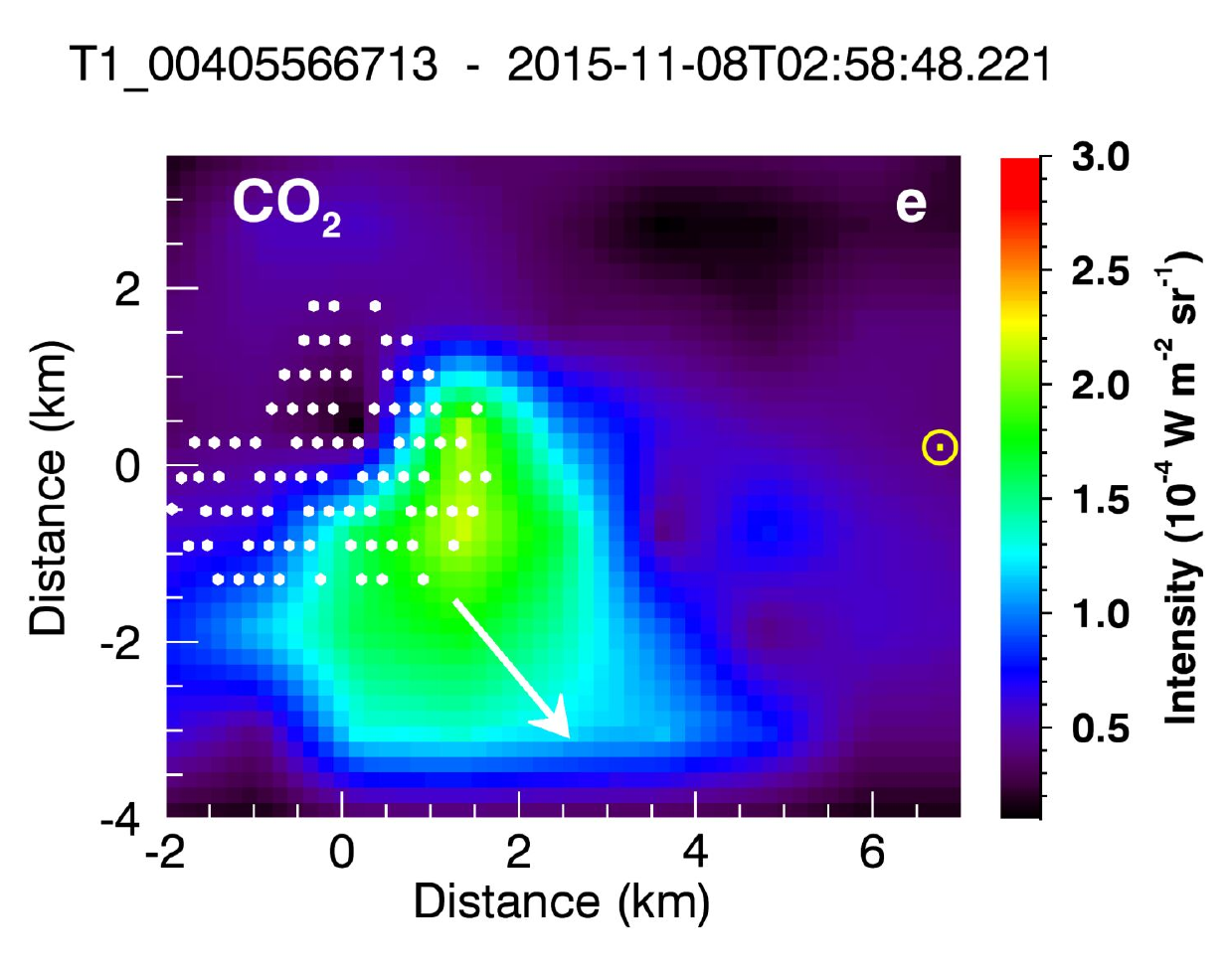}  &
\includegraphics[width=0.3\textwidth,trim=4.5cm 9cm 4.5cm 9cm,clip]{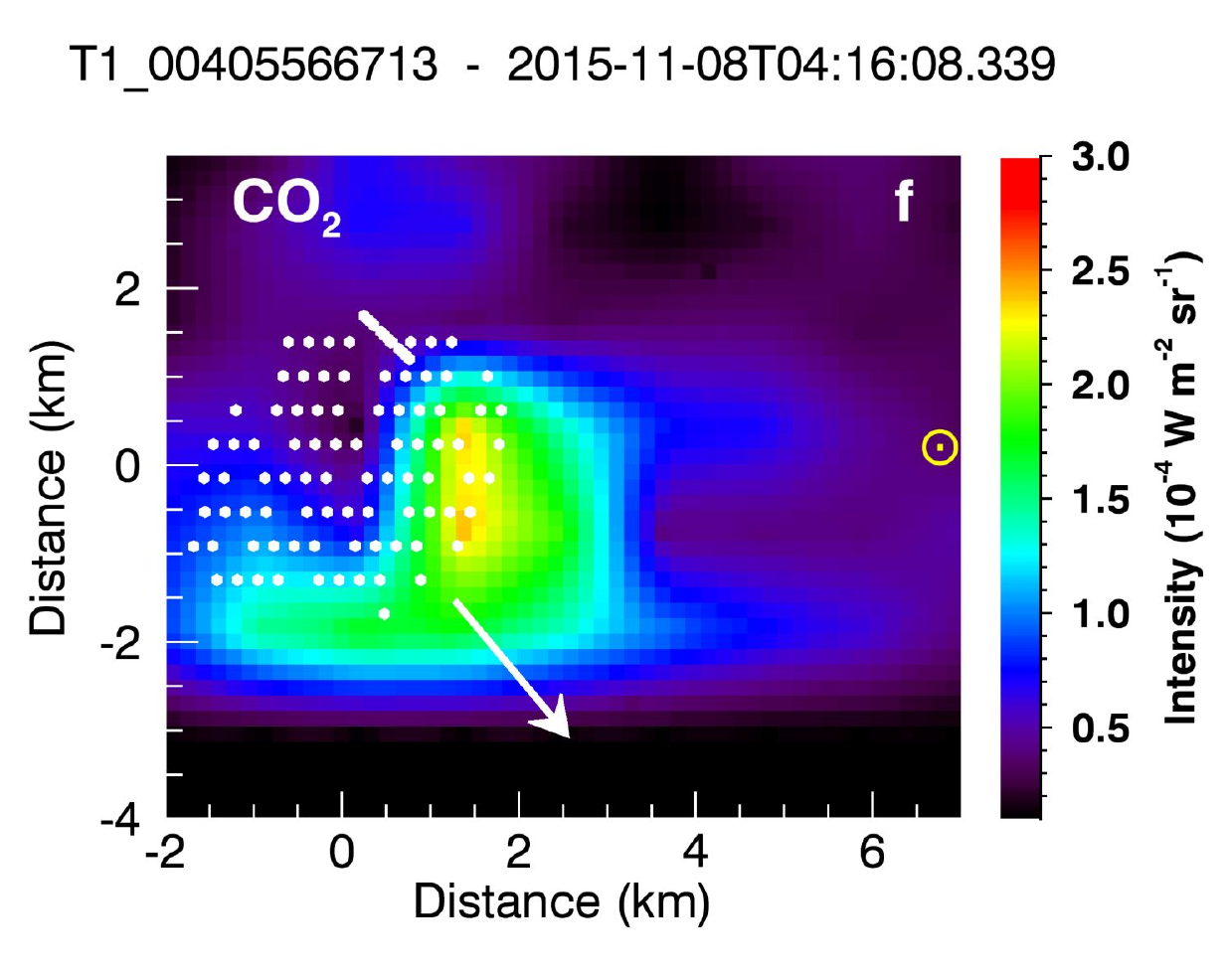} \\
\includegraphics[width=0.3\textwidth,trim=4.5cm 9cm 4.5cm 9cm,clip]{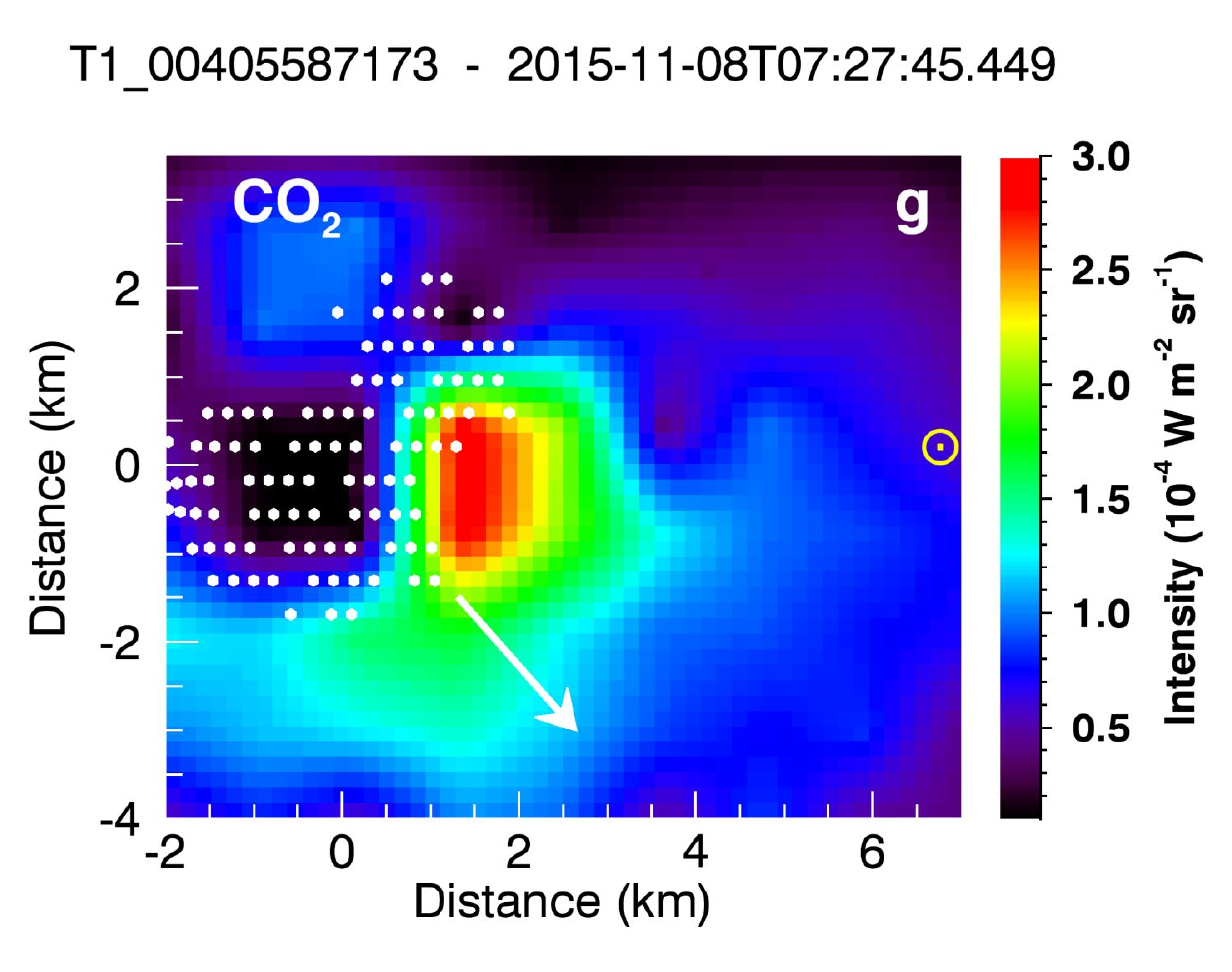}  &
\includegraphics[width=0.3\textwidth,trim=4.5cm 9cm 4.5cm 9cm,clip]{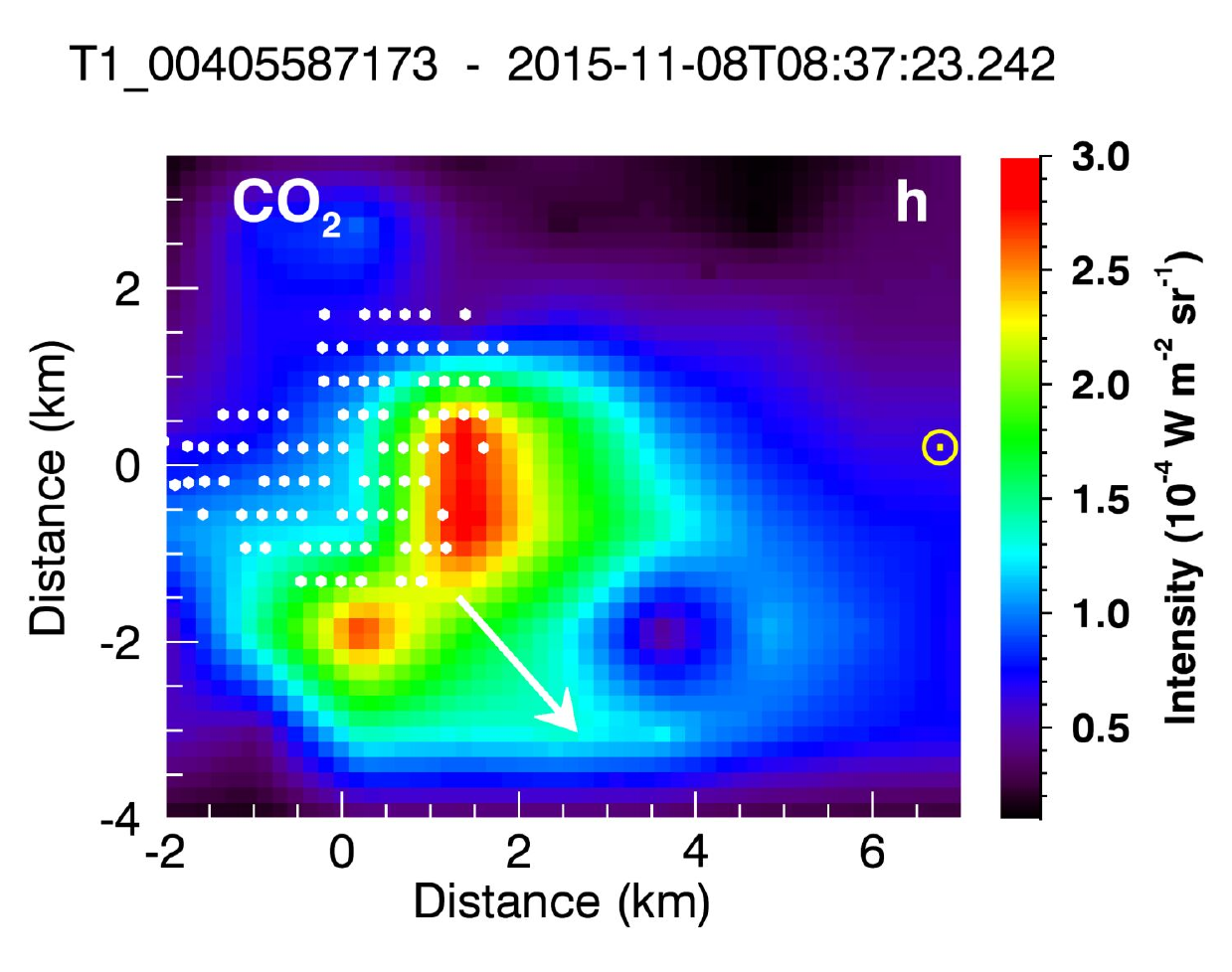}  &
\includegraphics[width=0.3\textwidth,trim=4.5cm 9cm 4.5cm 9cm,clip]{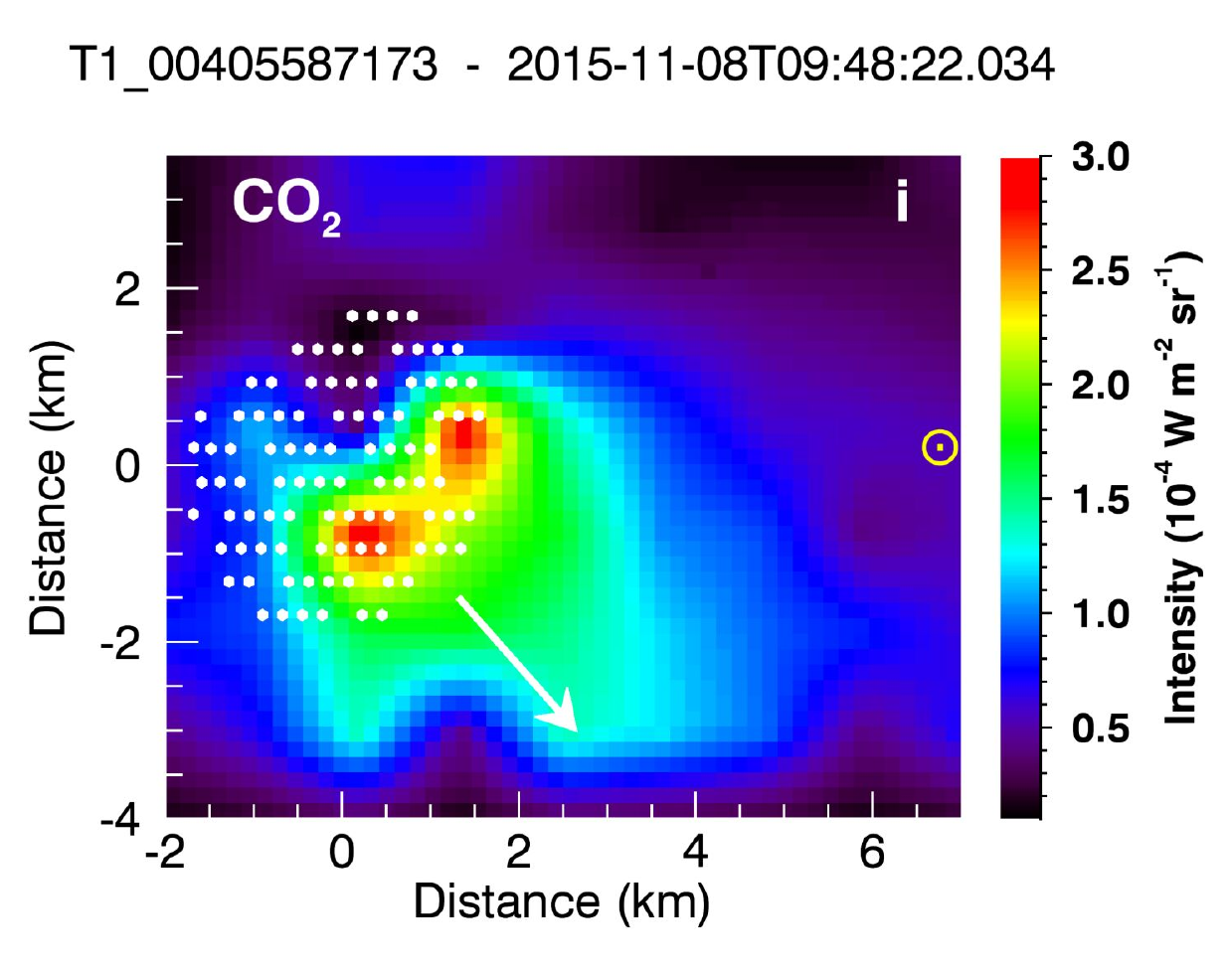} \\
\end{tabular}
}\caption{VIRTIS-H - \ce{CO2} 4.27 $\mu m$ band maps. Data on nucleus are not included to minimize signal from the nucleus and are shown as white dots. The colorbar shows the intensity of the emission feature. The white line
extends the rotation axis from the south pole. The sunward direction is to the right (see yellow sun).}
\label{fig:VIRTIS CO2 Maps}
\end{figure*}

Maps made by the VIRTIS-H team from observations during the raster scan have a similar time resolution to that of the Alice maps, with a $\Delta t$ between maps of approximately one hour. Spatial resolution is improved relative to the Alice maps, providing context for the morphology seen in the Alice maps. Both the \ce{H2O} (Figure \ref{fig:VIRTIS H2O Maps}) and \ce{CO2} maps (Figure \ref{fig:VIRTIS CO2 Maps}) exhibit an intensity 2.5$\times$ stronger at an angle of approximately $-$45$\degree$ relative to the Sun-comet line, near the south pole direction displayed with a white solid line, than at +60$\degree$ for the same time period as Alice observing block 2 (Figure \ref{fig:VIRTIS H2O Maps},\ref{fig:VIRTIS CO2 Maps} upper left map; Figure \ref{fig:map_1}; respectively). However, the molecular emission does not display the same extension as the Alice atomic emission. As the comet proceeds to rotate in the next maps, covering from  22:36:06 UTC November 7 to 09:48:22 UTC November 8, the intensity remains large within 5 km of the nucleus (all panels) and occasionally begins to approach 30$\degree$ to 45$\degree$ relative to the Sun-comet line (panels b, c, d, and e). Panel b) of Figures \ref{fig:VIRTIS H2O Maps} and \ref{fig:VIRTIS CO2 Maps} coincides with observing block 3 (Figure \ref{fig:map_2}, which lacks coverage in the same area where the relative intensity is highest in the VIRTIS-H maps.

The VIRTIS-H \ce{CO2} intensity has the greatest variability between maps. The strongest periods of relative emission are seen at 21:26 UTC of November 7 and 04:16 UTC on November 8 (Fig. \ref{fig:VIRTIS CO2 Maps}). The earlier instance shows the most extension from the nucleus along the south pole axis, while the latter is rotated approximately 30$\degree$ anti-sunward from the south pole axis. Only for the 21:26 UTC map is there a matching Alice component map that can be correlated to the VIRTIS-H \ce{CO2} intensity measurements. Both the Alice \ion{C}{1} 1657 \AA\ and VIRTIS-H \ce{CO2} maps from 21:00-22:00 UTC show a significant extension from the southern hemisphere.  

\subsubsection{MIRO Maps}
\begin{figure*}
\centering
\subfloat{%
\begin{tabular}{c c c}
\includegraphics[width=0.3\textwidth,trim=3cm 1cm 3cm 1cm,clip]{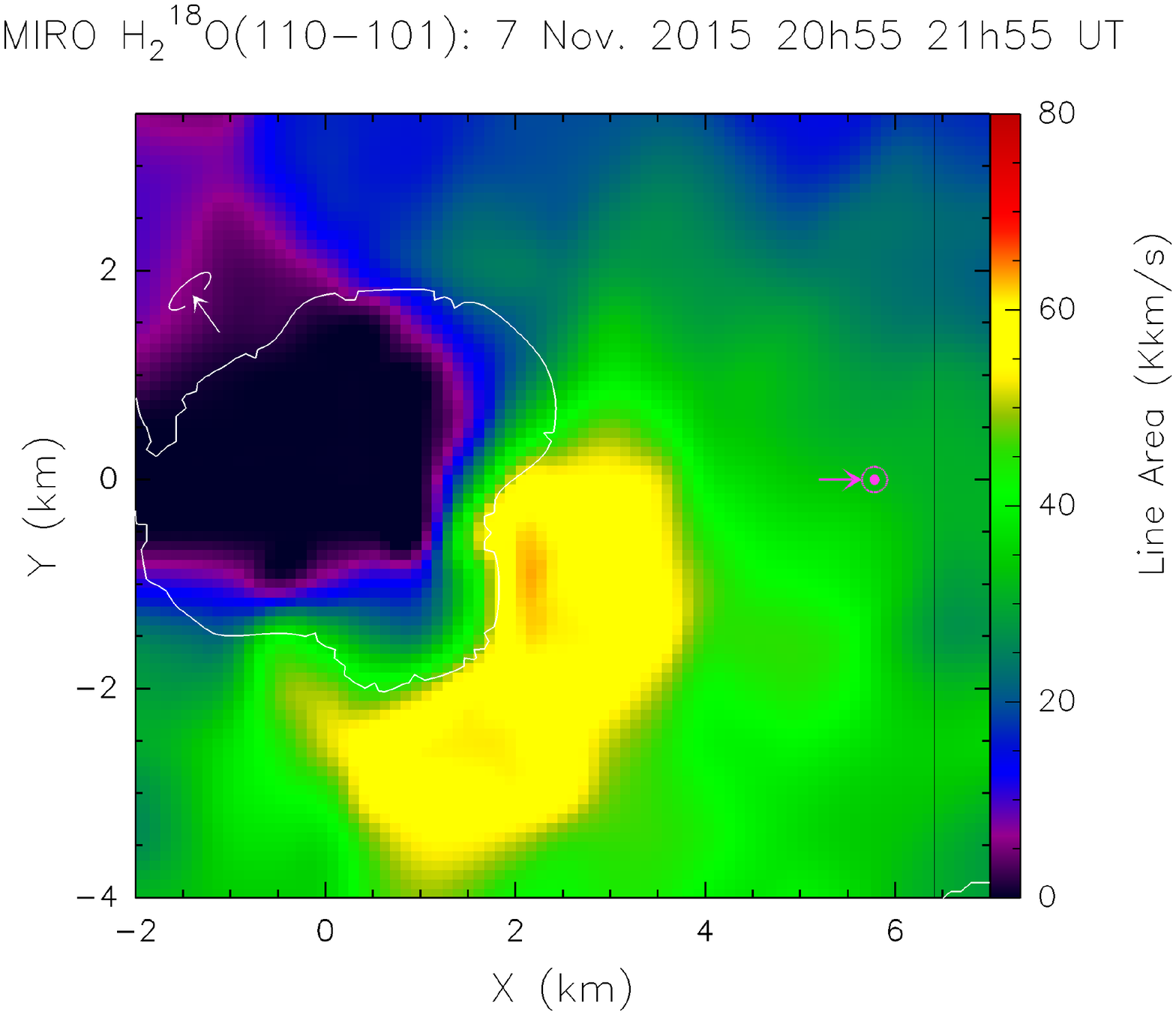} &
\includegraphics[width=0.3\textwidth,trim=3cm 1cm 3cm 1cm,clip]{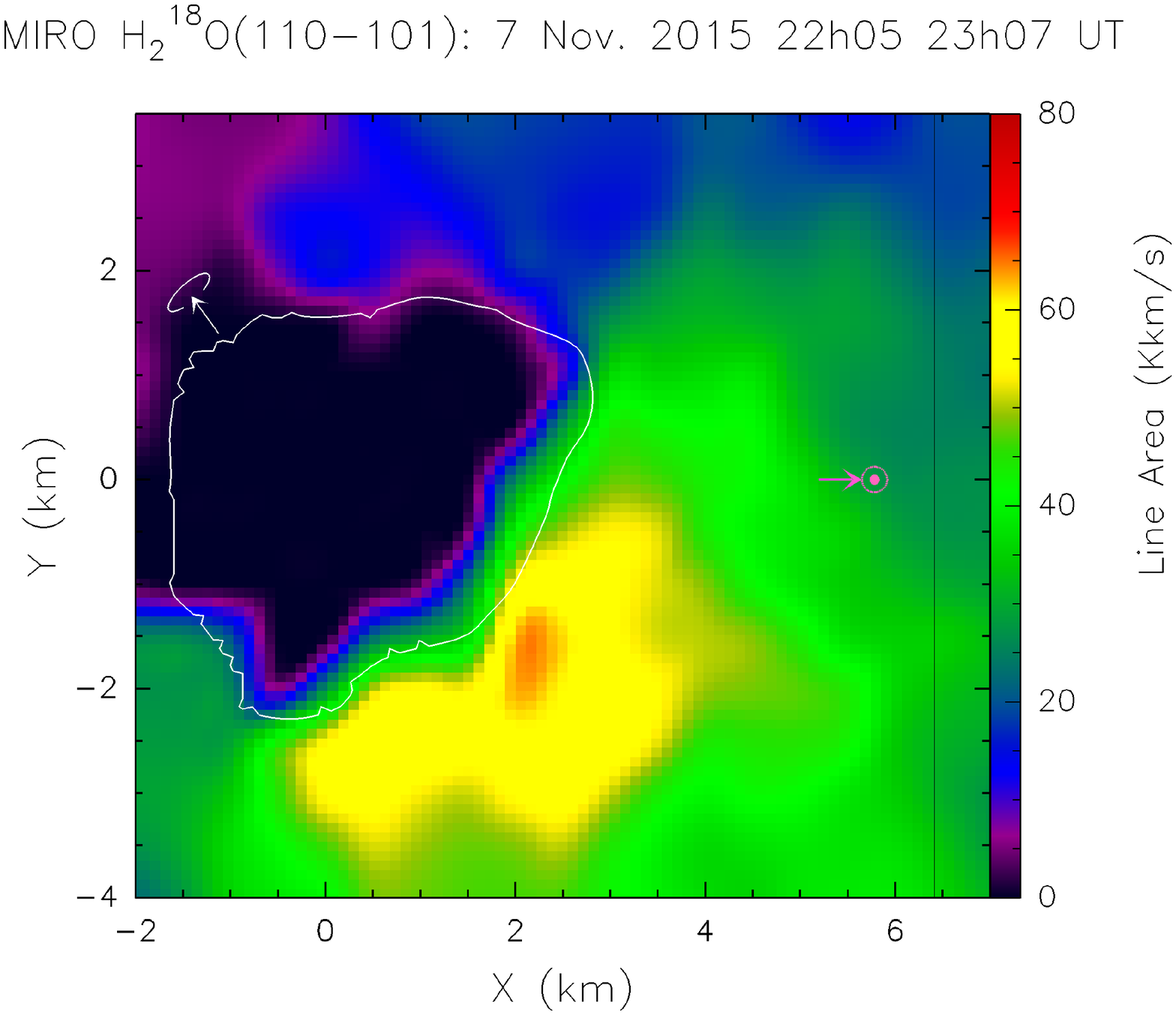} &
\includegraphics[width=0.3\textwidth,trim=3cm 1cm 3cm 1cm,clip]{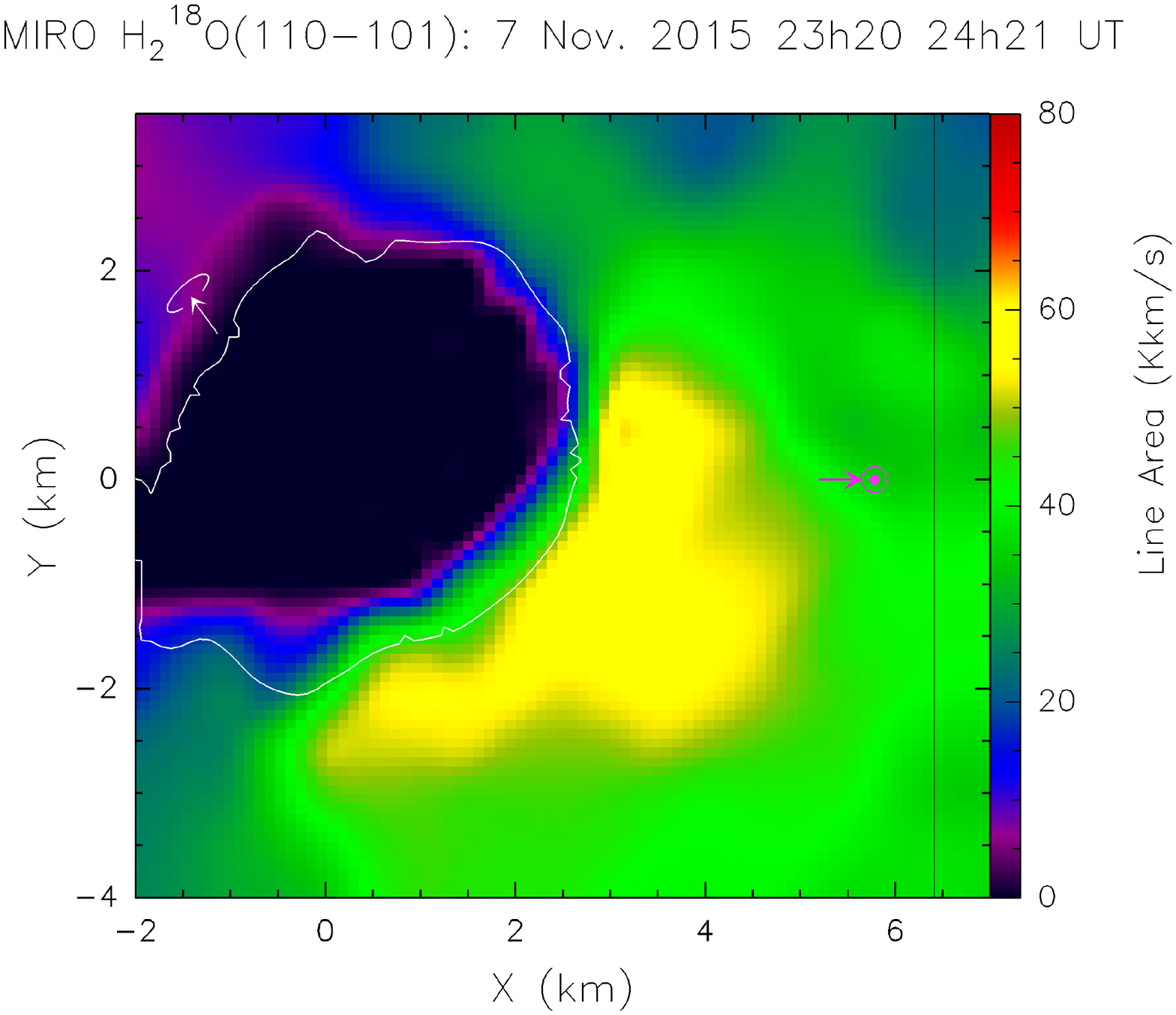} \\
\includegraphics[width=0.3\textwidth,trim=3cm 1cm 3cm 1cm,clip]{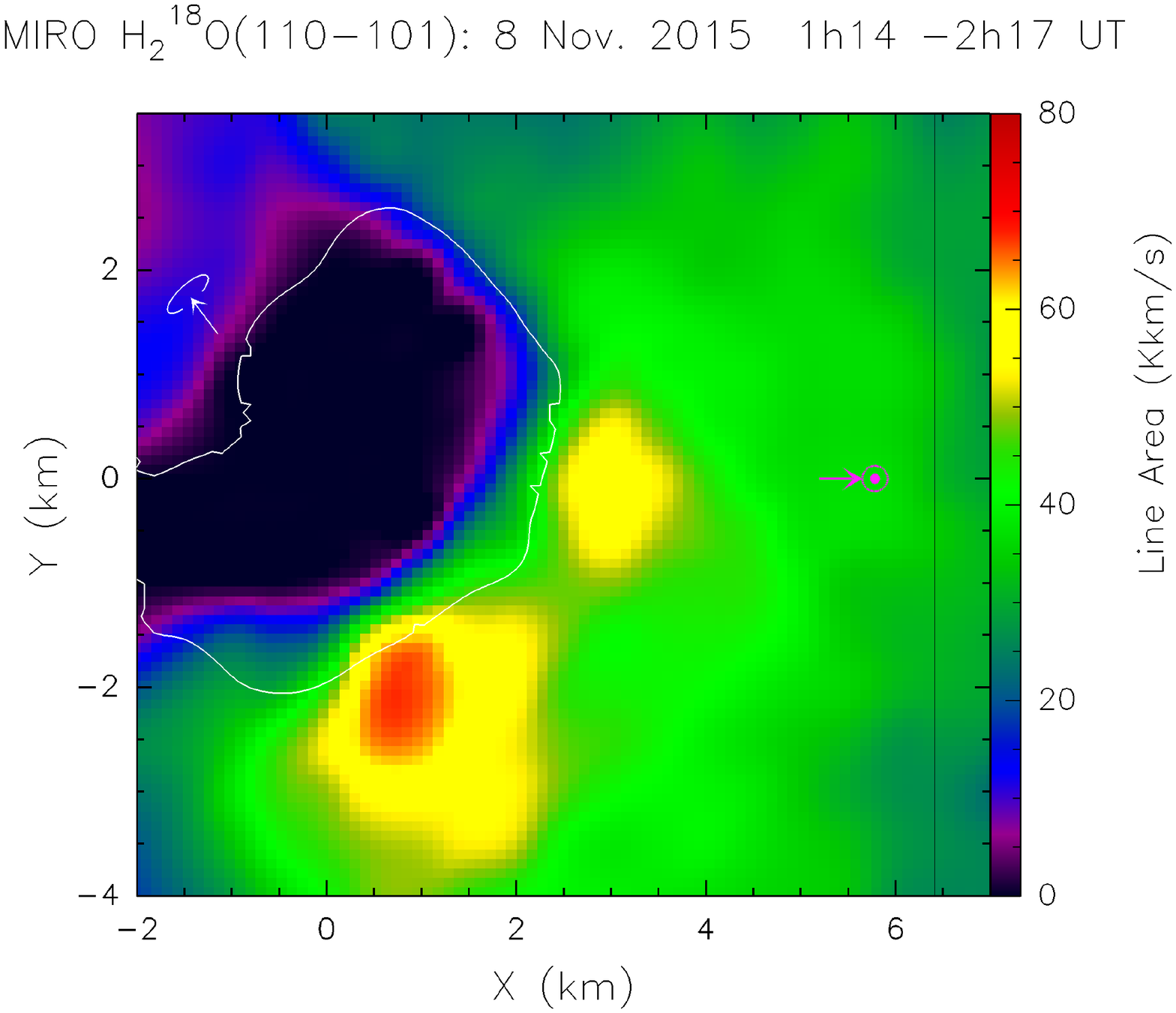}  &
\includegraphics[width=0.3\textwidth,trim=3cm 1cm 3cm 1cm,clip]{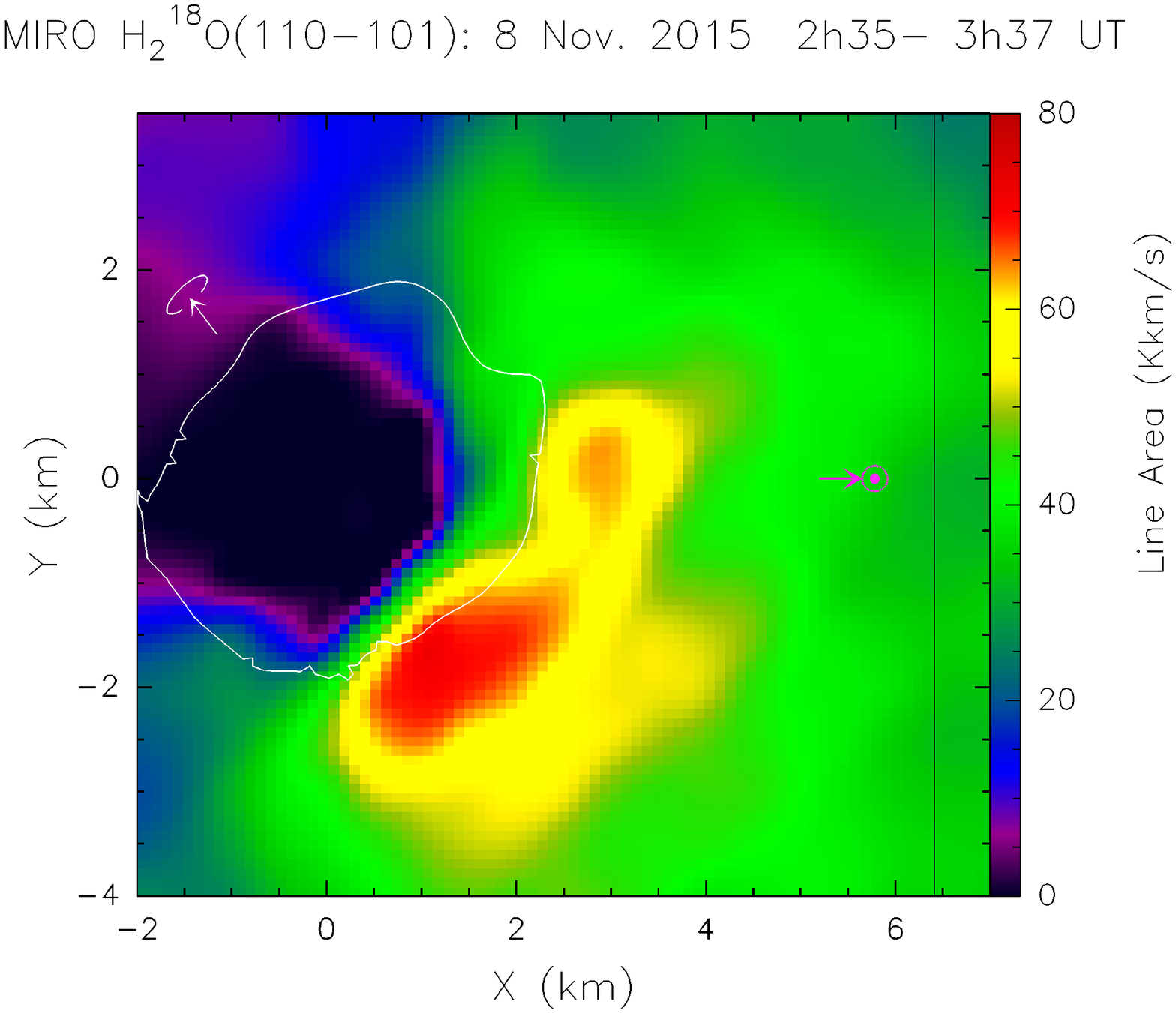}  &
\includegraphics[width=0.3\textwidth,trim=3cm 1cm 3cm 1cm,clip]{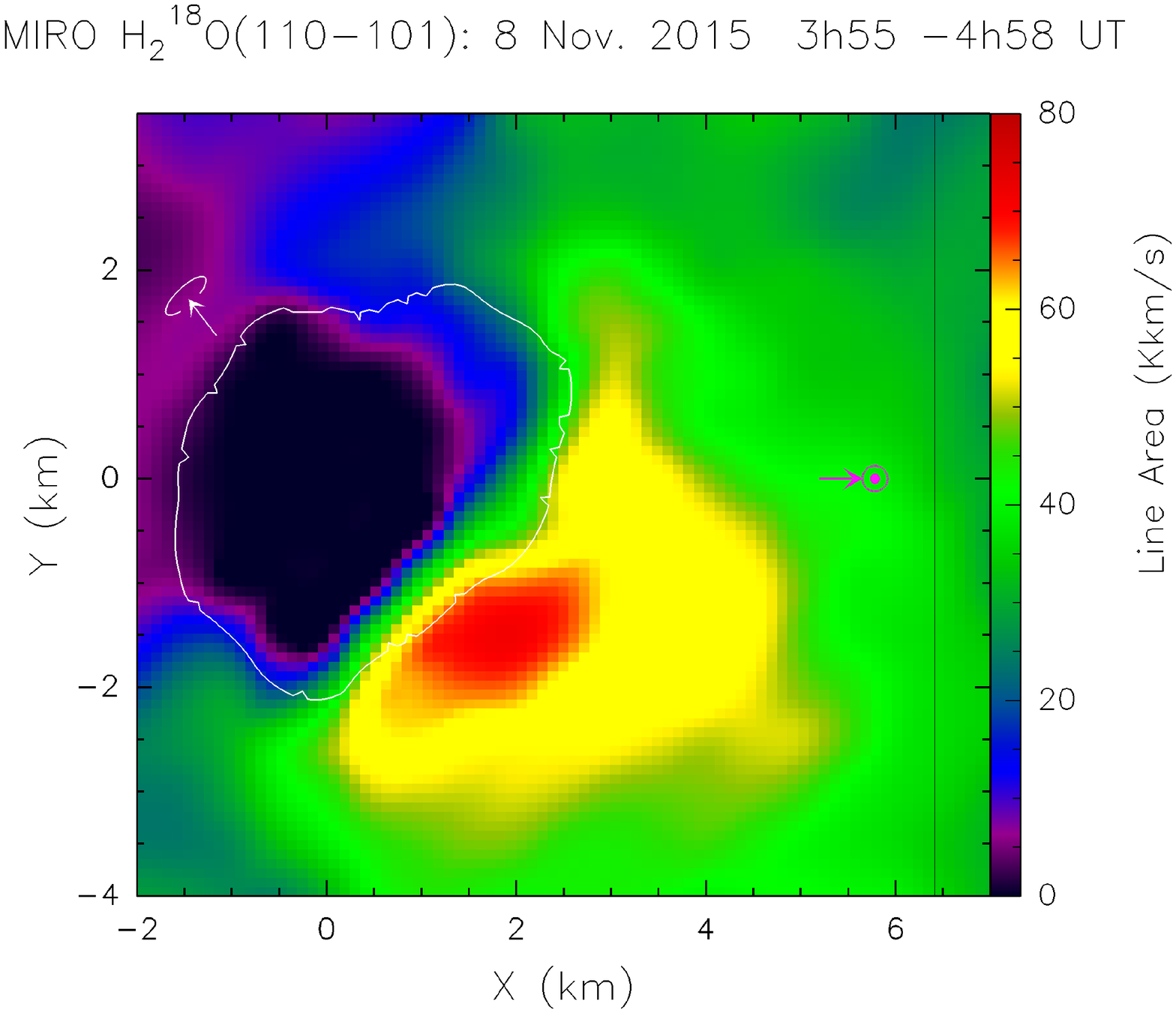} \\
\includegraphics[width=0.3\textwidth,trim=3cm 1cm 3cm 1cm,clip]{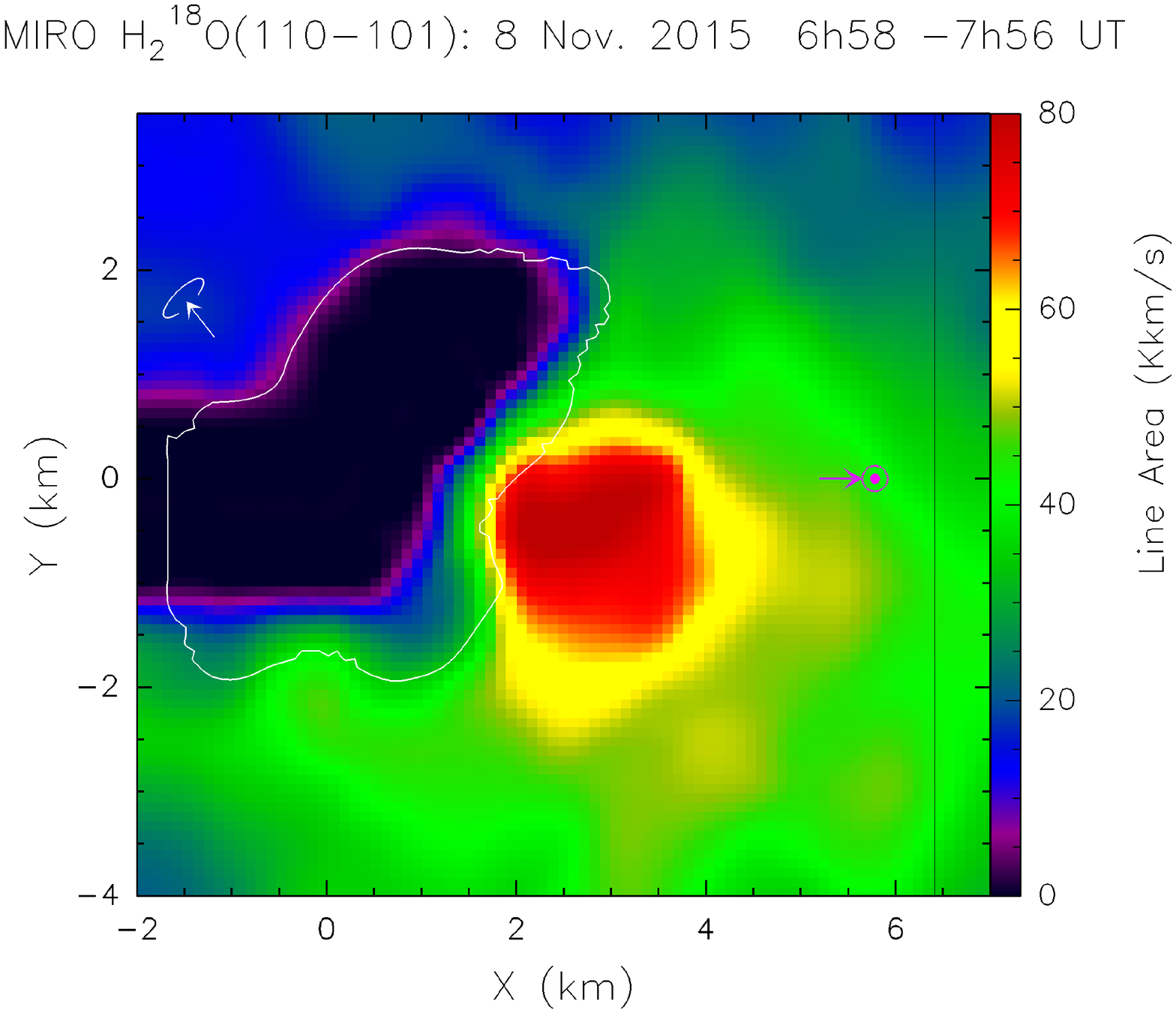}  &
\includegraphics[width=0.3\textwidth,trim=3cm 1cm 3cm 1cm,clip]{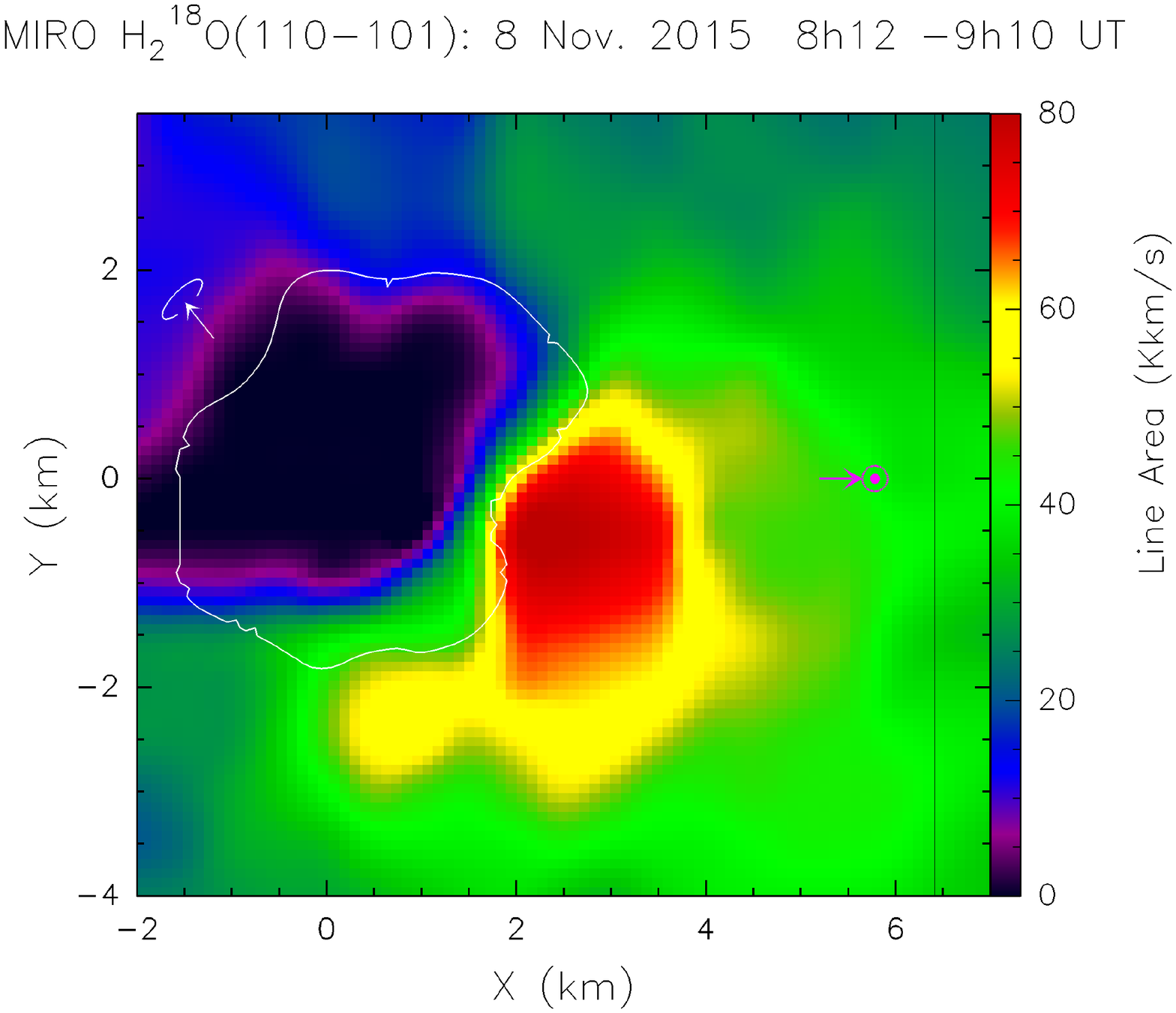}  &
\includegraphics[width=0.3\textwidth,trim=3cm 1cm 3cm 1cm,clip]{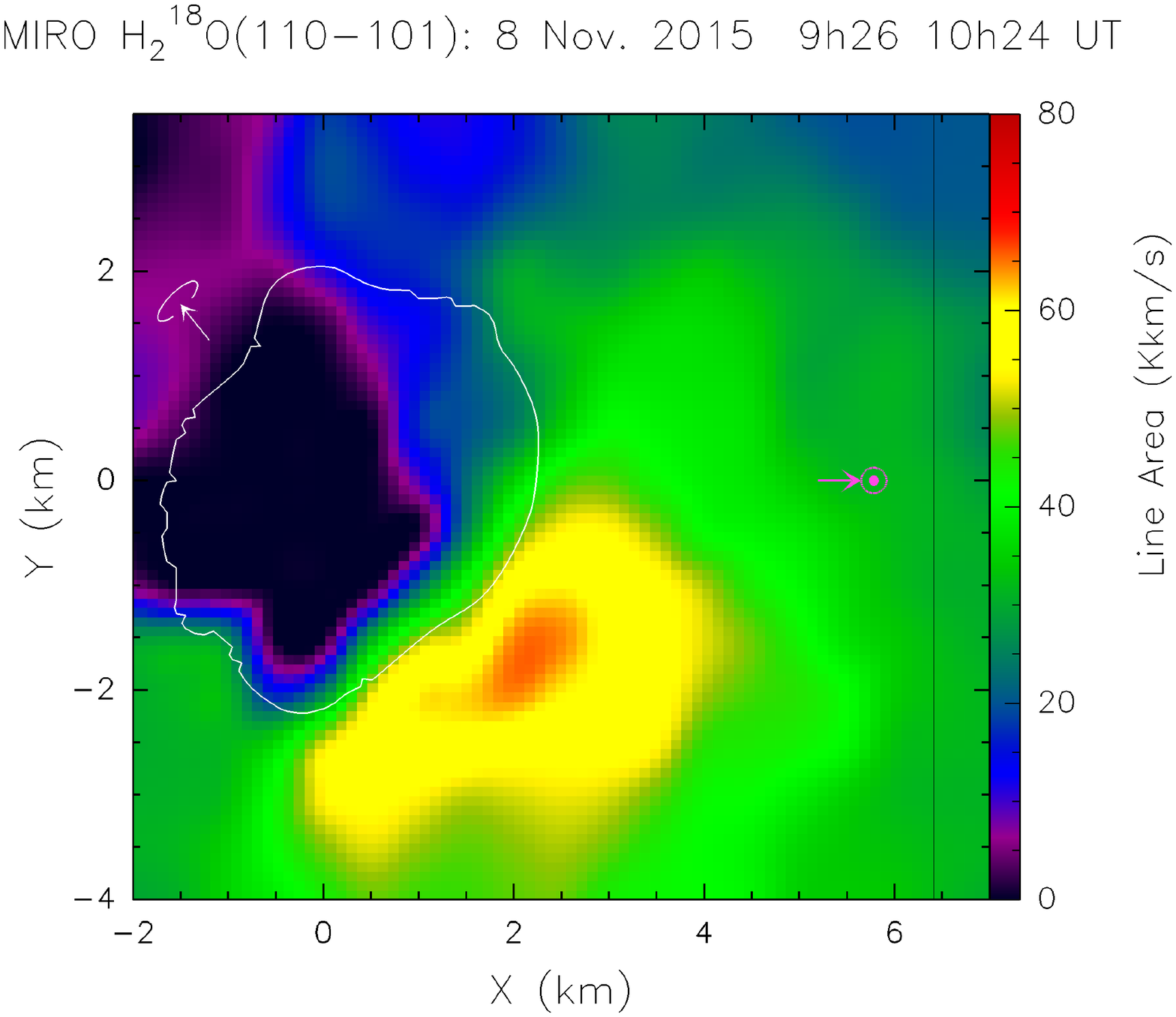} \\
\end{tabular}
}\caption{Maps derived from MIRO line areas of the (110-101) transition of \ce{H_2 ^1^8 O} for November 7 20:55 UTC to November 8 10:24 UTC. An outline of the nucleus is shown in white at the correct position and scale for the midpoint of the scan. The line is mostly in absorption against the nucleus; black regions correspond to negative line areas where water appears in absorption. Discrepancy between the white outline and black region is the result of the long duration for the raster scan to acquire the full frame, during which the nucleus will rotate. The sunward direction is to the right in each panel.}
\label{fig:MIRO Maps 1}
\end{figure*}

The data available from MIRO for this period are more sparsely sampled in terms of spatial resolution compared to VIRTIS-H, but allow for useful comparison to both the Alice and VIRTIS-H data. MIRO maps display line area in K km s$^{-1}$, derived from the \ce{H^1^8_2 O} emission feature at 547.676 GHz, which was optically thin at the time of observation (Figure \ref{fig:MIRO Maps 1}), yielding a qualitative idea of the relative water column density. Due to the short integration times and regions with mixed absorption and emission of \ce{H2O} it is difficult to create precise water column density maps as described in \cite{gulkis2015subsurface}, \cite{biver2015distribution}, and \cite{Biver2019}, with the model detailed in \cite{zakharov2007radiative} and \cite{lee2011non}. In all MIRO maps there is an increased line area in the south pole direction in agreement with the VIRTIS-H maps of the same time period. In the first three maps the integrated intensity is as high as 80 K km/s. This elevated line area is a factor of 2-3 higher than areas of the coma perpendicular to the south pole at the time of the activity increase, but the low resolution makes interpretation from these maps alone difficult. To an order of magnitude this upper value is consistent with VIRTIS-H post-perihelion water column densities of $\sim$2.7$\times$10$^{16}$ cm$^{-2}$ \citep{bockelee2016evolution, Biver2019}. All three instrument maps show evidence of large scale activity increase on the southern hemisphere that begins to decrease substantially after 1:15 UTC on November 8. 

\section{Discussion}\label{Discussion}
The addition of two-dimensional spectral maps for this period of activity provides a new analysis tool for understanding cometary activity in the UV. The alignment of the initial jet seen in the NAVCAM images with the Alice slit, the additional activity that takes place just before the raster pointing scheme, and the availability of the MIRO and VIRTIS-H datasets for context invite us to address the morphology, composition, and emission mechanisms of the cometary activity.

\begin{figure}
\centering
\gridline{\fig{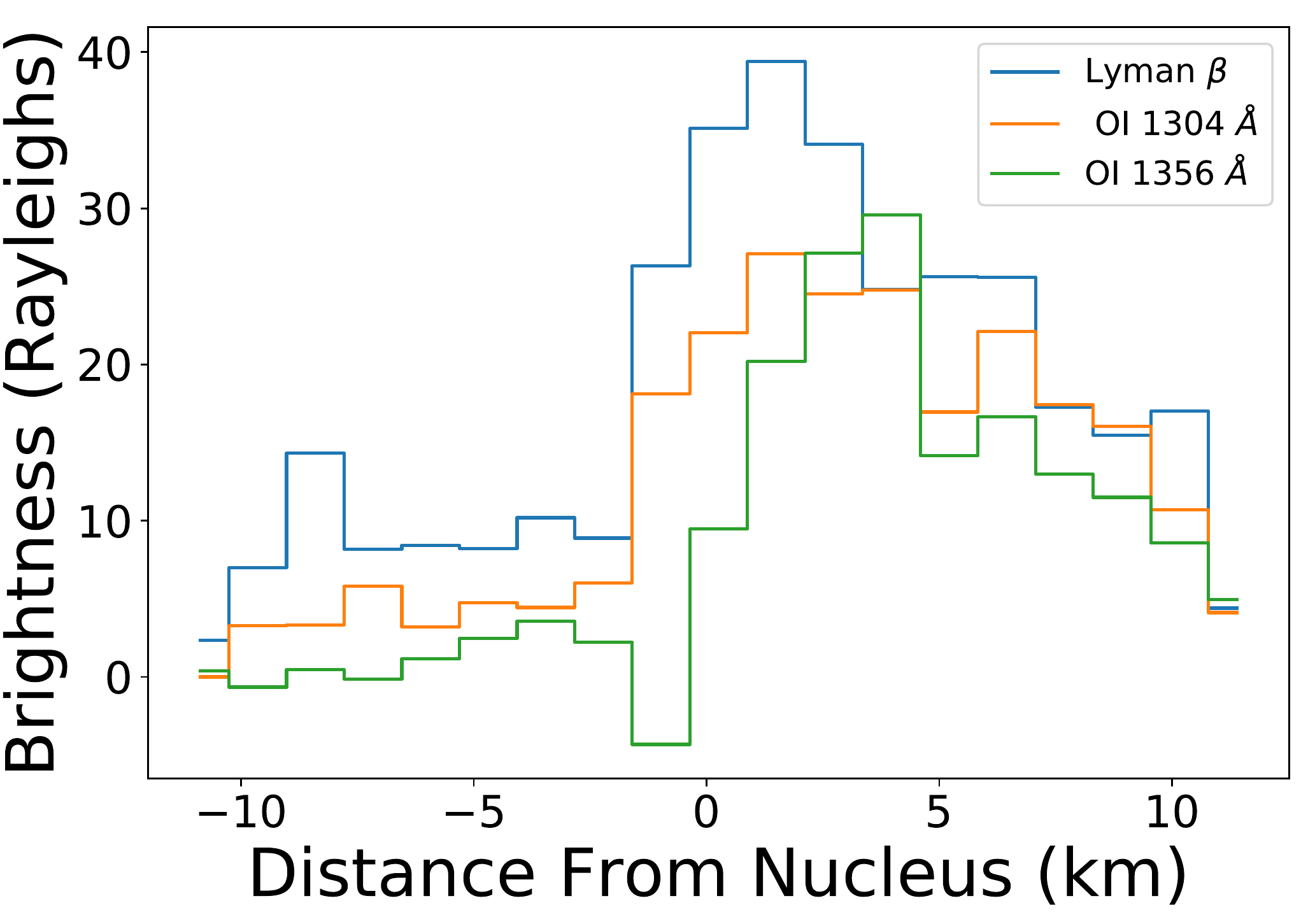}{0.5\linewidth}{2015/11/07 19:18 UTC}
\fig{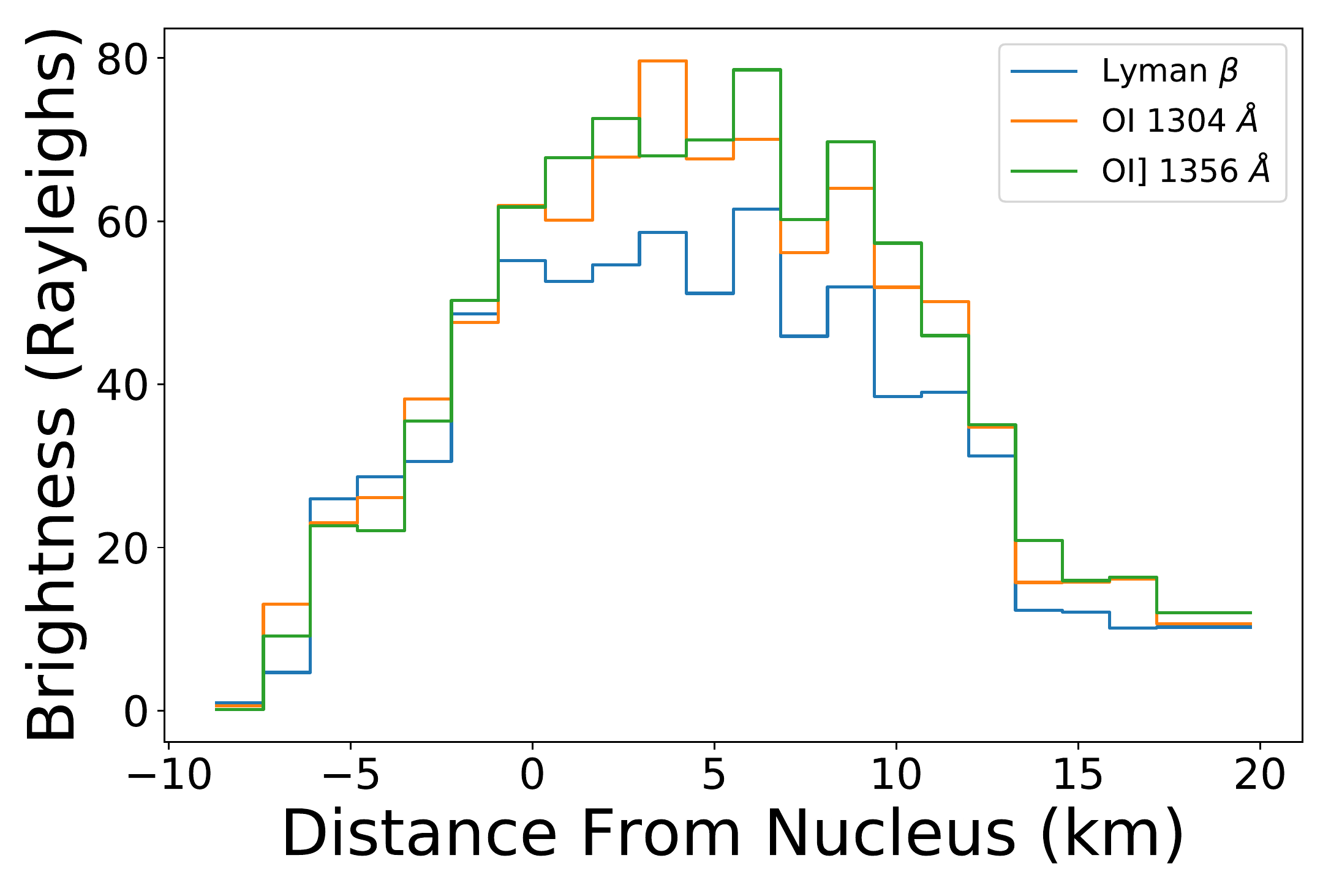}{0.5\linewidth}{Map 1, 2015/11/07 21:16-22:07 UTC}}
\caption{Spatial profiles of an earlier jet (a) and of map 1 (b). The spatial profile derived from map 1 is the X=0.5 km row of the emission map, shown in Fig. \ref{fig:map_1}. The sunward direction is in the $+$X direction.  Each detector row subtends approximately 1.2 km. Of particular interest is the extension of enhanced \ion{O}{1}] 1356 \AA\ emission to approximately 10 km in the activity maps, with brightnesses approaching 60 Rayleighs, and the slope similarity of the three emission features which indicates they share the same emission process, dissociative electron impact emission. Compare to Figure 4 of \citet{feldman2016nature}. Note that 1$\sigma$ errors on each brightness are less than 5\%. }
\label{fig:map_profile}
\end{figure}

\begin{figure}
\centering
\gridline{\fig{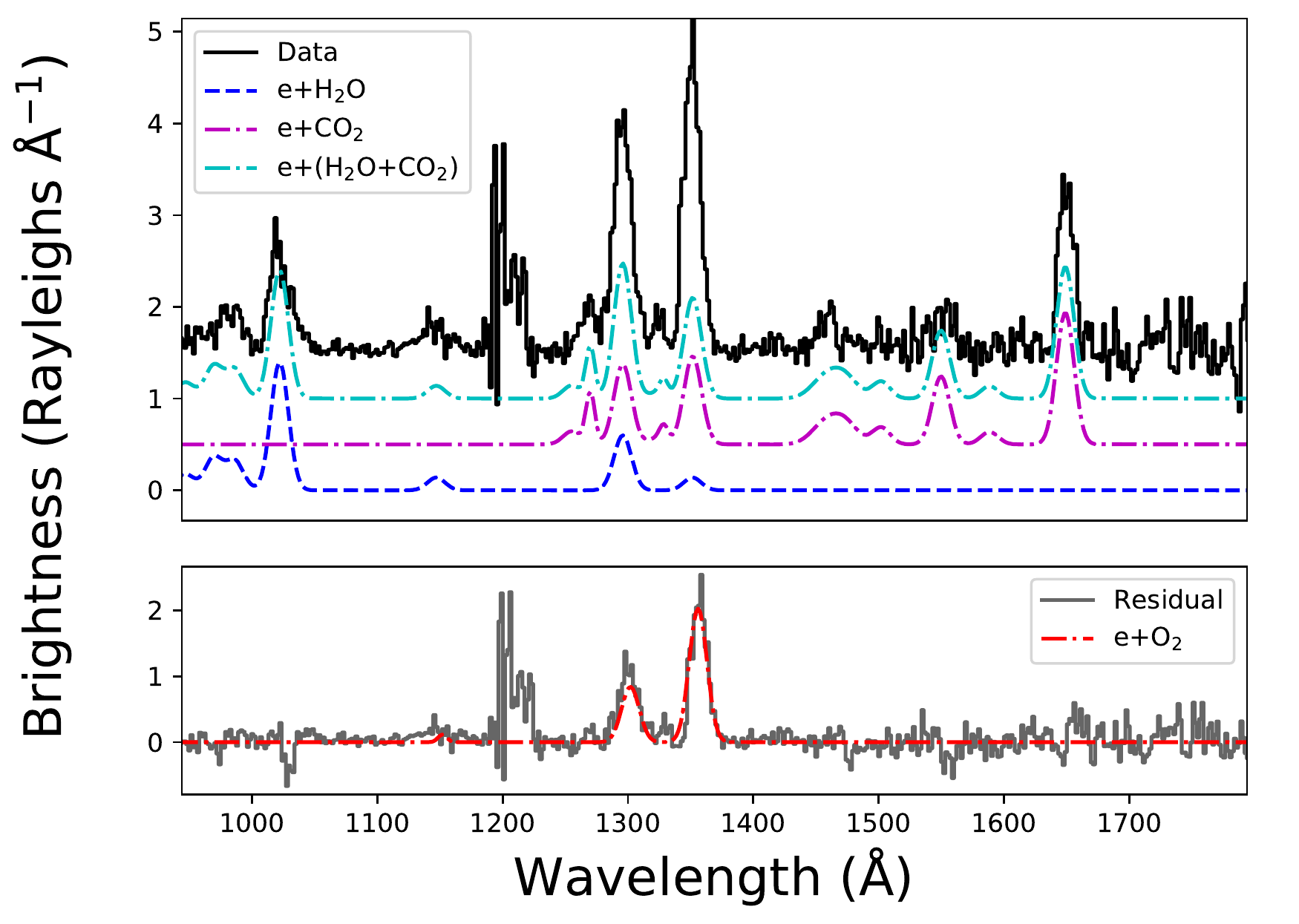}{0.5\linewidth}{a) 21:44 UTC 7 Nov.}
\fig{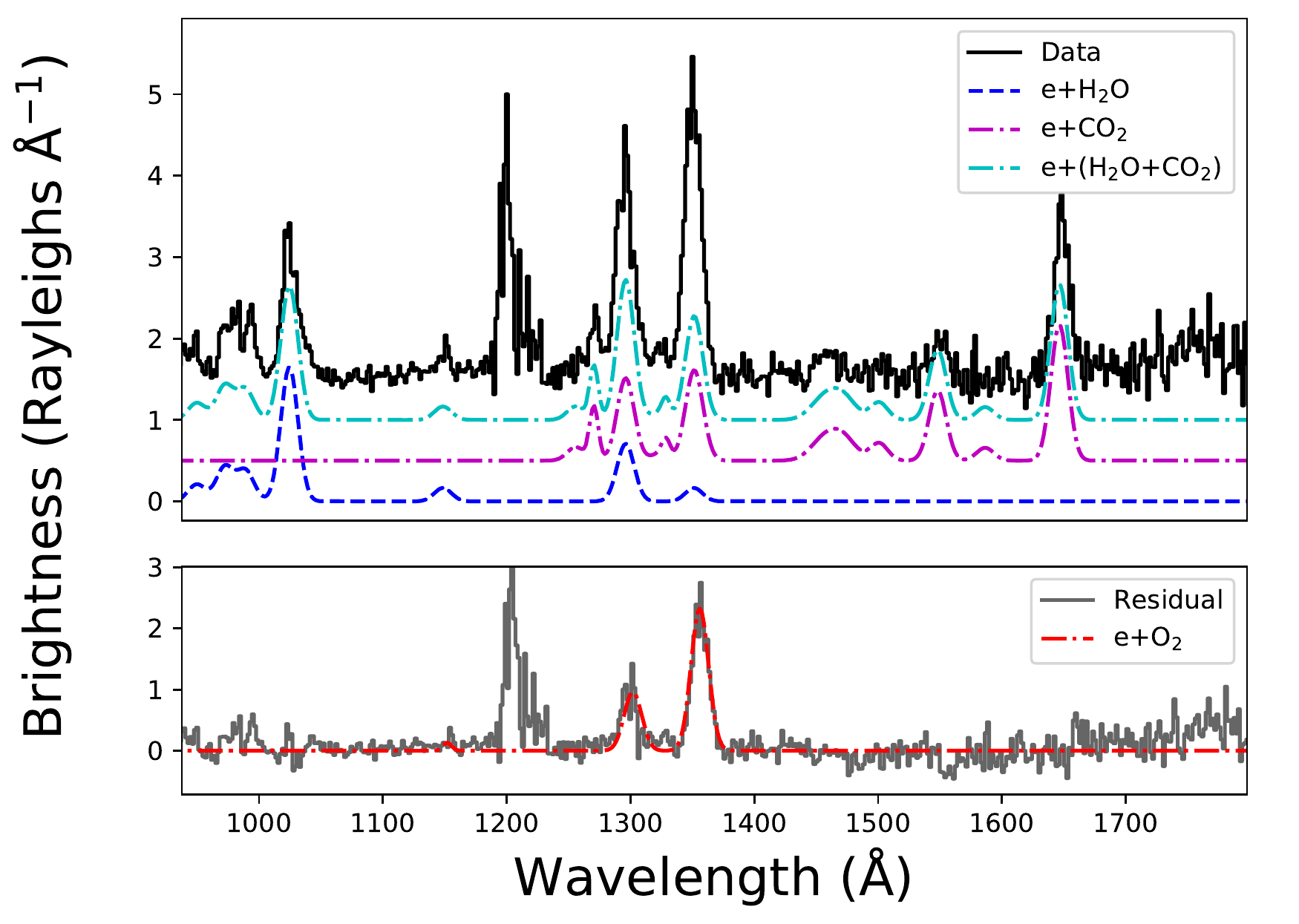}{0.5\linewidth}{b) 22:34 UTC 7 Nov.}}
\gridline{\fig{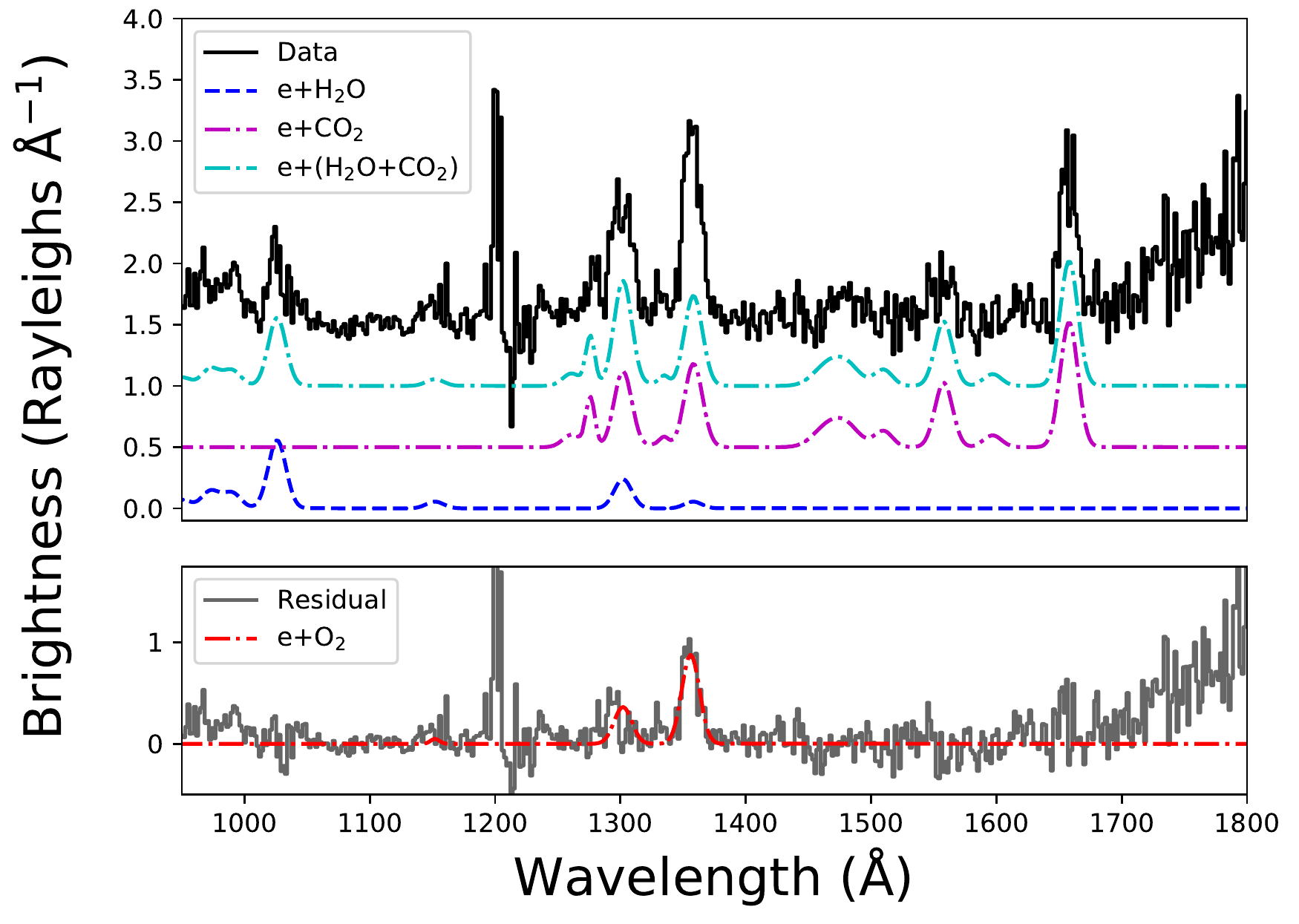}{0.5\linewidth}{c) 23:45 UTC 7 Nov.}
\fig{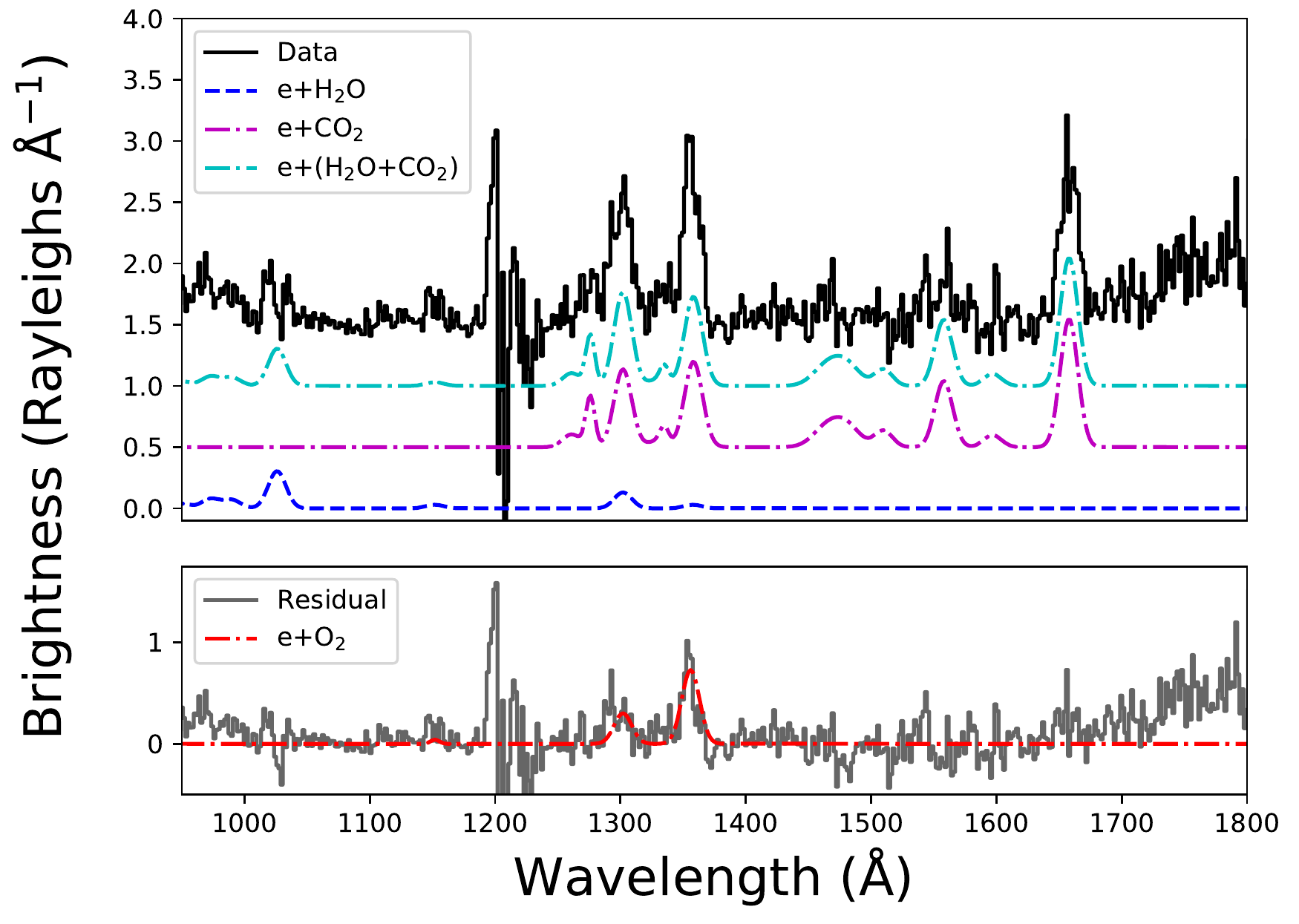}{0.5\linewidth}{d) 01:47 UTC 8 Nov.}}
    \caption{Alice quiescent-subtracted spectra showing the post-activity onset near-nucleus coma. The spectrum from 15:04 UTC shown in Figure \ref{fig:stable_spectra} is used as the quiescent spectrum. The relative abundances of \ce{CO2} and \ce{O2} are listed in Table \ref{table:compositions}. Note that Lyman-$\alpha$ emission is omitted from fitting due to gain sag of the detector. }
\label{fig:e+mol_maps}
\end{figure}
\subsection{Morphology}
Information from the NAVCAM images suggests that there was a jet active prior to the activity onset and before the raster pointing started. The Alice slit was aligned with this jet during the stare pointing, but subsequent rotation of the comet would have prevented the jet from producing the consistent morphology shown in the four instrument maps. The NAVCAM activity shown in Figure \ref{fig:NAVCAM_Jet} is seen along the Sun/comet line (taken to be 0$\degree$) as opposed to the $\sim-$40 to $-$45$\degree$ angle seen in the later Alice, MIRO, and VIRTIS-H maps. Further, the emission strengths change between the stable Alice spectra and the raster maps, from $\sim$30 Rayleighs to $\sim$50 Rayleighs, implying a change in production rate of the active area as well as location as the nucleus rotated approximately 60$\degree$ between 19:18 UTC and 21:18 UTC (Figure \ref{fig:time_vs_brightness}). 

The projection of a jet perpendicular to the Wosret region is unable to explain the emission that is seen extending along the south pole, as the region is nearly antipodal to the the instrument line-of-sight at the time of mapping (21:00 UTC and on). Barring an outburst at the south pole of 67P, which was the site of only one recorded outburst prior to November 7 \citep{vincent2016summer}, the source of the outburst would be on the southern portion of the neck or small lobe, with a normal vector that when seen in profile from \textit{Rosetta} appears to be parallel to the south pole of 67P. Given these constraints we can discuss several candidate areas that have been identified as possible outburst locations and the major clues favoring them.

The decline in elevated emissions after Alice observations restarted at 21:16 UTC suggests that a sharp increase in activity had occurred during the Alice observation gap. During this observing gap (19:18-21:16 UTC) a small area, previously in shadow owing to an overhang in the Neith region \citep{el2016regional}, is fully illuminated. This overhang area previously had an outburst identified in \citet{vincent2016summer} (outburst 28 in Table 1) that saw a relative increase in brightness of 10\% on 2015 September 10, just under two months before the activity increase discussed in this paper. However, five outbursts were identified with similar timestamps on that date in Sobek and Anukhet as well, making it difficult to attribute any observed ultraviolet emission in Alice observations on that day to a specific region \citep{Fornasier2019b}. There are two outburst sites located farther south in the Geb region that produce outburst vectors that do not have a south pole parallel component and will be eliminated as a possibility. The cluster of possible outburst sites is contained between 270$\degree$ and 315$\degree$ longitude and $-$15$\degree$ to $-$45$\degree$ in latitude and experienced local noon at the time of the activity increase. We note that the possible outburst sites are between 60$\degree$ and 120$\degree$ west of the likely outburst sites identified by VIRTIS-M in \citet{Noonan2021hybrid}. 

We cannot rule out that the activity that occurred during the gap period may be due to rapid sublimation driven by the quick change in illumination of Neith region, which was shielded by an overhang on Wosret prior to the Alice observing gap. The vector normal to this region would extend at an angle approximately 35-40$\degree$ clockwise from the Sun-comet line as seen from \textit{Rosetta}. However, the direct comparison between the jets and the later activity shown in Figure \ref{fig:map_1} is a strong indicator that the activity drove the electron impact emission dominated region surrounding the nucleus three times farther from its boundaries at 15:04 UTC, which is difficult to explain with insolation-driven sublimation alone. 


 We can isolate spatial profiles from the activity emission maps (Figure \ref{fig:map_profile}), which we can then compare to the earlier observations from 19:18 UTC (See \citet{Noonan2021hybrid}). The spatial profile from the X=0.5 km row of the observing block 2 emission map shows significant emission of Lyman-$\beta$, \ion{O}{1} 1304, and \ion{O}{1}] 1356 \AA\ between 0 and 13 km in the sunward direction from the nucleus. All three emission features show a similar slope with respect to distance from the nucleus, and the presence of \ion{O}{1}] 1356 \AA\ at or above a 1:1 ratio with \ion{O}{1} 1304 \AA\ suggests dissociative electron impact as a dominant mechanism out to 13 km from the nucleus, distinctly farther than in the earlier profiles (Figure \ref{fig:map_profile}a). 

\subsection{Composition}
 Ideally we would be able to derive the composition using a method similar to \citet{galand2020far} and \citet{stephenson2021multi}; combining IES electron data with measured water column densities from MIRO or VIRTIS and relative coma abundances of \ce{CO2}, \ce{O2}, and \ce{CO} from ROSINA. However, this same method is unfortunately not applicable for the set of observations considered in this paper for numerous reasons: the nucleus is slightly more active, outbursts (and therefore rapid changes to the near-nucleus coma composition) are prevalent, the spacecraft was over twice as far away from the nucleus, precise water column measurements are unavailable from VIRTIS and MIRO, and it's unclear if the accelerated solar wind electrons are still the dominant source for the inner coma during outburst emissions. The change in atomic line ratios is a clear indicator that the the composition of the near-nucleus coma has changed, as this is not possible through changes in plasma energy or density alone. This leaves us with two options: assume the plasma density and distribution along the Alice line of sight and use the absolute dissociative electron impact cross sections from the literature to derive column densities of neutrals, or fit the relative intensity spectra from dissociative electron impact excitation of \ce{H2O}, \ce{CO2}, and \ce{O2} to the observations and derive relative abundances of the neutrals. Due to the large distance from the nucleus during this period, and the possibility that the spacecraft was looking through different plasma environments on different sides of the diamagnetic cavity \citep[]{madanian2016plasma}, we will use the latter method.
 
Compositional analysis of several key times during observing blocks 1 and 2 was completed to determine the dominant components of the initial jet and activity increase using spectra that had been background and quiescent-subtracted using the quiescent average spectrum from earlier on November 7 shown in Fig. \ref{fig:stable_spectra}. The remaining emission lines show significant dissociative electron impact emission that can be fit with relative intensity spectra using laboratory electron impact emission data for \ce{H2O}, \ce{CO2}, \ce{CO}, and \ce{O2} \citep{makarov2004kinetic,ajello2019uv,mumma1972dissociative, ajello1971dissociative, ajello1971emission,kanik2003electron}. As described in detail by \citet{stephenson2021multi} the excitation cross sections for each emission feature as a function of energy are limited, but every emission feature in our data is characterized at electron impact energies of 100 eV. By fitting the 100 eV electron impact \ce{H2O} and \ce{CO2} models to Ly-$\beta$ and \ion{C}{1} 1657, 1561, and 1278 \AA\ as well as \ion{C}{2} 1335 \AA\ emissions first and subtracting off both molecules' contributions to the \ion{O}{1} 1304 and 1356 \AA\ emission features, the 100 eV \ce{O2} electron impact spectrum can be fit to the residual (Figures \ref{fig:e+mol_maps}). In the absence of in-plume plasma energy distribution data the assumption that the atomic cross section ratios taken at 100 eV prevail at the range of electron energies above their threshold energies must be made in order to compute these relative abundances. This assumption was used by \citet{feldman2015measurements}, but comes with the errors on those measurements in addition to the uncertainties introduced by any deviation from the assumed constant ratio in the experimental data. Based on the dissociative electron impact excitation rates as a function of energy for \ce{H2O}, \ce{CO2}, \ce{CO}, and \ce{O2} \citep{makarov2004kinetic,ajello2019uv,mumma1972dissociative, ajello1971dissociative, ajello1971emission,kanik2003electron} we find this value to be between 5 and 20\%, depending on which transitions are compared. As noted in Table 1 of \citet{stephenson2021multi} many of these features are assumed to have the same shape as other measurements in the source literature (ex. \ion{O}{1} 1304 and \ion{O}{1} 1356 \AA\ in \citet{makarov2004kinetic}), which would render our assumed constant ratio, somewhat optimistically, perfect. We will thus taken the upper limit error of 20\%, and sum in quadrature with the known error in the excitation measurements at 100 eV to find that our relative abundance calculations should be within $\pm$30-36 \% of the actual relative abundance. For transitions with significantly higher threshold energies, like \ion{C}{2} 1335 \AA, a scaling factor has been added to the fitting algorithm. Physically this represents the depletion of electrons with energies over 44 eV from the population of electrons over the lower threshold energies of the other transitions, which are between $\sim$15 and 25 eV. For example, if all electrons in the environment were $\geq$ 44 eV, this factor would be 1, and no depletion of the \ion{C}{2} 1335 \AA\ feature relative a ratio with \ion{C}{1} 1657 \AA\ expected at 100 eV would be present. The scaling factors are allowed to vary between 0 and 1, with the model finding fits between 0.1 and 0.25 to produce the observed emission, indicating that the near-nucleus environment has a significantly lower number of electrons over 44 eV than the number of electrons over 15 eV. This parameter does not steer the relative abundance model, which is driven by the \ion{C}{1} 1657 \AA: Ly-$\beta$ ratio to determine \ce{CO2}/\ce{H2O}, but is interesting to note.  
\begin{table*}
\begin{center}
\begin{tabular}{c c c c c c}
\hline
Observation ID & observing block & UTC Time & \ce{CO2}/\ce{H2O} & \ce{O2}/\ce{H2O} \\
\hline
ra\_151107214444\_hisa\_lin & 2 & 21:44 & 0.6 & 0.3\\
ra\_151107223043\_hisa\_lin & 3 & 22:30 & 0.7 & 0.3 \\
ra\_151107234505\_hisa\_lin & 4 & 23:45 & 1.2 &  0.3\\
ra\_151108014708\_hisa\_lin & 6 & 01:47 & 2.4 & 0.4 \\

\end{tabular}
\caption{Modelled abundances for \ce{CO2} and \ce{O2} relative to \ce{H2O} derived from quiescent-subtracted spectra shown in Figure \ref{fig:e+mol_maps}. Errors for \ce{CO2}/\ce{H2O} are approximately 30\% for observing blocks 2, 3, and 4 and closer to 45\% for observing block 6 due to the decreased signal in Lyman-$\beta$. Errors for \ce{O2}/\ce{H2O}} are higher due to errors induced by residual subtraction and are near 40-45\% for observing blocks 2, 3, and 4 and closer to 55\% for observing block 6. We conservatively adopt the error to be 50\% for observing blocks 2, 3 and 4, accordingly. The errors on the ratios are calculated from the root mean square of the error in the data, fitting error, and error in laboratory measurements of the emission line used for relative intensity calibration.  These are added in quadrature with the deviation in line ratios as a function of impacting electron energy. The last two sources of error are the same for each observation.	
\label{table:compositions}
\end{center}    
\end{table*}

 The observations at 21:44 UTC, in observing block 2, yield a spectrum in rows 18-23 with a small change in \ce{CO2}/\ce{H2O} composition relative to the earlier spectrum at 19:18 UTC, resulting in \ce{CO2}/\ce{H2O} = 0.6$\pm$0.2, a 1.2-$\sigma$ difference from the value of 1.2$\pm$0.4 measured earlier during jet activity at 19:18 UTC (\citet{Noonan2021hybrid}). These values are somewhat higher than the typical coma values between 0.3-0.5 observed by the VIRTIS instrument \citep{bockelee2016evolution}. 

The strengths of \ion{C}{1} 1561 and 1657 \AA\ are slightly inconsistent with e+\ce{CO2} as the sole source of the atomic carbon emissions. For e+\ce{CO2} the expected line ratio would be 2:1 for \ion{C}{1} 1657:1561 \AA\ \citep{ajello2019uv}, but the value in these data is often near 3 following quiescent spectrum subtraction (Fig. \ref{fig:e+mol_maps}). This excess is most clearly seen in Figure \ref{fig:e+mol_maps} a) and b), where the electron impact fit has difficulty matching both emission features with the e+\ce{CO2} model. Given that e+\ce{CO} is not a significant source of either \ion{C}{1} 1657 or 1561 \AA\ in this particular dataset, this discrepancy is likely associated with a significant column of C atoms. We can quantify the necessary atomic carbon column by multiplying the \ion{C}{1} 1561 \AA\ emissions by 2, subtracting the resulting value from the measured \ion{C}{1} 1657 \AA\ emission, and using the $g$-factor, or fluorescence efficiency, for the \ion{C}{1} 1657 \AA\ feature to calculate the requisite atomic carbon column. For the spectra displayed in Figure \ref{fig:e+mol_maps} a) and b) we find residual emissions of $\sim$5 Rayleighs, and an atomic carbon column density of approximately 5$\times$10$^{11}$ cm$^{-2}$ is required to explain the observed line ratios.The source of this atomic carbon column is likely caused by photodissociation of \ce{CO2}. To explain the residual atomic \ion{C}{1} 1657 \AA\ emission with a combined photodissociation and excitation rate for \ion{C}{1} 1657 \AA\ from \ce{CO2} at 1.6 au of 5.3$\times$10$^{-9}$ s$^{-1}$ \citep{robert1978atomic} requires a \ce{CO2} column of $\sim$1$\times$10$^{15}$ cm$^{-2}$, approximately 10\% the water column measured by the VIRTIS instrument three days later on 8 November 2015 \citep{Biver2019}. This comparison is not robust, as our determination is from quiescent subtracted spectra in an actively outbursting coma, while the later VIRTIS measurements are not. However, this comparison does show that the derived \ce{CO2} column to explain excess atomic carbon emissions is on the same order of what would be expected.

By fitting an e+\ce{O2} spectrum to the quiescent- and e+(\ce{H2O}+\ce{CO2})-subtracted spectra from the secondary activity from observing block 2, we are able to calculate a relative abundance of \ce{O2}/\ce{H2O} of 0.3 ($\pm$50\%). This is elevated with respect to the early mission relative abundance for \ce{O2}/\ce{H2O} of $\sim$0.05 described by the ROSINA team \citep{bieler2015abundant}, but falls within the range of 0.05-0.3 reported in other Alice observations \citep{feldman2016nature,noonan2018ultraviolet,keeney2017h2o,keeney2019occultation}. The appearance of strong \ion{O}{1} emission has previously been attributed to gaseous outbursts, and we cannot rule out an outburst occurring during the gap in observations as the cause of the activity change as discussed in Section \ref{lightcurves}. Notably, the residual spectrum at 19:18 UTC of the jet does not show similar \ion{O}{1} emission, and is poorly fit by \ce{O2}/\ce{H2O} $\sim$0.1 \citep{Noonan2021hybrid}, the detection limit for the routine. This is determined by adding a synthetic e+\ce{O2} spectrum to Alice data and attempting to retrieve that signal, with a \ce{O2}/\ce{H2O} ratio of 0.1 being the lowest for the data in question. This suggests that the period captured in the Alice, VIRTIS, and MIRO is new onset activity, compositionally distinct from earlier sustained activity. 

If the outgassing velocity of neutrals is assumed to be $\sim$600-800 m s$^{-1}$, as modeled in \citet{fougere2016direct} and \citet{lai2017gas}, neutrals would have crossed the slit in approximately 15 seconds, much less then the duration of an Alice exposure at this time. This implies that there was continuous outgassing of \ce{O2} and other volatiles for this period. However, it is important to note that this slit-crossing timescale may vary by as much as a factor of two during the observations due to the scanning motion of the slit. This same scanning behavior makes it difficult to co-add spectra reliably to lower the upper limit of detection. 

\subsection{Excitation Processes}\label{subsec:ex_pro}
The extension of the electron impact environment out to distances of 13 km from the nucleus during the raster despite the lack of large changes in measured molecular emissions from MIRO and VIRTIS-H, and the lack of similar extension during the earlier Alice jet observations, is intriguing. Electron impact is important in the near-nucleus coma, but the changes in spatial profiles for \ion{O}{1}] 1356 \AA\ between Figures \ref{fig:map_profile}a) and \ref{fig:map_profile}b), which can be interpreted as an indicator of the dissociative electron impact dominated region, then suggest an increase to dissociative electron impact emission without substantial changes to molecular column density. However, it is difficult to explain the changes in relative molecular abundances if only the near-nucleus plasma has experienced increases in average energy and/or density, as this does would not drastically change the expected emission line ratios. In short, dissociative electron impact emission in the jet only appears to be dominant out to 5$\pm$1.2 km while in the spatial profile taken from the raster maps it extends all the way out to 13$\pm$1.3 km. All 3 emission features shown in Figure \ref{fig:map_profile}b) show the same sharp decrease at this point, similar to the 15 Rayleigh drop for \ion{O}{1}] 1356 \AA\ shown in Figure \ref{fig:map_profile}a). Further interpretation is difficult owing to the odd/even effect of the Alice detector, but it is clear that the profile of the cometary jet does not match that of the later activity. Spectroscopically and compositionally the activity appears to be more similar to outburst B identified earlier on November 7 \citep{Noonan2021hybrid}, while the extension out to 13 km also suggests changes to the near-nucleus plasma environment that extended the typical region for dissociative electron impact. Without plasma data from that region of the coma it is difficult to disentangle these combined effects. 

Further, it is useful to compare the activity in question to the dusty outburst of 2016 February 19 discussed in \citet{grun20162016} and \citet{hajraimpact}. \cite{hajraimpact} found that the plasma environment experienced a 3-fold increase in electron density while simultaneously experiencing a 2-9 fold decrease in electrons with energies over 10 eV. In the case where neutral density remains constant while the average energy decreases and the ``cold" electron density increases would result in the calculation of a lower limit for column density if the 100 eV cross sections from the literature are implemented. These changes to the plasma environment would produce a dramatic decrease in dissociative electron impact excitation in Alice data, providing a qualitative case study to compare with data taken on November 7-8. 
Referring back to Figure \ref{fig:time_vs_brightness}, we see that the semi-forbidden \ion{O}{1}] 1356 \AA\ emission remains elevated by a factor of five relative to quiescent values until 02:00 UTC on November 8. The observed activity may have a higher energy and/or density electron distribution than the earlier quiescent period. When compared to the dusty outburst discussed in \citet{hajraimpact} the observed activity spectra is inconsistent with an order of magnitude decrease in electron energy, especially given the presence of high-threshold energy \ion{C}{2} 1335 \AA\ emission at 21:44 UTC. This, in tandem with the dust outburst observations in \citet{Noonan2021hybrid}, suggests that gas and dust outbursts experience different levels of dissociative electron impact emission and may have unique spectral signatures in the UV. Even without direct observations of the dust contribution during this period of high activity Alice observations are not consistent with a dusty outburst or outbursts as the driver of activity.

\section{Summary}\label{Summary}
Cometary activity in 67P/Churyumov-Gerasimenko was observed by Alice, VIRTIS-H, and MIRO instruments on November 2015 7-8. Over the course of 12 hours Alice observations were taken in a ride-along scheme during a series of VIRTIS raster observations. In this paper we discussed the following results:
\begin{enumerate}
    \item Alice spectra acquired during the stable observations show an elevated level of dissociative electron impact emission relative to quiescent activity levels, evident from an \ion{O}{1}] 1356/\ion{O}{1} 1304 \AA\ brightness ratio of $\sim$1, and have different spectral signatures than cometary activity measured just prior to the start of the raster observations.
    \item Observations taken during the raster scan show a similar composition to those taken during the earlier outbursts of the stare scheme, with a moderate \ce{CO2}/\ce{H2O} ratio of $\sim$0.6-1.2 and significant presence of \ion{O}{1}] 1356 \AA\ emission out to 13 km, indicative of an extended dissociative electron impact dominated emission region. Due to the large spacecraft - comet distance ($\sim$230 km) we are unable to determine if the extension of this region is due to enhanced neutral column density, electron density, electron energies, or a combination of the three. 
    \item A spectrum taken at the peak emission period at 21:44 UTC on November 7 shows a relative abundance for \ce{O2}/\ce{H2O} of $\sim$0.3 ($\pm$50\%), which is increased relative to both earlier outbursts at 16:07 and 17:32 UTC and the cometary activity at 19:18 UTC. The \ce{O2}/\ce{H2O} ratio remains elevated until 01:47 UTC on November 8, at which point a decreased Ly-$\beta$ signal makes it difficult to determine the relative abundance.
    \item Alice, MIRO, and VIRTIS-H maps all show significant atomic and molecular emissions from the southern hemisphere of 67P between 21:00 UTC on November 7 and 02:00 UTC on 2105 November 8. The direction of activity observed in the Alice maps, between 35-40$\degree$ clockwise from the Sun-comet line, is consistent with that observed in the MIRO and VIRTIS-H maps. 
   \item Alice \ion{C}{1} 1657 \AA\ and VIRTIS-H \ce{CO2} maps between 21:00 and 23:00 UTC on November 7 show similar morphology, and given the reasonable, though not perfect, fit of e+\ce{CO2} to Alice spectra we find that \ce{CO2} is likely the dominant contributor to carbon emission features at this time. 
\end{enumerate}   
   The addition of two-dimensional emission maps to the Alice dataset provides a useful tool for understanding the near-nucleus coma environment and the regions where emission mechanisms dominate, and future work will apply this same mapping method to other similar raster observations in the Alice dataset. Future UV comet observations capable of detecting the inner coma region may produce similar observations and allow comparisons to a wider range of comets and activity, ultimately improving interpretation of observations with larger spatial scales. 

\begin{acknowledgements}
\textit{Acknowledgements.} This work was made possible thanks to the ESA/NASA \textit{Rosetta} mission with contributions from ESA member states and NASA. The Alice team would like to acknowledge the support of NASA's Jet Propulsion Laboratory, specifically through contract 1336850 to the Southwest Research Institute. Parts of this research were completed by LESIA, Observatoire de Paris, with financial support from CNRS/Institut des sciences de l'univers. A component of this research was carried out at the Jet Propulsion Laboratory, California Institute of Technology, under a contract with the National Aeronautics and Space Administration. JWN would like to acknowledge Peter Stephenson for his helpful discussions regarding dissociative electron impact modeling. The team would also like to acknowledge the anonymous reviewer for initiating insightful discussions within the team that strengthened the paper. 
\end{acknowledgements}

\bibliographystyle{aasjournal}

\bibliography{uv_mapping}

\end{document}